%% file: threehadrons.tex
\documentclass[a4paper,11pt]{article}
\pdfoutput=1 \usepackage{amssymb}
\usepackage{amsmath}
\usepackage{braket}
\usepackage{mathtools}
\usepackage[usenames,dvipsnames,svgnames,table]{xcolor}
\usepackage [utf8] {inputenc}
\usepackage{color}
\usepackage{cite}
\usepackage{subcaption}
\usepackage{lmodern}
\usepackage{footnote}
\usepackage{glossaries-extra}
\setabbreviationstyle[acronym]{long-short}
\glssetcategoryattribute{acronym}{nohyperfirst}{true}

\usepackage{slashed}
\usepackage[pdftex] {graphicx}
\usepackage{multirow}
\usepackage{jheppub} 
\usepackage{here}
\usepackage{hyperref}
\usepackage{verbatim}
\usepackage[justification=centering,singlelinecheck=false]{caption}
\usepackage{booktabs}
\usepackage{cleveref}
\usepackage{listings}
\usepackage{fancyvrb}

\Crefname{equation}{eq.}{eqs.}
\Crefname{figure}{figure}{figures}
\Crefname{section}{section}{sections}
\Crefname{table}{table}{tables}
\Crefname{appendix}{appendix}{appendices}

\newcommand{\SU}{\rm SU}

\newcommand{\cI}[0]{\mathcal I}
\newcommand{\cJ}[0]{\mathcal J}
\newcommand{\cK}[0]{\mathcal K}
\newcommand{\cL}[0]{\mathcal L}
\newcommand{\cM}[0]{\mathcal M}
\newcommand{\cO}[0]{\mathcal O}

\newcommand{\cV}[0]{\mathcal V}

\newcommand{\df}[0]{\mathrm{df}}
\newcommand{\iso}[0]{{\rm iso}}

\newcommand{\Kiso}[0]{{\cK_{\df,3}^{\iso}}}

\newcommand{\Kisoone}[0]{{\cK_{\df,3}^{\iso,1}}}
\newcommand{\Kdf}[0]{{\cK_{\df,3}}}

\newcommand{\kdf}{\mathcal{K}_{\text{df},3} }

\newcommand{\bm}[0]{\boldsymbol}

\newcommand{\Luscher}[0]{Luscher:1986n2,Luscher:1991n1}

\newcommand{\HSQCa}[0]{Hansen:2014eka}
\newcommand{\HSQCb}[0]{Hansen:2015zga}

\newcommand{\HSTH}[0]{Hansen:2016fzj}

\newcommand{\BHSnum}[0]{Briceno:2018mlh}

\newcommand{\dwave}[0]{Blanton:2019igq}
\newcommand{\largera}[0]{Romero-Lopez:2019qrt}

\newcommand{\HHanal}[0]{Blanton:2019vdk}
\newcommand{\isospin}[0]{Hansen:2020zhy}

\newcommand{\threepithreeK}[0]{Blanton:2021llb}
\newcommand{\BRSimplement}[0]{Blanton:2021eyf}

\newcommand{\BSequiv}[0]{Blanton:2020gha}
\newcommand{\BSnondegen}[0]{Blanton:2020gmf}
\newcommand{\BStwoplusone}[0]{Blanton:2021mih}

\newcommand{\Akakia}[0]{Hammer:2017uqm}
\newcommand{\Akakib}[0]{Hammer:2017kms}

\newcommand{\DDK}[0]{Pang:2020pkl}
\newcommand{\AkakiRel}[0]{Muller:2021uur}

\newcommand{\MDpi}[0]{Mai:2018djl}

\newcommand{\HSrev}[0]{Hansen:2019nir}
\newcommand{\Akakirev}[0]{Rusetsky:2019gyk}
\newcommand{\MDRrev}[0]{Mai:2021lwb}
\newcommand{\Frev}[0]{Romero-Lopez:2021zdo}
\newcommand{\Bijnensb}[0]{Bijnens:2004bu}
\newcommand{\Bijnensa}[0]{Amoros:1999dp}
\newcommand{\phifourresonance}[0]{Garofalo:2022pux}
\newcommand{\ThreeQCDNumerics}[0]{%
Beane:2007es,
Detmold:2008fn,
Detmold:2008yn,
Detmold:2011kw,
Mai:2018djl,
Horz:2019rrn,
Blanton:2019vdk,
Culver:2019vvu,
Mai:2019fba,
Fischer:2020jzp,
Hansen:2020otl,
Alexandru:2020xqf,
Brett:2021wyd,
Blanton:2021llb,
Garofalo:2022pux,
NPLQCD:2020ozd}

\newcommand{\ThreeBody}[0]{%
Detmold:2008gh,
Beane:2007qr,
Briceno:2012rv,
Polejaeva:2012ut,
Hansen:2014eka,
Hansen:2015zga,
Briceno:2017tce,
Hammer:2017uqm,
Hammer:2017kms,
Mai:2017bge,
Briceno:2018aml,
Briceno:2018mlh,
Pang:2019dfe,
Jackura:2019bmu,
Blanton:2019igq,
Briceno:2019muc,
Romero-Lopez:2019qrt,
Pang:2020pkl,
Blanton:2020gha,
Blanton:2020jnm,
Romero-Lopez:2020rdq,
Blanton:2020gmf,
Muller:2020vtt,
Blanton:2021mih,
Muller:2021uur,
Muller:2022oyw}

\newacronym{CMF}{CMF}{center-of-momentum frame}

\preprint{MIT-CTP/5536}

\title{Interactions of $\pi K$, $\pi \pi K$ and $KK\pi$ systems at maximal isospin from lattice QCD}
\author[1]{Zachary T. Draper}
\affiliation[1]{Physics Department, University of Washington, Seattle, WA 98195-1560, USA}
\author[2]{, Andrew D. Hanlon}
\affiliation[2]{Physics Department, Brookhaven National Laboratory, Upton, New York 11973, USA}
\author[3]{, Ben H\"orz}
\affiliation[3]{Intel Deutschland GmbH, Dornacher Str. 1, 85622 Feldkirchen, Germany}
\author[4]{, Colin Morningstar}
\affiliation[4]{Department of Physics, Carnegie Mellon University, Pittsburgh, Pennsylvania 15213, USA}
\author[5]{, Fernando Romero-L\'opez}
\affiliation[5]{CTP, Massachusetts Institute of Technology, Cambridge, MA 02139, USA}
\author[1]{, and Stephen R. Sharpe}

\emailAdd{ztd@uw.edu}
\emailAdd{ahanlon@bnl.gov}
\emailAdd{ben.hoerz@intel.com}
\emailAdd{cmorning@andrew.cmu.edu}
\emailAdd{fernando@mit.edu}
\emailAdd{srsharpe@uw.edu}

\abstract{ 
We study the interactions of systems of two and three nondegenerate mesons composed of pions and kaons 
at maximal isospin using lattice QCD, specifically $\pi^+K^+$, $\pi^+\pi^+K^+$ and $K^+K^+\pi^+$. 
Utilizing the stochastic LapH method, we determine the spectrum of these systems on two CLS $N_f=2+1$ 
ensembles with pion masses of $200$ MeV and $340$ MeV, and include many levels in different momentum frames.  
We constrain the K matrices describing two- and three-particle interactions by fitting the spectrum to
the results predicted by the finite-volume formalism, including up to $p$ waves.
This requires also results for the  $\pi^+\pi^+$ and $K^+ K^+$  spectrum, which have been obtained previously
on the same configurations.
We explore different fitting strategies, comparing fits to energy shifts with fits to energies
boosted to the rest frame, and also comparing simultaneous global fits to all relevant two- and three-particle channels to those where we first fit two-particle channels and then add in the three-particle information.
We provide the first determination of the three-particle K matrix in $\pi^+\pi^+K^+$ and $K^+ K^+ \pi^+$ systems, 
finding statistically significant nonzero results in most cases. 
We include $s$ and $p$ waves in the K matrix for $\pi^+ K^+$ scattering, 
finding evidence for an attractive $p$-wave scattering length. 
We compare our results to Chiral Perturbation Theory, 
including an investigation of the impact of discretization errors,
for which we provide the leading order predictions obtained using Wilson Chiral Perturbation Theory. 
}

\allowdisplaybreaks
\begin{document}

\maketitle
\flushbottom
\clearpage

\section{Introduction}
\label{sec:intro}
\input{intro}

\input{lattice}
\input{QC}

\input{fits}

\input{results}

\section{Conclusion}
 \label{sec:conc}
 \input{conclu}

\acknowledgments

We thank A. Rodas for useful discussions.

The work of ZTD and SRS is supported in part by the U.S. Department of Energy (USDOE) 
grant No. DE-SC0011637.
The work of ADH is supported by
The U.S. Department of Energy, Office of Science, Office of Nuclear Physics through \textit{Contract No. DE-SC0012704},
and within the framework of Scientific Discovery through Advanced Computing (SciDAC) award \textit{Fundamental Nuclear Physics at the Exascale and Beyond}.
CJM acknowledges support from the U.S.~NSF under award PHY-2209167.
FRL has been supported in part by the USDOE, Office of Science, Office of Nuclear Physics, under grant Contract Numbers DE-SC0011090 and DE-SC0021006. FRL acknowledges financial support by the Mauricio and Carlota Botton Fellowship.

Calculations for this project were performed on the HPC clusters ``HIMster II'' 
at the Helmholtz-Institut Mainz and ``Mogon II'' at JGU Mainz. 
We are grateful to our colleagues in the CLS consortium for sharing ensembles.

\appendix
\clearpage
\section{Energy levels used in fits}
\label{app:A}
\input{AppA}
\clearpage
\section{ChPT result for $\kdf$ including $\cO(a^2)$ errors}
\label{app:B}
\input{AppB}

\section{NLO ChPT result for the $I=3/2$ $p$-wave $\pi K$ scattering length}
\label{app:pK}
\input{app_pK}

\input{app_operators}

\clearpage

\bibliographystyle{JHEP}      
\bibliography{ref.bib}

\end{document}

%% file: intro.tex

Multihadron dynamics emerge nonperturbatively from the strong interactions between quarks and gluons 
described by Quantum Chromodynamics (QCD). 
Processes involving several hadrons have important implications for Nature, 
such as the properties of hadronic resonances
and the emergence of nuclei as multi-particle systems. 
Thus, understanding these processes from first principles is an important goal for lattice QCD (LQCD)~\cite{Bulava:2022ovd}.

The study of hadron spectroscopy using LQCD has progressed rapidly in recent years;
 see Refs.~\cite{Briceno:2017max,Hansen:2019nir,Rusetsky:2019gyk,Horz:2022glt,Mai:2021lwb,Mai:2022eur,Romero-Lopez:2021zdo,Romero-Lopez:2022usb} for recent reviews. 
One of the current frontiers is the systematic computation of three-hadron processes from LQCD. 
Indeed, there has been a concerted effort by several collaborations to understand three-particle processes 
in finite volume~\cite{\ThreeBody}. Using these theoretical tools in conjunction with numerical simulations of LQCD,
several applications to simple systems have been carried out~\cite{\ThreeQCDNumerics}.

A class of three-particle systems that has not been extensively explored in simulations is that which involves
nondegenerate particles. 
Two examples of resonances that decay to 
such three-hadron systems are 
(i) the doubly-charmed tetraquark, $T_{cc} \to D D \pi$~\cite{LHCb:2021vvq}, 
and (ii) the Roper resonance $N(1440) \to N\pi\pi$~\cite{Roper:1964zza}.
Both are exotic resonances for which there is a great deal of interest in obtaining predictions from LQCD.
The formalism for such processes is rather complicated (and, indeed, has not yet been fully developed), 
and the LQCD computations are technically involved. 
Thus, for now, we leave aside the complications of resonances and spin quantum numbers, 
and focus on simpler systems of nondegenerate, spinless particles. 
Specifically, 
in this work we study systems of kaons and pions at maximal isospin, 
in particular $\pi^+ \pi^+ K^+$ and $K^+ K^+ \pi^+$,
which we refer to as ``2+1'' systems. 
We use two ensembles with different values of the pion and kaon masses,
which allows a rough extrapolation to the physical values. 

The formalism for these systems was developed in Ref.~\cite{\BStwoplusone}, 
following the relativistic field theory approach (RFT). 
Further details of the implementation of the formalism, as well as useful theoretical results,
were presented in Ref.~\cite{\BRSimplement}. 
These 2+1 systems exhibit new features compared to three identical particles, 
namely the presence of interactions in odd partial waves
and of multiple two-particle subchannels, 
e.g. both $\pi\pi$ and $\pi K$ subchannels contribute to $\pi\pi K$ scattering. 
One of the goals of this work is to determine the three-particle K matrix, denoted $\kdf$,
that describes short-range three-body interactions.\footnote{%
We note that results for zero-momentum three-particle interactions for $2+1$ systems 
were obtained in Ref.~\cite{Detmold:2011kw} by fitting the ground state energy shifts of systems of multiple $\pi^+$ and $K^+$ mesons to the results of a $1/L$ expansion. The results used heavier pion and kaon masses than we consider. 
The relation of the interactions so obtained to $\Kdf$ is not known, but we expect it to dominantly involve the
leading terms in the threshold expansion, namely $\cK_0$ and $\cK_1$.
}
The only previous work that considers this quantity (to our knowledge) is 
Ref.~\cite{\BRSimplement}, in which the
leading order prediction of chiral perturbation theory (ChPT) is obtained.
Compared to the corresponding quantity for three identical particles,
$\kdf$ has reduced particle-interchange symmetry,
and thus the number of independent terms in the threshold expansion of $\kdf$ is increased.
This makes the determination of their coefficients more challenging.

A second goal of this work is to address the timely question of fitting strategies for multihadron systems. 
As the complexity of the system grows, the number of quantities to constrain from LQCD becomes larger. 
Indeed, a qualitative step can already be seen in this work compared to that for identical particles: 
we require more terms in $\kdf$, and two different two-meson amplitudes. 
In particular, the question arises whether it is preferable to
determine the two-meson amplitudes from
results for the two-particle spectrum (e.g. for $\pi^+\pi^+$ and $\pi^+ K^+$), and then use the results
in a fit to the three-particle spectrum (e.g. for $\pi^+\pi^+ K^+$),
or, instead, do a global fit to all channels at once.
This is but one example of a general issue. An extreme case is provided by the determination of the
$\gamma^\ast \gamma^\ast \to \pi \pi$ amplitude using finite-volume methods:
the formalism requires also input from the finite-volume processes
$\pi \pi \rightarrow \pi \pi$, $ \pi \gamma^{\star} \rightarrow \pi$, 
$\gamma^{\star} \rightarrow \pi \pi$, and $ \pi \pi \gamma^{\star} \rightarrow \pi \pi$~\cite{Briceno:2022omu}. 
To determine which fitting procedures are preferable for more complex fits, we investigate 
and compare several fitting strategies.

Byproducts of this work are well-determined two-particle amplitudes. 
In particular, we compute the isospin $I=3/2$ $\pi K$ scattering amplitude in both $s$ and $p$ waves. 
By comparing our results for the $s$-wave scattering length to ChPT, we are able to extract low energy constants (LECs) in a threshold expansion.
Our results for the  $I=3/2$ $p$-wave scattering length are at lower pion masses than that obtained
previously in Ref.~\cite{Wilson:2014cna}.
We compare our result with this LQCD calculation, with the ChPT prediction,
and with the results from a dispersive analysis.

This work also contains the first estimates of discretization errors in the three-particle K matrix.
To achieve this, we present a new calculation of $\kdf$ using an extension
of continuum chiral perturbation theory (ChPT) in which
the effects of discretization errors are included, so-called Wilson ChPT (WChPT)~\cite{SS, BRS03}.
This allows us to determine the dependence of $\kdf$ on the lattice spacing in terms of low energy constants.
We also calculate the corresponding dependence for two-particle amplitudes,
which allows a determination of the relevant LECs by fitting to our results for these amplitudes.
Thus we can estimate the magnitude of discretization effects in $\kdf$.

This paper is organized as follows. In \Cref{sec:lattice} we provide details of the LQCD simulation, 
describe our choice of interpolating operators, and discuss the fitting and results for
the single and multiple-hadron spectra.
Then, in \Cref{sec:FVformalism}, we review the finite-volume formalism required for this work, present the
parametrizations of the K matrices that we use, and collect the results from ChPT that we need.
Next, in \Cref{sec:fits}, we describe the strategies that we use to fit to the spectra, and the results obtained,
comparing different approaches.
We collect our final results for infinite-volume scattering parameters in \Cref{sec:discussion}, and compare
them to ChPT, extracting several low-energy coefficients.
We conclude in \Cref{sec:conc}.
 We include four appendices. Appendix~\ref{app:A} displays a table with the energy levels used in fits, 
 \Cref{app:B} provides details on the calculation of discretization effects using WChPT, \Cref{app:pK} sketches the derivation of the analytical result for the $p$-wave $\pi K$ scattering length at NLO in ChPT, 
 and \Cref{app:operators} summarizes the sets of operators used in this work.

%% file: lattice.tex
\section{Finite-volume Spectrum Extraction}
\label{sec:lattice}

Here we present the details regarding the extraction of the finite-volume spectrum from the two-point correlation functions computed using LQCD.
The computational methods and strategies used follow those laid out in Ref.~\cite{Blanton:2021llb}.
However, for convenience, we review the pertinent details.

\subsection{Computation of Correlators}

The extraction of finite-volume energies proceeds by first calculating two-point temporal correlation functions,
which can be seen to contain all the information on the spectrum through their spectral decomposition
\begin{equation}
  C_{ij}(t_{\rm sep}) \equiv \braket{\mathcal{O}_i (t_{\rm sep} + t_0) \overline{\mathcal{O}}_j(t_0)} = \sum_{n=0}^{\infty} \bra{\Omega} \mathcal{O}_i \ket{n} \bra{\Omega} \mathcal{O}_j \ket{n}^\ast e^{-E_n t_{\rm sep}} ,
  \label{eq:spec_decomp}
\end{equation}
where $\mathcal{O}_i(t)$ and $\overline{\mathcal{O}}_j(t)$
are annihilation and creation operators,\footnote{%
The operators $\cO_j$ and $\overline{\mathcal{O}}_j$ are the analytic continuation to Euclidean space of the
Minkowski-space operators $\cO^{(M)}$ and $\mathcal{O}^{(M) \dagger}_j$, respectively.
}
 respectively, $\ket{\Omega}$ corresponds to the vacuum state, and $E_n$ is the energy of the $n$th eigenstate $\ket{n}$ of the Hamiltonian.
The indices on the operators indicate the possibility of a set of linearly independent interpolators that all share the same quantum numbers.
Details on the procedure for extracting the spectrum from these correlators will be given in later sections,
and we now turn to how the correlators themselves are computed.

The multi-hadron interpolating operators we consider in this work (see \Cref{sec:operators} and \Cref{app:operators}) involve definite-momentum projections for the individual hadrons, which in turn necessitates quark propagators from all spatial sites at a given source time to all other lattice sites.
One such method, referred to as distillation, acheives this with a smaller computational cost by utilizing 
a particular smearing based on the covariant Laplacian that cuts off higher-lying modes within the quark fields~\cite{Peardon:2009gh}.
The smeared quark propagators can then be obtained by performing inversions within the much smaller subspace spanned by the $N_{\rm ev}$ retained eigenvectors of the covariant Laplacian.
However, the number of required eigenvectors needed to keep a constant smearing radius grows proportionally with the volume $L^3$ and can quickly become prohibitively expensive for large volumes.
To mitigate this issue, rather than use distillation directly, we use a stochastic variant, referred to as stochastic LapH which has a better cost scaling~\cite{Morningstar:2011ka}.
This strategy combines stochastic sources within the distillation subspace and the method of dilution~\cite{Foley:2005ac} to estimate the smeared quark propagators in terms of source $\varrho^{(r,d)}$ and sink $\varphi^{(r,d)}$ functions
\begin{align}
  \varrho^{(r,d)} &= V_s P^{[d]} \varrho^{(r)} , \\
  \varphi^{(r,d)} &= \mathcal{S} D^{-1} \varrho^{(r,d)} ,
\end{align}
where $r$ labels the $N_r$ stochastic sources, $d$ labels the dilution partition, the columns of $V_s$ contain the $N_{\rm ev}$ retained eigenvectors, $P^{[d]}$ is a dilution projector, $\mathcal{S} = V_s V_s^\dagger$ is the smearing kernel, $D$ is the Dirac matrix, and $\varrho^{(r)}$ is a noise vector within the distillation subspace.

From the quark sinks and sources we form the meson sinks and sources, respectively, which are rank-2 tensors in the dilution indices.
The final correlators are constructed from contractions of these tensors over the dilution indices.
We make use of common subexpression elimination~\cite{10.1007/11758501_39} to reduce the number of needed contractions
and diagram consolidation to speed up the optimization.
These algorithms have also been utilized for two-baryon~\cite{Horz:2020zvv} and meson-baryon~\cite{Bulava:2022vpq} systems.
Further details on our implementation of these contraction speedups can be found in Ref.~\cite{Horz:2019rrn},
the code for which has been made publicly available~\cite{contractionop}.

\subsection{Interpolating Operators}
\label{sec:operators}

The single-hadron interpolators we use, which annihilate the single-hadron states, are 
\begin{align}
  H_{\pi^+} (\textbf{p}, t) &= \sum_{\textbf{x}} e^{-i \textbf{p} \cdot \textbf{x}} \; \overline{d}(x) \gamma_5 u(x) , \\
  H_{K^+} (\textbf{p}, t) &= \sum_{\textbf{x}} e^{-i \textbf{p} \cdot \textbf{x}} \; \overline{s}(x) \gamma_5 u(x) ,
\end{align}
where $u,d,s$ are up, down, and strange quark fields, respectively.
These are then substituted into our two- and three-hadron interpolating operators, which are of the form
\begin{gather}
  [H_{f_1} H_{f_2}]_{\Lambda} (\textbf{P},t) = \sum_{\textbf{p}_1, \textbf{p}_2} c^{\textbf{P}, \Lambda}_{\textbf{p}_1,f_1;\textbf{p}_2,f_2} H_{f_1}(\textbf{p}_1, t) H_{f_2}(\textbf{p}_2, t) , \\
  [H_{f_1} H_{f_2} H_{f_3}]_{\Lambda} (\textbf{P},t) = \sum_{\textbf{p}_1, \textbf{p}_2, \textbf{p}_3} c^{\textbf{P}, \Lambda}_{\textbf{p}_1,f_1;\textbf{p}_2,f_2;\textbf{p}_3,f_3} H_{f_1}(\textbf{p}_1, t) H_{f_2}(\textbf{p}_2, t) H_{f_3}(\textbf{p}_3, t),
\end{gather}
respectively, where $\textbf{P} = \sum_{i} \textbf{p}_i$ is the total momentum, $\Lambda$ is the irrep of the little group of $\textbf{P}$, $f_i$ labels the flavor of the individual hadrons, and $c^{\textbf{P},\Lambda}$ are  Clebsch-Gordan coefficients.
The Clebsch-Gordan coefficients are determined by requiring the overall operator transform according to the irrep $\Lambda$.
The sum on the right-hand-side includes all momenta related via rotations within the little group of $\textbf{P}$, with the constraint that $\sum_{i} \textbf{p}_i = \textbf{P}$.
A table listing the multi-hadron interpolating operators used in this work is given in \Cref{app:operators}.

\subsection{Lattice details}

Our calculations are performed on two ensembles generated by the CLS (Coordinated Lattice Simulations) consortium~\cite{Bruno:2014jqa}.
The ensembles use $N_f = 2 + 1$ flavors of nonperturbatively $\mathcal{O}(a)$-improved Wilson fermions and the tree-level $\mathcal{O}(a^2)$-improved L\"uscher-Weisz gauge action.
The bare quark masses are tuned such that they follow a chiral trajectory in which the sum of the quark masses is held fixed.
The practical effect of this choice is that the kaon mass approaches its physical value from below as the pion mass approaches its physical value from above.
The two ensembles used in this study, named N203 and D200, both share a lattice spacing of $a \approx 0.06426(76)$ fm,
which was determined from the linear combination $\frac{2}{3}(F_K + \frac{1}{2}F_\pi)$ of decay constants~\cite{Bruno:2016plf}.\footnote{
  The scale setting of Ref.~\cite{Bruno:2016plf} was recently updated, also using the pseudoscalar decay constants, giving a value of $a \approx 0.0633(4)(6)$ fm~\cite{Strassberger:2021tsu}.
Additionally, a scale setting using baryon masses was recently reported, giving a value $a \approx 0.06379(37)$ fm~\cite{RQCD:2022xux}.}
Other details, including the stochastic LapH setup, on these two ensembles can be found in \Cref{tab:ensems}.

\begin{table}
  \centering
  \tabcolsep=0.15cm
  \begin{tabular}{c c c c c c c c c}
    \toprule
    & $(L/a)^3 \times (T/a)\phantom{^3}$ & $M_\pi \, [\mathrm{MeV}]$ & $M_K \, [\mathrm{MeV}]$ & $N_\mathrm{cfg}$ & $t_\mathrm{src}/a$ & $N_\mathrm{ev}$ & dilution & $N_r(\ell/s)$ \\
    \midrule
    N203 & $48^3 \times 128$ & 340 & 440 & 771 & 32, 52 & 192 & (LI12,SF) & 6/3 \\
    D200 & $64^3 \times 128$ & 200 & 480 & 2000 & 35, 92 & 448 & (LI16,SF) & 6/3 \\
    \bottomrule
  \end{tabular}
  \caption{Specific details on the ensembles used in this work, including the name, geometry,
  approximate pseudoscalar masses, number of configurations $N_{\rm cfg}$, source positions $t_{\rm src}$
  used, number of eigenvectors $N_{\rm ev}$ of the covariant Laplacian retained, dilution scheme
  (see Ref.~\cite{Morningstar:2011ka} for details), and
  number of noises $N_{\rm r}$ used for the light ($l$) and strange ($s$) quark sources.
  Both ensembles have the same lattice spacing $a \approx 0.063$ fm.
  }
  \label{tab:ensems}
\end{table}

In order to prevent topological charge freezing at fine lattice spacings, open temporal boundary conditions are employed~\cite{Luscher:2012av}.
This requires care when choosing the temporal locations of the source and sink times used for the correlators, as we must make sure that they are sufficiently far from the temporal boundaries in order to suppress any effects from the boundary.
As there was no need to produce additional quark sinks beyond what was used in our previous study~\cite{Blanton:2021llb},
the arguments used there to justify the source and sink positions carry over here.
Essentially, evidence for sufficient suppression of boundary effects on the D200 ensemble was given in Ref.~\cite{Andersen:2018mau},
where it was found that a temporal distance of $\sim 32 a$ from the boundary was enough for the exponentially decaying boundary effects to be negligible.
Further, it is expected that the boundary effects are more severe on D200 than N203, as the leading contribution comes from the lowest state with quantum numbers of the vacuum, which should be a two-pion state for the quark masses considered here, and therefore has a smaller energy on D200.
Thus, as the source positions considered for N203 are even further from the boundary than D200, our choices should be safe from the effects of the open boundary conditions.
Note that the source position of $t_{\rm src} = 92 a$ for D200 only has sink times smaller than $92a$ associated with it 
(i.e. the correlators go backward in time, see Ref.~\cite{Blanton:2021llb} for more details).

Finally, autocorrelations, which lead to underestimated errors, can be checked for by observing dependence on the error estimates from averaging $N_{\rm rebin}$ successive configurations across all the original measurements into $N_{\rm cfg}/N_{\rm rebin}$ new bins.
While there is evidence that values as high as $N_{\rm rebin} = 20$ are needed for D200 to completely remove autocorrelations~\cite{Bulava:2022vpq},
this is not plausible for our use-case, as the number of energies used in our fits in \Cref{sec:fitresults} is too high to reliably estimate the covariance matrix with so few bins.
However, we have found little to no dependence on the final results for N203 when using $N_{\rm rebin} = 1$ or $N_{\rm rebin} = 3$, suggesting the observables of interest here are not affected significantly by autocorrelations.
We therefore use $N_{\rm rebin} = 1$ for N203, while using $N_{\rm rebin} = 3$ for D200 in order to still obtain reliable estimates for the covariance matrix while removing some autocorrelation.
Additionally, we note that the configurations used on N203 are separated in Markov time by 
twice the distance used for D200, which is why we use a conservative choice for the rebinning on D200.

\subsection{Finite-volume energies from correlators}

As can be seen from the spectral decomposition in \Cref{eq:spec_decomp}, 
in principle one can extract any energy so long as the operators used have non-zero overlap onto the corresponding eigenstate.
However, with finite statistics, reliably determining the states beyond the first few terms from fits to a single correlator is difficult.
As we are only interested in the ground states for the single-hadron particles, we can obtain these from single-exponential fits to the correlators starting after all higher-lying states are exponentially suppressed.
Fortunately, for the single hadron correlators, the signal-to-noise ratio is either constant or slowly decaying
which allows for a good signal after all higher-lying states have decayed away.
The needed single-hadron masses were determined in our previous work~\cite{Blanton:2021llb}
and are reproduced in \Cref{tab:masses_decay_constants} for convenience, along with
the decay constants needed for the chiral extrapolations performed later.

\begin{table}
   \centering
   \begin{tabular}{c c c c c c c}
     \toprule
     & $a M_\pi$ & $a M_K$ & $M_\pi L$ & $M_K L$ & $ M_\pi / F_\pi$  & $M_K/F_K$ \\
     \midrule
     N203 & 0.11261(20) & 0.14392(15) & 5.4053(96) & 6.9082(72) & 3.4330(89) & 4.1530(72) \\
     D200 & 0.06562(19) & 0.15616(12) & 4.200(12)  & 9.9942(77) & 2.2078(67) & 4.5132(93) \\
     \bottomrule
   \end{tabular}
   \caption{Pion and kaon masses and decay constants for the two ensembles considered in this work.
     The masses were determined in our previous work~\cite{Blanton:2021llb}.
     The decay constants were determined in Ref.~\cite{Ce:2022eix}.}
   \label{tab:masses_decay_constants}
\end{table}

Our eventual goal is to obtain the multi-hadron interactions, which are constrained by the multi-hadron finite-volume energies.
Each of these energies provides a constraint on these interactions, and, therefore, including as many energies as possible will typically improve the reliability of the extracted interactions.
In general, the gaps between the multi-hadron states are smaller than those of the single-hadron states, making a reliable extraction more challenging, 
especially coupled with the increased statistical errors of the multi-hadron fits.
Thus, we need a method to reliably determine several energies from the multi-hadron correlators.
We utilize a variational method~\cite{Luscher:1990ck,Blossier:2009kd}, which relies on solving a generalized-eigenvalue problem (GEVP) using a correlator matrix built from sets of interpolating operators that all transform in the same way.
This is the same method used in Ref.~\cite{Blanton:2021llb}, but we repeat most of the details here to make the discussion self-contained.

For each overall flavor, irrep, and total momentum squared, we construct a set of $N$
operators and use them to calculate a correlation matrix $C_{ij}(t_{\rm sep})$ as in \Cref{eq:spec_decomp}.
We then form a GEVP as
\begin{equation}
  C(t) \upsilon_n (t, t_0) = \lambda_n(t, t_0) C(t_0) \upsilon_n (t, t_0) ,
\end{equation}
where $t_0$ is referred to as the metric time.
As long as $t_0 \geq t/2$, one can show that the generalized eigenvalues behave as
\begin{equation}
  \lambda_n(t, t_0) = |A_n|^2 e^{-E_n (t - t_0)} \big[1 + \mathcal{O}(e^{-\Delta_n t})\big] ,
\end{equation}
where $n = 0,\ldots,N-1$,  $E_n$ is the $n$th eigenenergy, and $\Delta_n \equiv E_N - E_n$.
Thus, the method provides a straightforward way of extracting the excited-state energies
 without having to perform a fit which includes many exponentials.
In fact, the eigenvalues of $C(t)$ also share this advantage, but with a gap $\Delta_n \equiv \text{min}_{m \neq n} |E_n - E_m|$ that is smaller than or equal to the gaps found in the case of the GEVP.
Therefore, the gap from the GEVP helps to further suppress unwanted contributions in the generalized eigenvalues.

One technical complication involves matching the eigenvectors at one time separation to another or one resampling to the next (\textit{i.e.} eigenvector pinning),
which can easily become problematic from ambiguous choices, especially at large time separations where the noise grows.
Instead, we solve the GEVP on the mean and at a single time separation $t_d$, where $t_0 < t_d \lesssim 2 t_0$,
and then use the eigenvectors to rotate the original correlator matrix on all resamplings and at all other times not equal to $t_d$
\begin{equation}
  \hat{C}_n (t) = (\upsilon_n (t_d, t_0), C(t) \upsilon_n (t_d, t_0)) ,
\end{equation}
where the parentheses indicate an inner product.
To avoid any systematics associated with only solving the GEVP for one time separation,
we look for stability in the extracted spectrum as the diagonalization time $t_d$ and metric time $t_0$ are varied.

To obtain the two- and three-hadron finite-volume energies, we use single-exponential correlated-$\chi^2$ fits to the ratios
\begin{equation}
  R_n(t) = \frac{\hat{C}_n (t)}{\prod_i C_{f_i}(\textbf{p}_i^2, t)} ,
\end{equation}
where $C_{f_i}(\textbf{p}_i^2, t)$ is an average of single-hadron correlators with flavor $f_i$ over all rotationally equivalent momentum with magnitude $\textbf{p}_i^2$.
The product of single-hadron correlators in the denominator is chosen based on the expected non-interacting energy level associated with the $n$th eigenstate, in which case the asymptotic behavior is
\begin{equation}
  \lim_{t \to \infty} R_n(t) \propto e^{-\Delta E_{\rm lab}^n t}, 
\end{equation}
where $\Delta E_{\rm lab}^n$ is the energy shift of the $n$th eigenstate from its non-interacting value in the lab frame.
The use of the ratio has a few advantages over fitting directly to $\hat{C}_n (t)$:
there is a strong cancellation of correlated fluctuations between the numerator and denominator, and 
the plateau in the effective energy of the ratio begins at earlier time separations and is more stable across several time separations.
However, despite the earlier plateau seen from the ratio, in order to not introduce any further systematics, we typically make a conservative choice for the beginning of our fit range $t_{\rm min}$ such that the single-hadron correlators have 
already attained their asymptotic behavior.

\begin{figure}
  \centering
  \includegraphics[width=.34\textwidth]{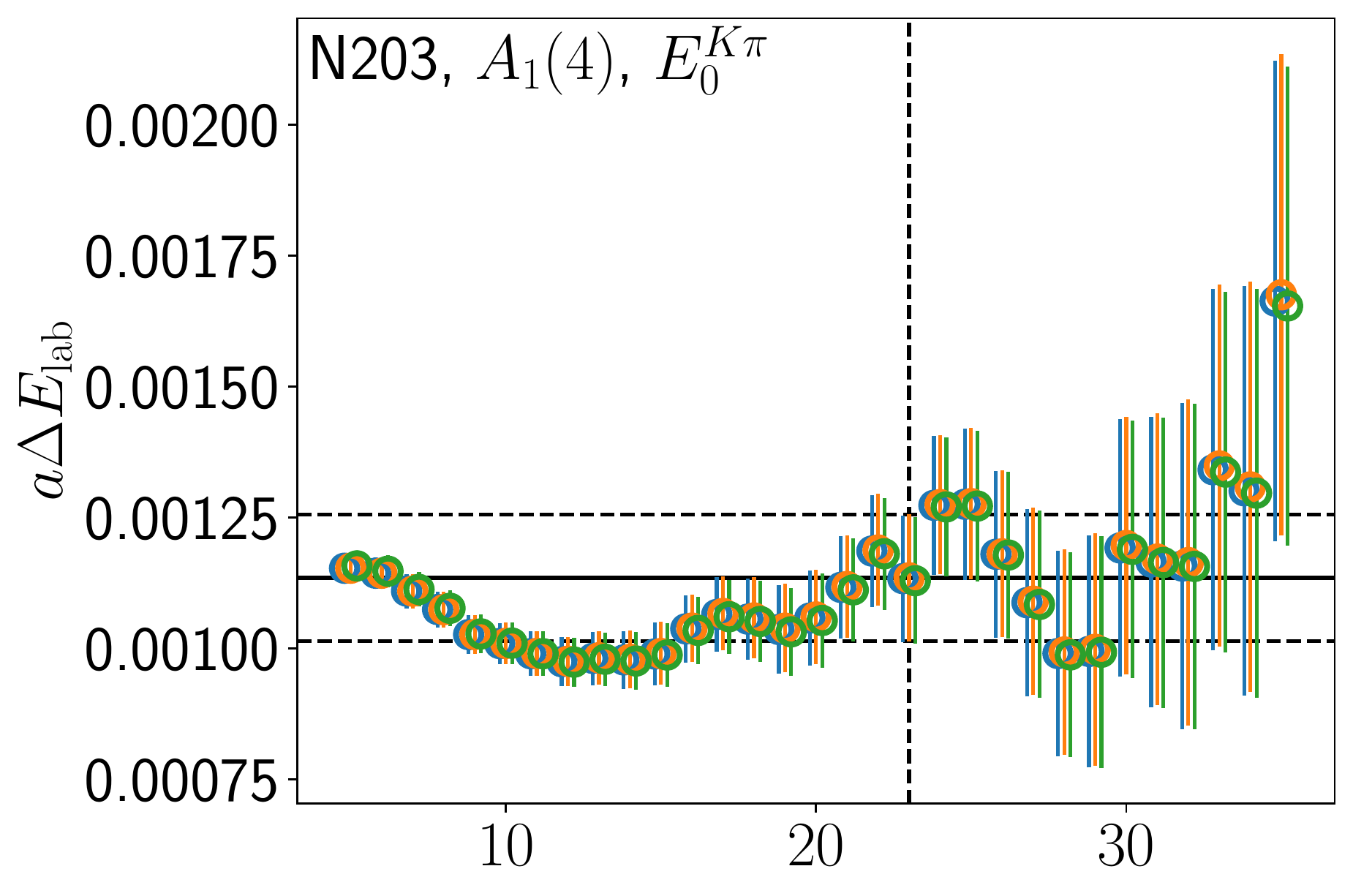}
  \includegraphics[width=.3175\textwidth]{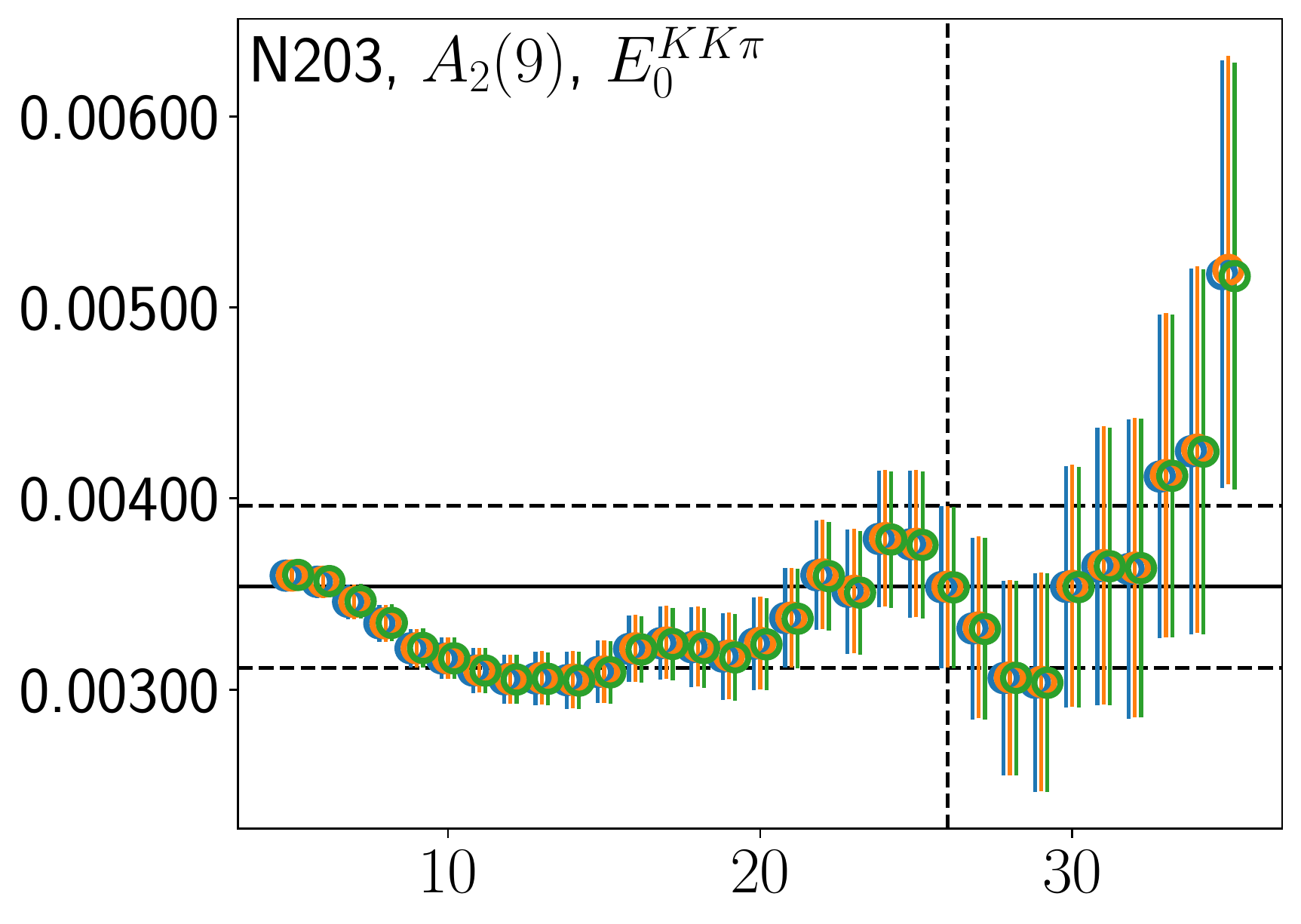}
  \includegraphics[width=.3175\textwidth]{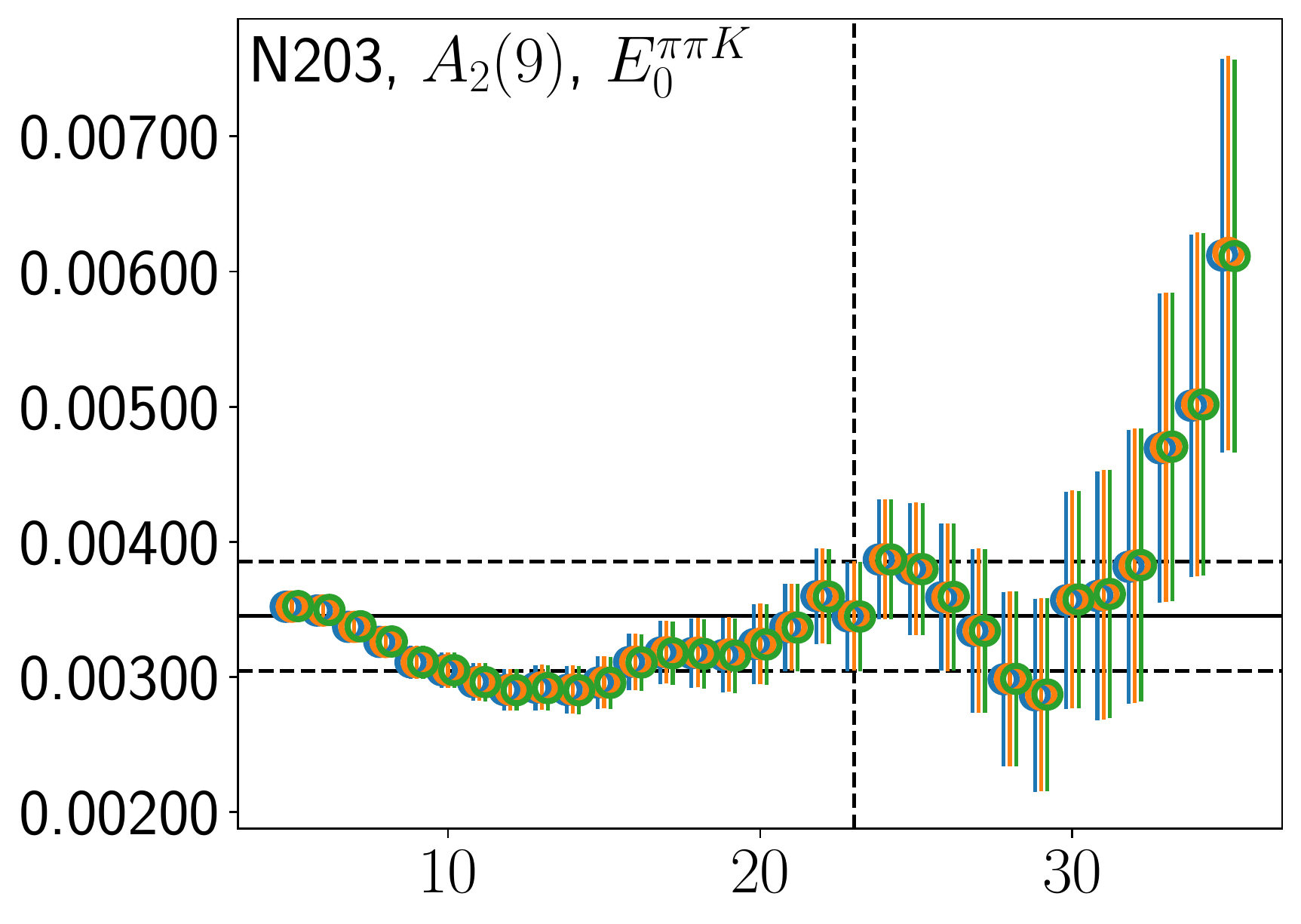}
  \includegraphics[width=.34\textwidth]{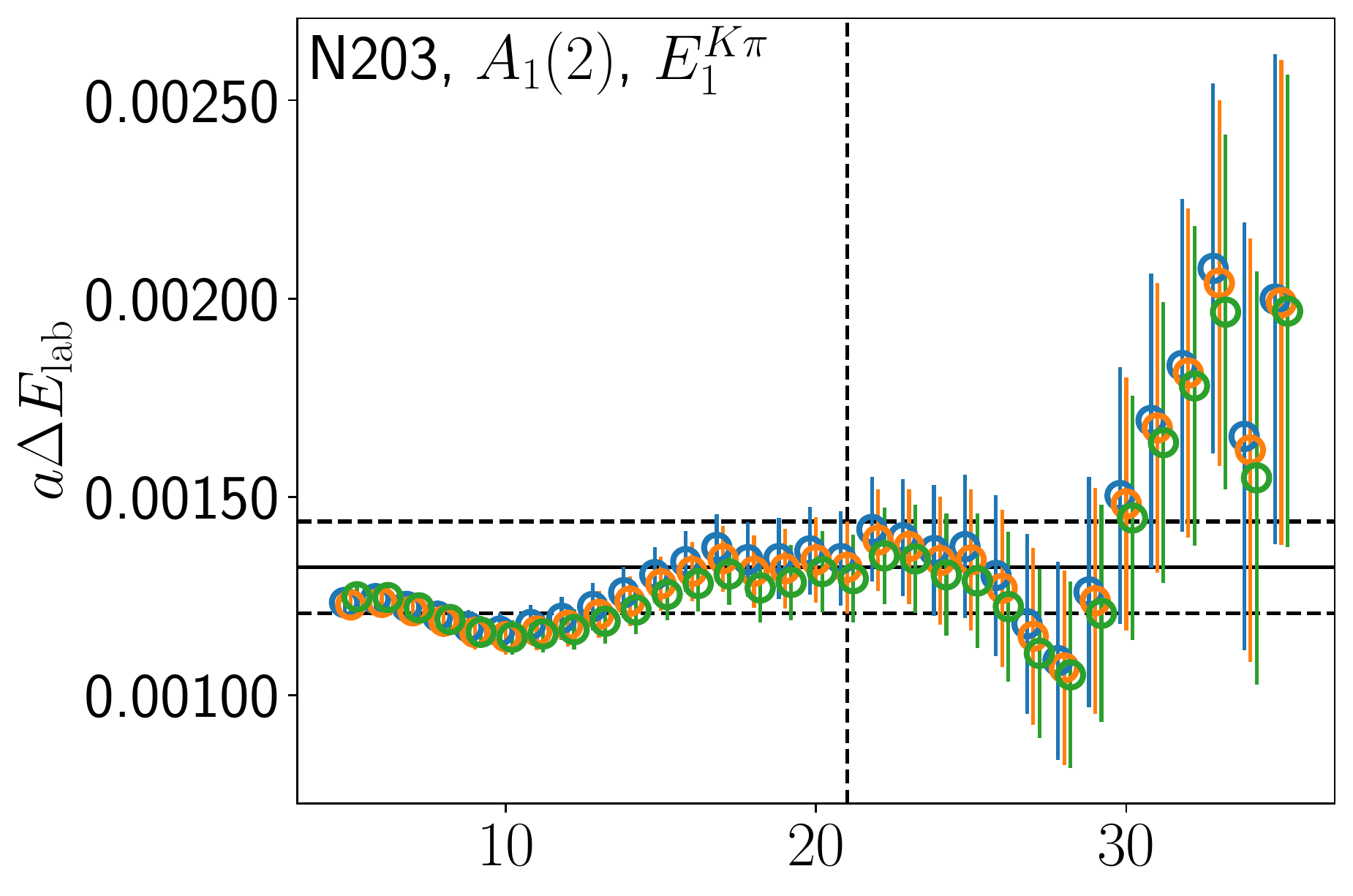}
  \includegraphics[width=.3175\textwidth]{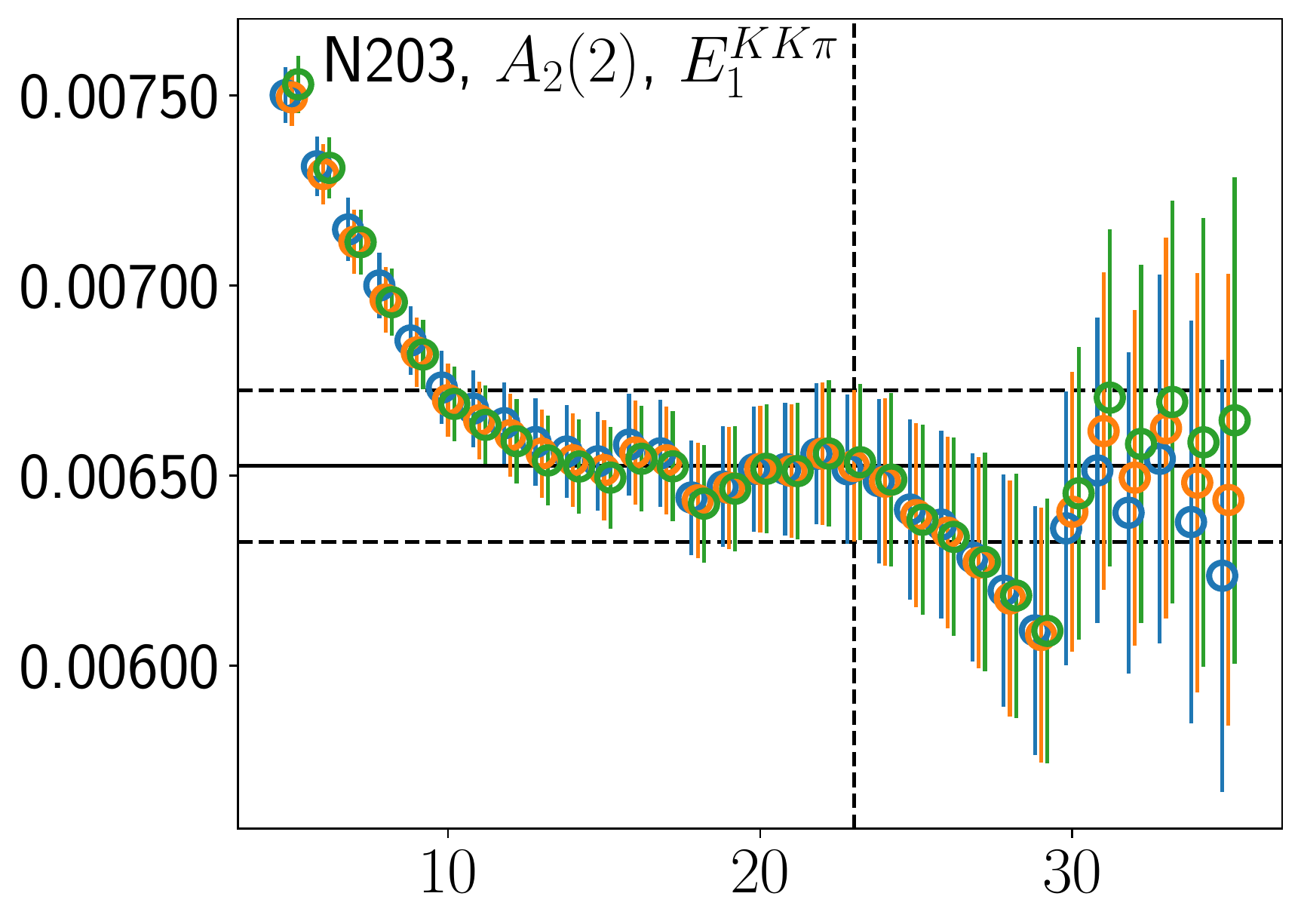}
  \includegraphics[width=.3175\textwidth]{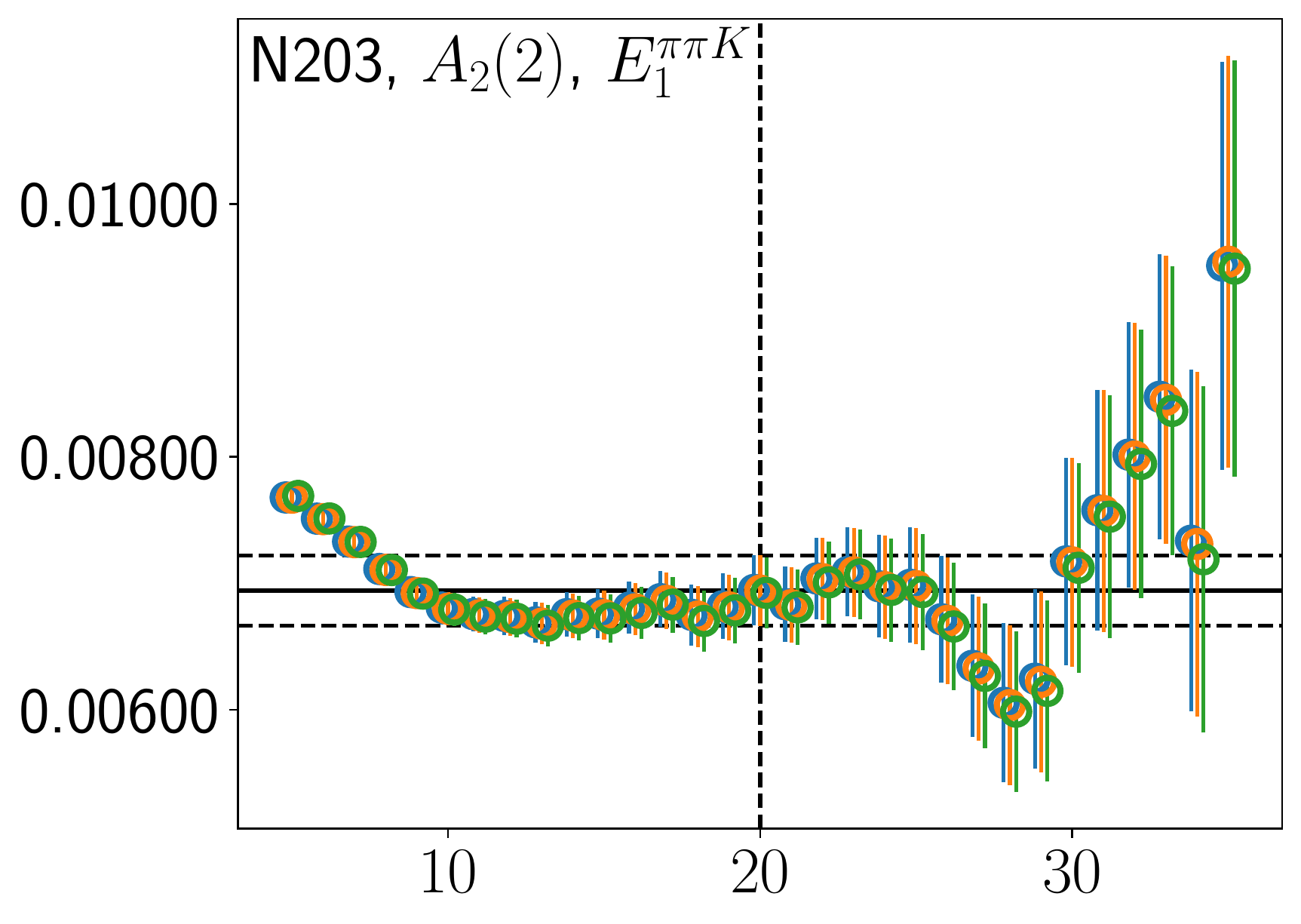}
  \includegraphics[width=.34\textwidth]{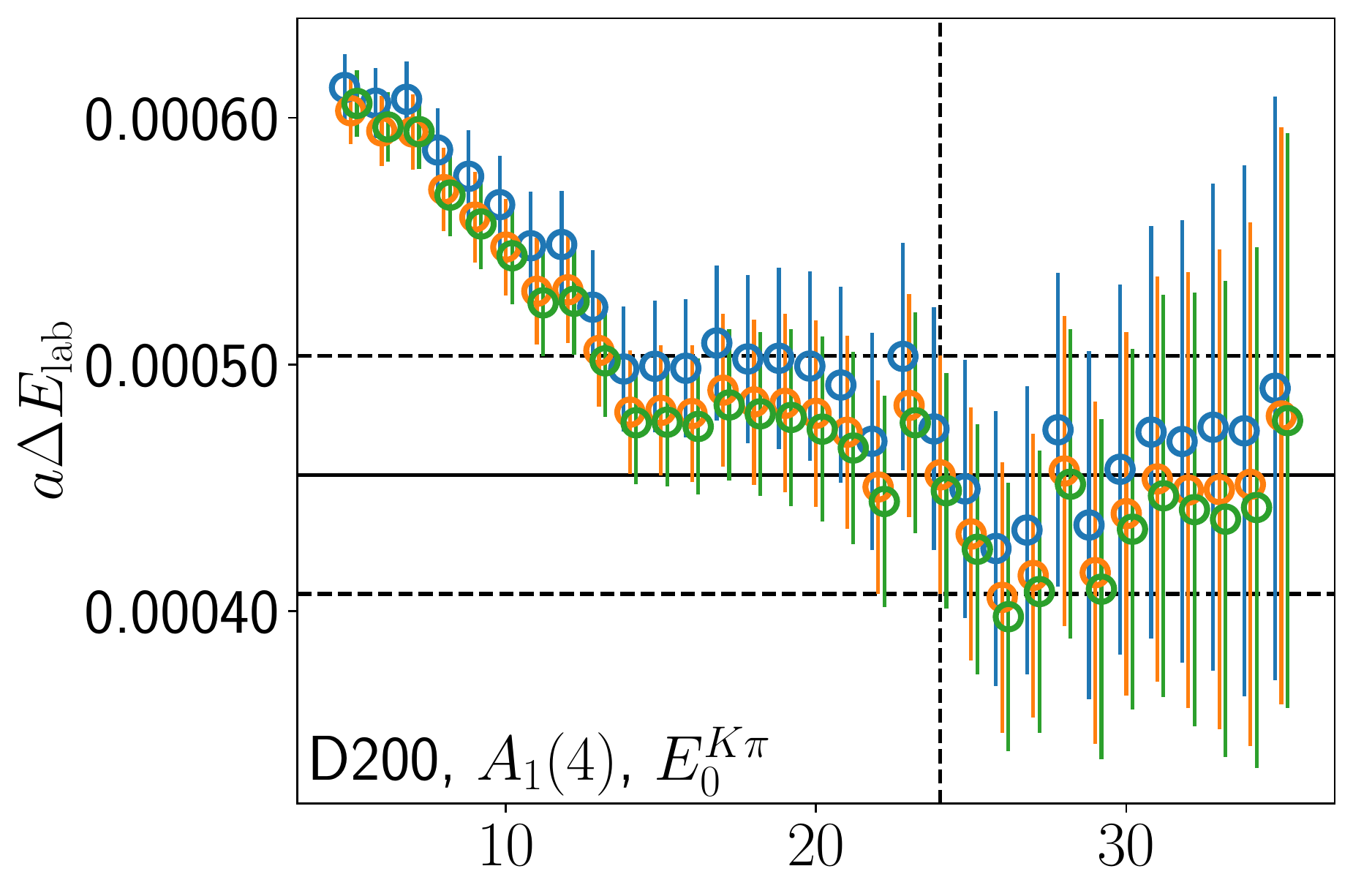}
  \includegraphics[width=.3175\textwidth]{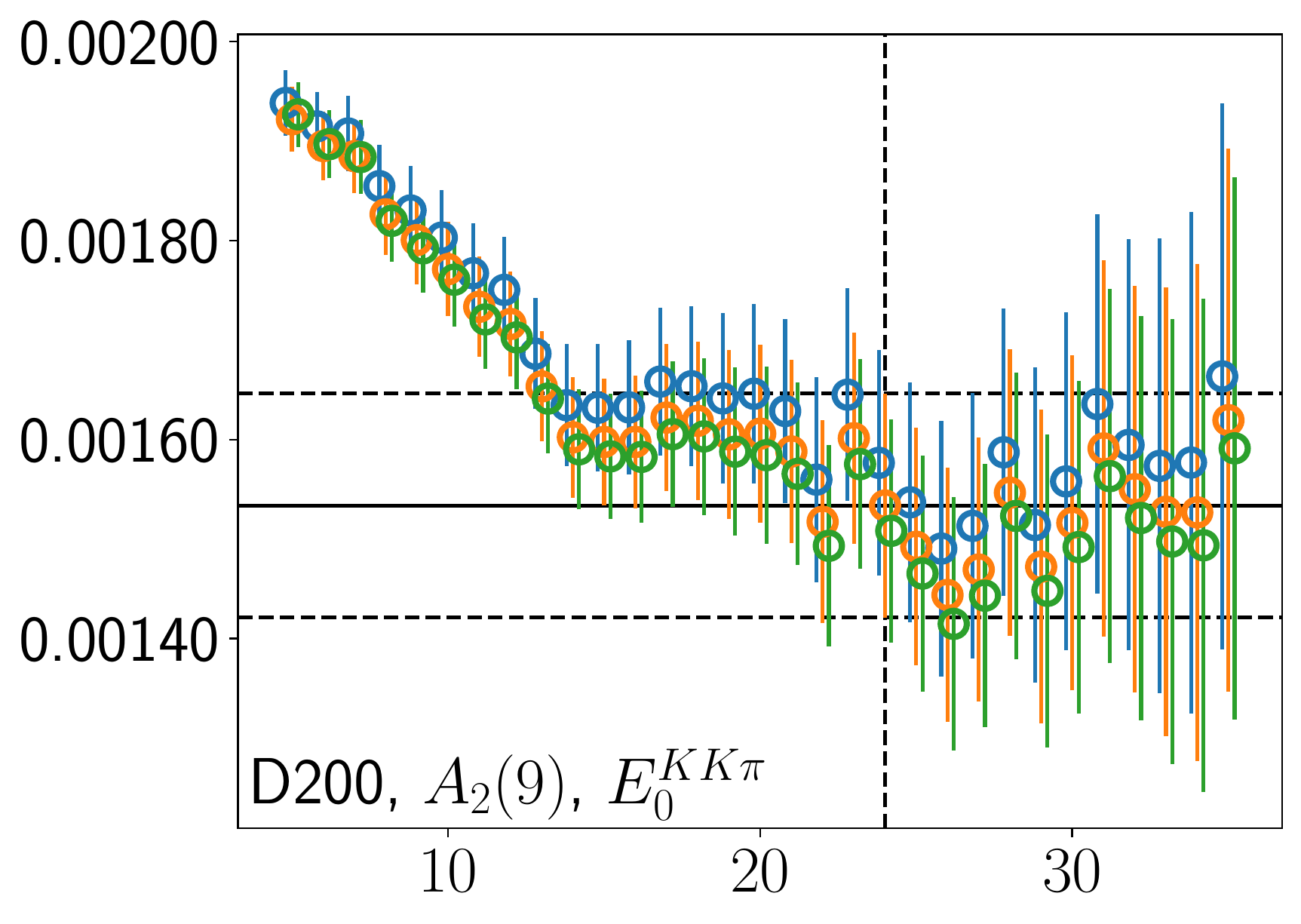}
  \includegraphics[width=.3175\textwidth]{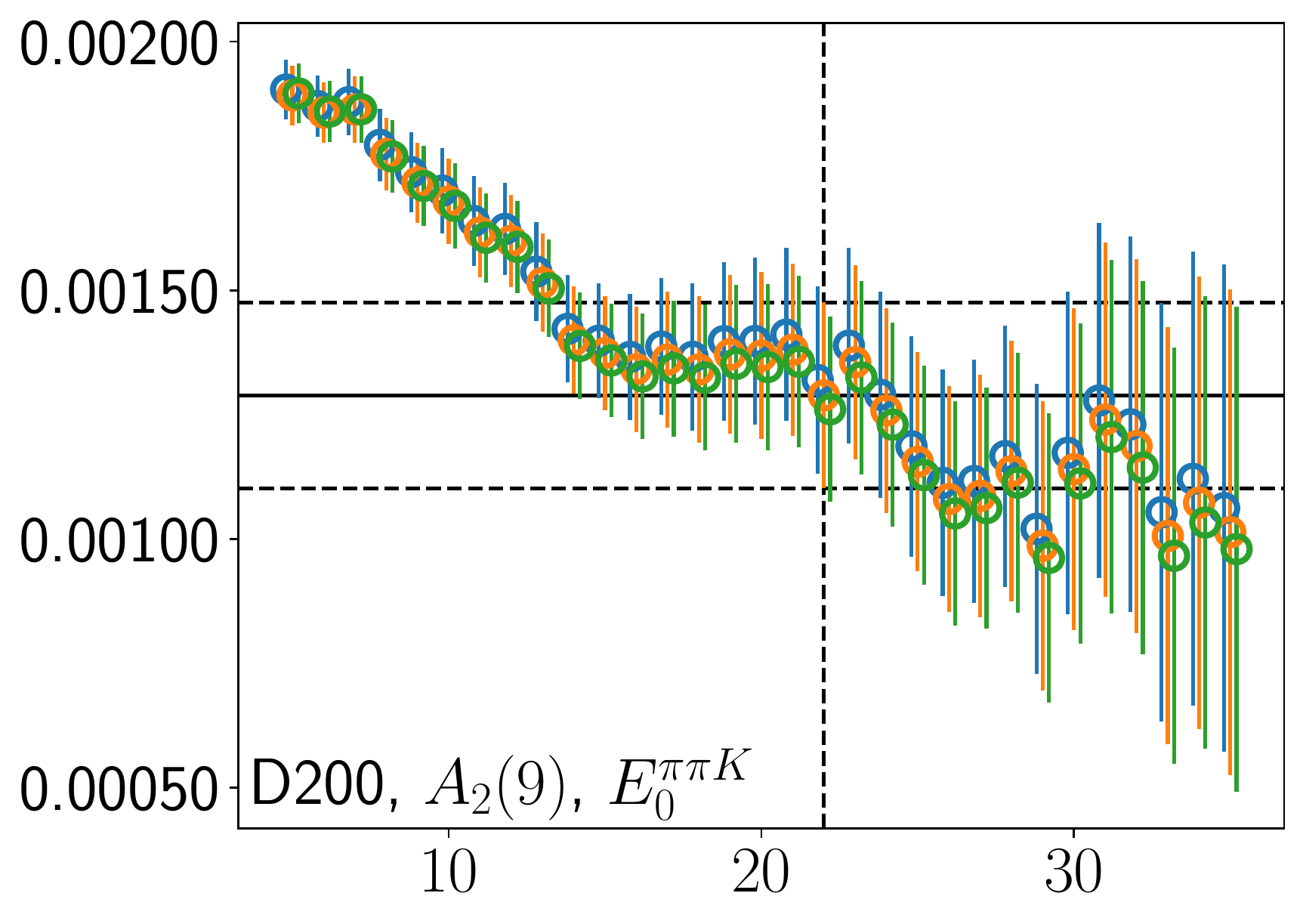}
  \includegraphics[width=.34\textwidth]{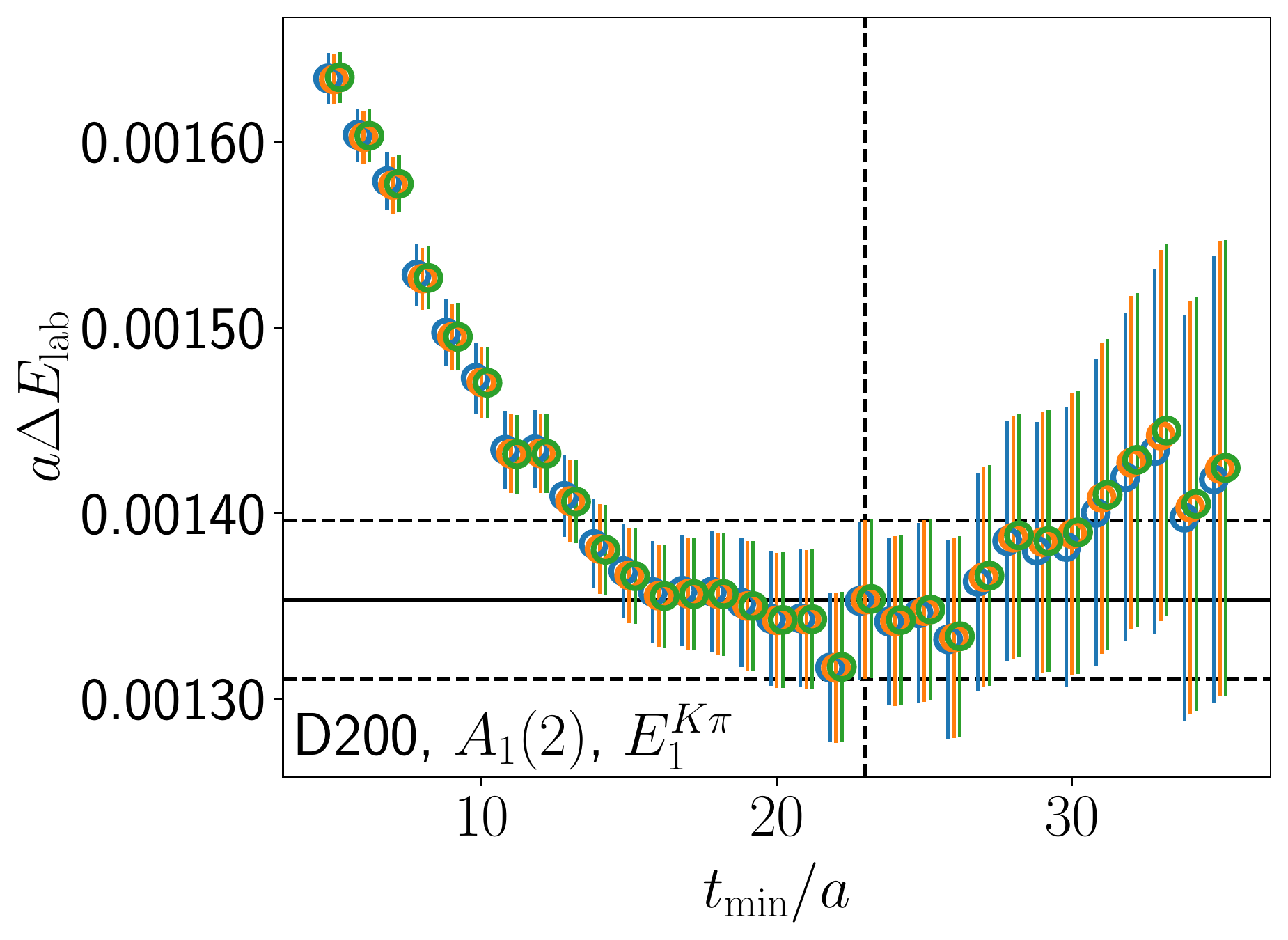}
  \includegraphics[width=.3175\textwidth]{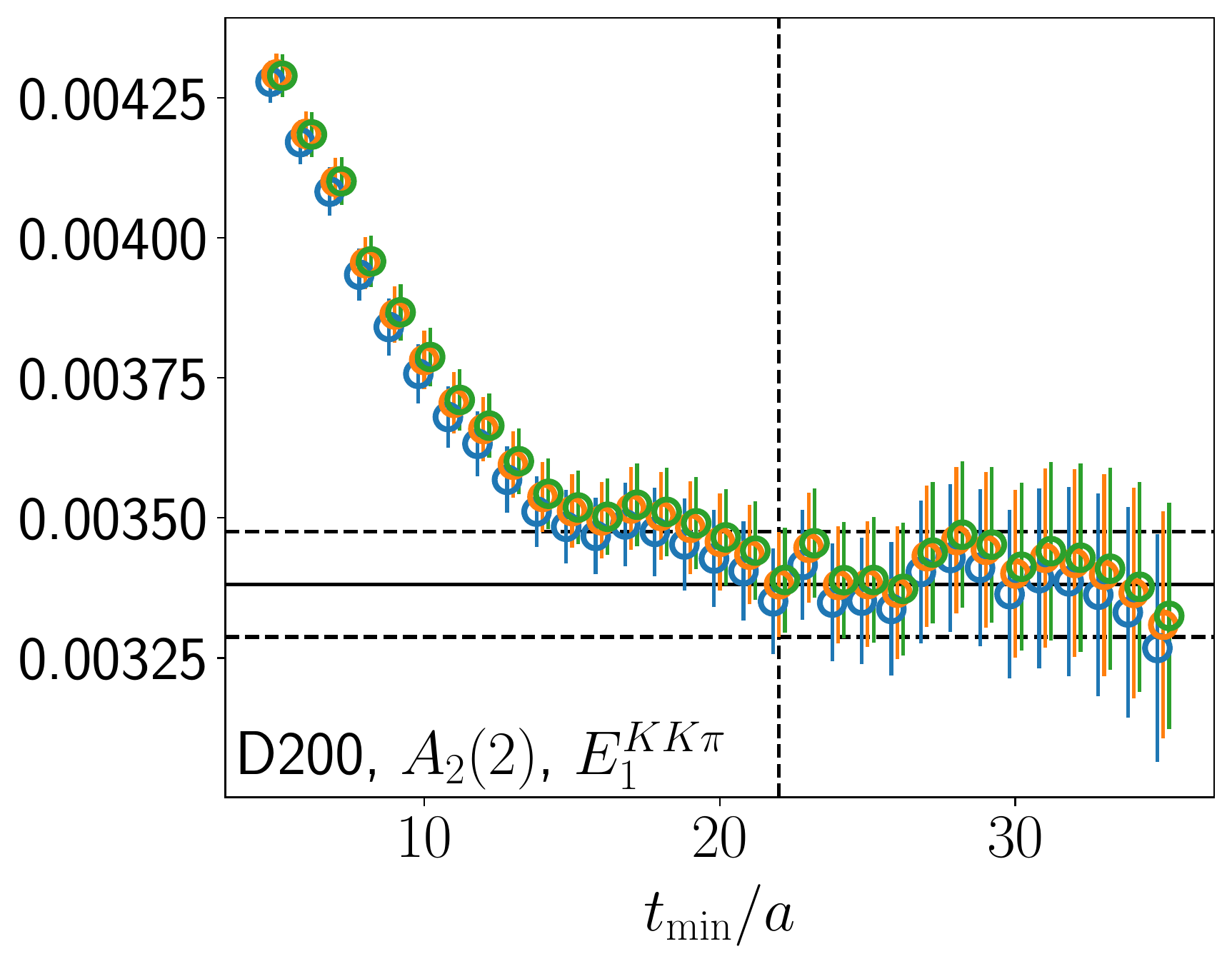}
  \includegraphics[width=.3175\textwidth]{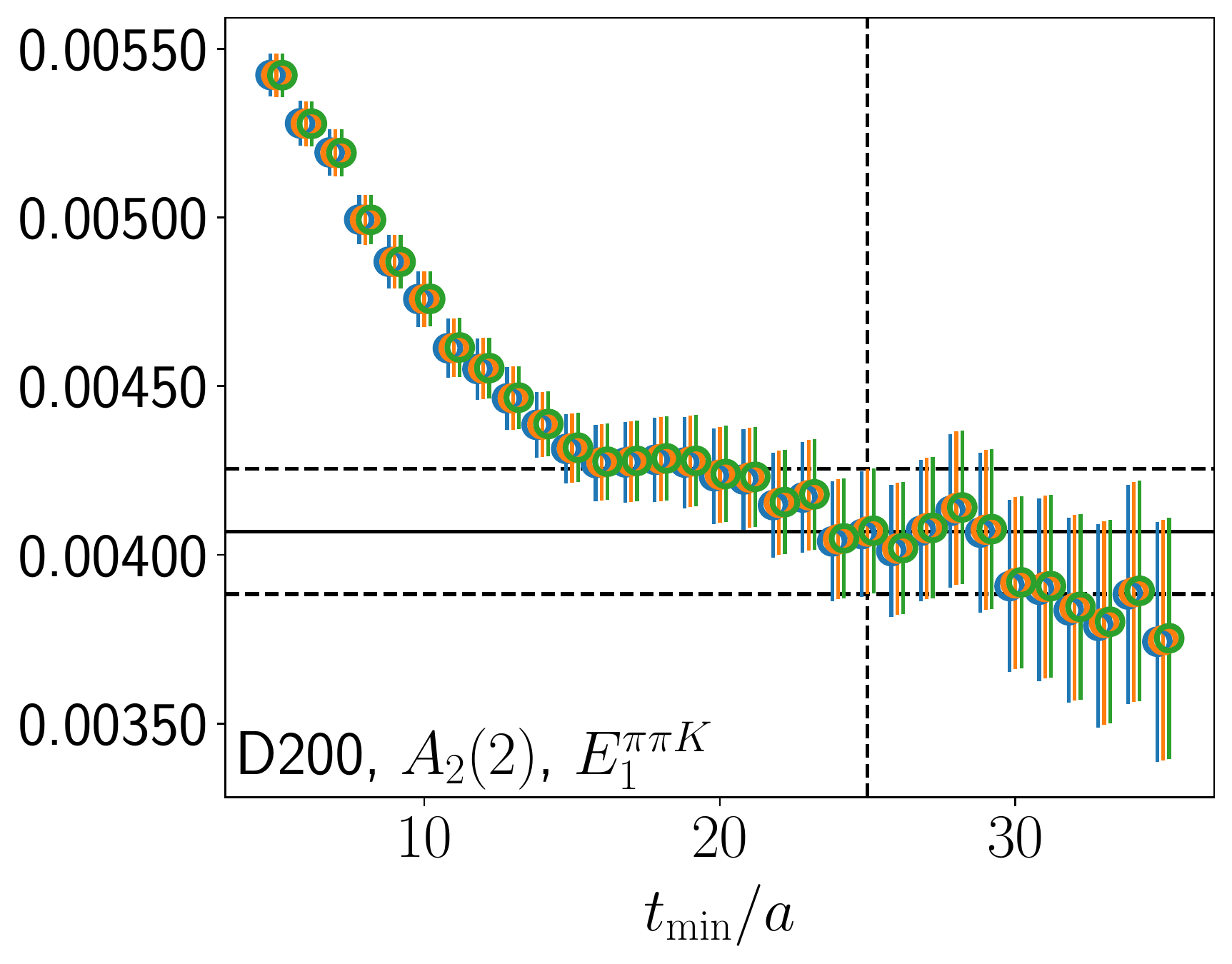}
  \caption{\label{fig:tmin} 
    The dependence of the energy shift in the lab frame extracted from fits to the ratio $R_n(t)$ on the smallest
    time separation $t_{\rm min}$ included in the fit. The colors correspond to different values of the GEVP metric time
    $t_0$ and diagonalization time $t_d$: $t_0/a=4,t_d/a=8$ (blue), $t_0/a=6,t_d/a=12$ (orange), 
    $t_0/a=12,t_d/a=24$(green).
    A small horizontal offset is applied to separate these three choices.
    Each plot indicates the
    ensemble, irrep, total momentum-squared (in units of $(2\pi/L)^2$, energy level, and flavor. 
    For example, in the bottom-right plot, the notation $A_2(2)$ indicates the $A_2$ irrep with
    momentum squared being $2(2\pi/L)^2$, while $E_1^{\pi\pi K}$ indicates that this is the first excited
    $\pi^+\pi^+ K^+$ level in this irrep.
    In each panel, the black horizontal solid and dashed lines indicate the mean and error, respectively,
    of the energy shift for the chosen fit, 
    for which the value of $t_{\rm min}$ is indicated by the vertical dashed line.
    }
\end{figure}

\begin{figure}
  \centering
  \includegraphics[width=\textwidth]{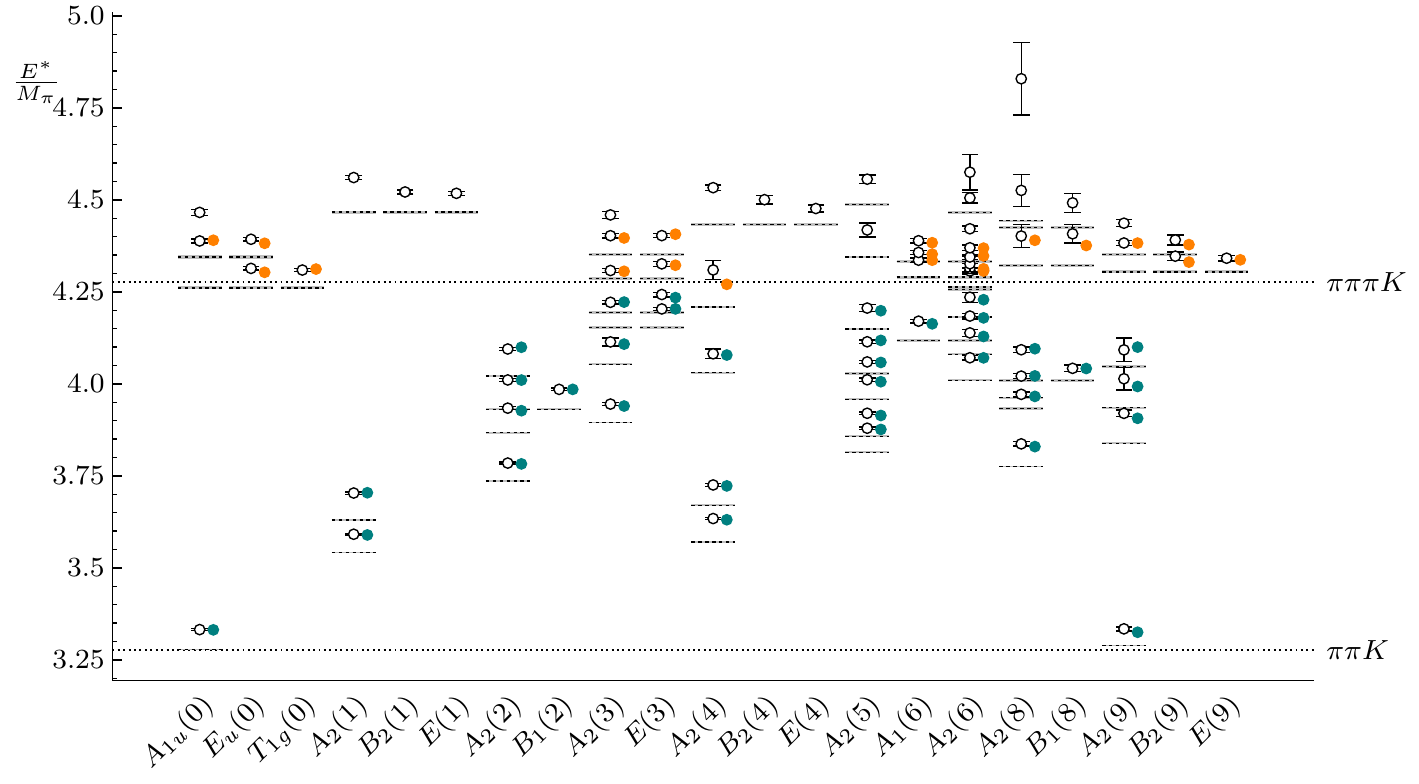}
  \caption{The $\pi\pi K$ center-of-mass frame energies, in units of $M_\pi$, on the N203 ensemble.
    The horizontal axis labels the irrep and (in parentheses) the total momentum squared in units of $(2\pi/L)^2$.
    The horizontal dashed lines and grey boxes (the latter barely visible)
    indicate the mean and error of the non-interacting energy levels,
    while the open circles with error bars correspond to the interacting energies.
    The colored symbols show the solutions of the quantization condition with the parameters found 
    in the 82-level fit in \Cref{tab:ppKN203}.
    Teal colored points are associated with energy levels included in the fit, while the orange points are for
    levels not in the fit.
    The horizontal dashed lines running across the entire plot show the ground state energy
    ($E^*=2 M_\pi+M_K$) and the first inelastic threshold ($E^*=3 M_\pi+M_K$).
    }
  \label{fig:n203_kpp_spectrum}
\end{figure}

In \Cref{fig:tmin}, we show the dependence of several extracted energy shifts on $t_{\rm min}$ and the GEVP parameters $t_0$ and $t_d$.
The energies shown correspond to the first excited state in an irrep with $\textbf{P}^2=2 (2\pi/L)^2$ and the ground state in an irrep with the largest momentum-squared used for a given flavor. We include all three mixed-flavor systems and both ensembles.
As can be seen in these $t_{\rm min}$ plots, typically there is a wide region of $t_{\rm min}$ with consistent energy shifts before correlated fluctuations take over when the signal starts to be lost.
The ability of the variational method to suppress contributions from unwanted nearby states is illustrated by
the results with $\textbf{P}^2=2 (2\pi/L)^2$, for in these cases there are usually several nearby energy levels.
Additionally, the dependence on the choices for $(t_0,t_d)$ is either very mild or not visible.\footnote{%
At a late stage in the fitting of the spectrum, one level was found to have a $\sim 3 \sigma$ variation in 
the shift away from the free level as $(t_d,t_0)$ were varied.
The specific level and the small effect it has on the final results is discussed in more detail in \Cref{sec:fitresults}.}
The value of $t_{\rm min}$ is chosen such that it is much larger than the onset of the stability in the extracted energy,
but not so large that correlated fluctuations begin to arise. Among the fits satisfying this critera, our final value
is based on making a conservative choice to ensure that 
any systematics are smaller than the statistical errors, while also making sure the fit quality is reasonable.
Examples of the choices of $t_{\rm min}$ are shown in the figure.

To illustrate the number of levels that we are able to determine, and the errors that we obtain,
we show in \Cref{fig:n203_kpp_spectrum} 
the energy levels for the $\pi \pi K$ system on the N203 ensemble. 
To better compare the levels from different momentum frames,
we show the center-of-mass frame (CMF) energies $E^*=\sqrt{E^2-\bm P^2}$.
Also shown (as teal dots) are the result of our standard fit to these levels, to be discussed below,
in which we fit only to levels that lie below the first inelastic threshold.
Although the formalism is not, strictly speaking, valid above this threshold,
we also display its predictions for some of the higher levels (as orange dots).
Analogous plots for the other three-meson systems that we consider are shown in \Cref{app:A}.
Detailed discussion of these  and other fits are provided below.

%% file: QC.tex

\section{Theoretical background}
\label{sec:FVformalism}

In this section we collect theoretical results needed to constrain the two- and three-particle K matrices
from the two- and three-particle spectra obtained using lattice QCD.
As discussed in the previous section, we have results for both degenerate and nondegenerate two- and
three-particle channels. The formalism for the degenerate cases has been reviewed in Ref.~\cite{\threepithreeK},
so we focus on the formalism needed for the new channels considered here,
i.e. $\pi K$ for the two-particle case, and $\pi\pi K/KK\pi$ for three particles.
We first provide a brief recapitulation of the quantization conditions and our fitting strategy,
then describe the parametrizations of K matrices that we use,
and finally collect the predictions of ChPT.

\subsection{Quantization conditions and fitting strategy}
\label{sec:QCs}

To extract infinite-volume scattering parameters from a finite-volume energy spectra, 
we make use of both two- and three-particle quantization conditions. 
The two-particle finite-volume formalism was first developed by L\"uscher~\cite{\Luscher}.
We will need the generalizations to moving frames
and to nondegenerate particles given in Refs.~\cite{Rummukainen:1995vs,Kim:2005gf,Gockeler:2012yj}.

The three-particle formalism has been developed using three approaches:
the relativistic field theory (RFT) approach~\cite{\HSQCa,\HSQCb}, 
which will be the basis of this analysis,
the non-relativistic effective field theory (NREFT) approach~\cite{\Akakia,\Akakib} 
(subsequently relativized in Ref.~\cite{\AkakiRel}),
and the finite-volume unitarity (FVU) approach~\cite{\MDpi}.
The formal equivalence (up to technical differences)
of the RFT and FVU formalisms was established in Ref.~\cite{\BSequiv},
and the equivalence of FVU and relativized NREFT formalisms is noted in Ref.~\cite{\AkakiRel}.
Reviews of these approaches and comparisons between them can be found in 
Refs.~\cite{\HSrev,\Akakirev,\MDRrev,\Frev}, and a direct comparison of their application
is given in Ref.~\cite{\phifourresonance}.
We will use the RFT result for three identical particles given in Refs.~\cite{\HSQCa,\HSQCb},
and the extension given
in Refs.~\cite{\DDK,\BSnondegen,\BStwoplusone} to so-called ``2+1"-systems, 
i.e. those involving one distinct and two identical spinless particles.

We will not recap the derivations here. We only note that the RFT method 
is based on an all-orders diagrammatic analysis in a generic relativistic effective field theory,
and applies in a kinematic regime in which only three-particle channels can go on shell. 
As one might expect from the name, this approach applies for relativistic particles; 
and if the relativistic forms of kinematic functions are used (see Ref.~\cite{\dwave}),
the formalism leads to Lorentz-invariant scattering amplitudes.

An extensive discussion of the implementation of the quantization condition for 2+1 systems is given
in Ref.~\cite{\BRSimplement}, and thus we present only an overview here. 
In particular, quantities not defined here can be found in that work.
We also make use of the formalism for identical particles derived in Ref.~\cite{\HSQCa}, and
in particular the implementation presented in Ref.~\cite{\dwave}.
A general feature of the three-particle formalism is a division into a spectator particle and the remaining
pair or dimer. 
For practical applications, one must impose a cutoff on the angular momentum, $\ell$, of the pairs.
In our previous work analyzing $3\pi^+$ and $3K^+$ spectra, we used $\ell_{\rm max}=2$, thus including $s$ and
$d$ waves ($p$ wave being forbidden for identical particles)~\cite{\threepithreeK}. However, working up to
$\ell_{\rm max}=2$ is not possible here, due to the proliferation of fit parameters, as discussed 
in Ref.~\cite{\BRSimplement} and recapitulated below.
Instead we use $\ell_{\rm max}=1$, with $s$ and $p$ waves in the $\pi^+ K^+$ subsystems, and only $s$ waves
for subsystems with identical particles ($2\pi^+$ and $2K^+$).

We begin by considering the two-particle quantization condition. 
The inputs are the total momentum, $\bm{P}$, the box size, $L$, a kinematic function denoted $F$, which contains the effects of finite-volume physics, and the two-particle K matrix, $\cK_2$. 
By solving the following two-particle quantization condition, 
\begin{equation}
\text{det} \left[ F\left( E_2, \bm{P}, L \right)^{-1} + \cK_2(E_2^*) \right] = 0\,,
\label{eq:QC2}
\end{equation}
we can determine the two-particle energies, $E_2$, and, from these, the corresponding center-of-mass frame (CMF) energies, $E_2^* = \sqrt{E_2^2 - \bm{P}^2}$, up to corrections suppressed by factors of $\exp(-M_\pi L)$ and $\exp(-M_K L)$.
As in the lattice simulations, we work in a cubic spatial box with side length $L$,
which restricts the allowed total momenta to the set $\bm{P} = (2 \pi / L) \bm{d}$, where $\bm{d} \in \mathbb{Z}^3$. 

The three-particle quantization condition determines the energies of three-particle states in a finite volume,
and is given by
\begin{equation}
\text{det} \left[ F_3\left( E, \bm{P}, L \right)^{-1} + \cK_{\text{df},3}(E^*) \right] = 0\,.
\label{eq:QC3}
\end{equation}
Here $E$ is the lab-frame energy of the three-particle state,
$E^* = \sqrt{E^2 - \bm{P}^2}$ is the corresponding CMF energy,
and $\kdf$ is a three-particle K matrix discussed further below.
Although, superficially, the three-particle quantization condition may look similar to that for two particles,
the former hides significant complexity. 
In particular, we note that whereas the quantity $F$ that appears in the two-particle quantization condition
is a purely kinematic function, 
$F_3$ depends on $F$, on an additional kinematic function $G$, 
as well as on the two-particle K matrix, $\cK_2$. 
The detailed form is given in Ref.~\cite{\BRSimplement}.
We also note that while the three-particle quantization in several different contexts takes exactly the same form
as in \Cref{eq:QC3} (see, e.g. Ref.~\cite{\isospin} for the case of degenerate but nonidentical scalars), 
the indices over which the determinant is taken differ.

We now describe the meaning of the determinants in \Cref{eq:QC2} and \Cref{eq:QC3}.
For the two-particle quantization condition, $F$ and $\cK_2$ are matrices in which the indices 
$\{\ell,m\}$ label the angular momentum of the two-particle state, 
and the determinant runs over these two labels. 
For the three-particle quantization condition, additional indices are required.
These are given by the spectator momentum $\bm{k}$, which is constrained to lie in the finite-volume set,
and the angular-momentum indices for the pair, $\{\ell,m\}$.
In addition, we need an index, $i$, to specify whether the spectator is identical to one of the other two particles
($i=1$) or whether it is distinct ($i=2$). For such a system, 
we therefore have a determinant which runs over the indices $\{k_i, \ell, m, i\}$, 
where $k_i$ labels the momentum of a spectator of particle species $i$.
The quantities $F_3$ and $\cK_{\text{df},3}$ are matrices with these indices.

The derivation of Refs.~\cite{\HSQCa,\BStwoplusone} requires a cutoff on the spectator momenta,
which must be implemented by a smooth function (rather than a sharp cutoff).
One way to understand the need for the cutoff is that, for a given total energy $E$, the pair is driven below threshold
as the spectator momentum increases. Far enough below threshold the two-particle interaction has a left-hand
cut, and this must be avoided as it introduces power-law volume dependence that is not controlled in the derivation.
The details of the required cutoff are discussed in Ref.~\cite{\BRSimplement}.
It implies that only a finite number of values of $\bm{k}_i$ contribute.
Combined with the cutoff on $\ell$ discussed above, this implies that all matrices in \Cref{eq:QC3} are 
of finite dimension.\footnote{%
We note that in the NREFT approach a hard cutoff can be used, and its value is not constrained by the
position of the left-hand cut~\cite{\Akakia,\Akakib,\AkakiRel}.}

We now return to the K matrices appearing in the quantization conditions.
Both $\cK_2$ and $\kdf$ are infinite-volume Lorentz-invariant quantities, but have important differences.
In particular, $\cK_2$ is algebraically related to the two-particle scattering amplitude $\cM_2$,
while $\cK_{\text{df},3}$ is 
related to the three-particle scattering amplitude $\cM_3$ through integral equations~\cite{\HSQCb}.
Furthermore, $\kdf$ depends on the cutoff function, whereas $\cK_2$ is cutoff independent above threshold.
However, both $\cK_2$ and $\kdf$ share the property of being real and smooth functions of Lorentz invariants
below the relevant inelastic thresholds.\footnote{%
The only exceptions are that $\cK_2$ and $\Kdf$ can have
poles in the presence of two- and three-particle resonances, respectively. 
These exceptions do not occur in this work.}
We make use of these properties to parametrize the K matrices in \Cref{sec:Parametrize}.

For given choices of $\cK_2$ and $\kdf$, the quantization conditions predict the two- and three-particle spectra.
Details of our numerical methods are given in Ref.~\cite{\BRSimplement}, and a basic python implementation
is publicly available~\cite{coderepo}.
We first block-diagonalize the quantization conditions by projecting them onto 
irreducible representations of the subgroup of the cubic group that
leaves the overall momentum, $\bm{P}$, invariant. This is the little group $\text{LG}(\bm{P})$. 
Within each block, we track the smallest eigenvalues of the matrix lying within the determinant,
and find zero crossings using a root-finding algorithm. 
A finite-volume energy level in the given irrep is predicted to occur for each such crossing.
We find that it is useful to have a good initial guess for the energy levels, and, for the most part,
this is provided by either the measured energy levels or the free levels.
We have implemented this methodology in three independent python codes, and all results presented
below have been obtained with at least two of these codes.
We find that, when running on about $\sim50$ cores, the convergence for the most challenging
cases (simultaneous fits to 2 two-particle and 1 three-particle channels) takes $2-3$ days.
The use of the Python compiler {\tt numba}~\cite{lam2015numba} to speed up core routines is essential to achieve this speed.

As discussed in Refs.~\cite{\BHSnum,\dwave,\largera}, one must ensure that the zero crossings are physical.
In particular, the eigenvalue must cross from negative to positive values as the energy is increased,
and no higher-order zeros are allowed.
The exception to the latter restriction is that there can be higher-order zeros at noninteracting energies,
but these are present only because of the truncation of the K matrices, as discussed extensively in
Ref.~\cite{\dwave}.
Such solutions can be dropped.
We find that all the crossings are physical.

\subsection{Parametrizing K matrices}
\label{sec:Parametrize}

We now turn to the parametrizations that we use for the matrices $\cK_2$ and $\cK_{\text{df},3}$.
The former is diagonal in angular momenta,
\begin{align}
\cK_2(E^{*}_{2})_{\ell' m'; \ell m} &= \delta_{\ell' \ell} \delta_{m' m} \cK_2^{(\ell)}(E^{*}_{2})\,,
\\
\left[{\cK_2^{(\ell)}(E^{*}_{2})}\right]^{-1} &= \frac{\eta}{8\pi E^{*}_{2}} \left\{q \cot{\delta_{\ell} (q)} + |q| [1 - H(q^2)] \right\}\,.
\label{eq:K2l}
\end{align}
We have written the expression so that it holds for both degenerate ($\pi\pi$ and $KK$) and 
nondegenerate ($\pi K$) channels: in the former case, $\eta=1/2$, while $\eta=1$ in the latter.
In both cases,  $q$ is the the magnitude of the three-momentum for each of the two particles 
in the CMF of the pair.
The function $H(q^2)$ plays the role of a cutoff. 
It equals unity for $q^2\ge 0$ [so that the $1-H$ term in \Cref{eq:K2l} vanishes above threshold], 
and smoothly drops to zero well below threshold (so that $\cK_2^{(\ell)}$ transitions into $\cM_2^{(\ell)}$).
The form of this function depends on the particle masses, as explained in Ref.~\cite{\BRSimplement}.
The $H$ function does not play a role in the two-particle quantization condition, \Cref{eq:QC2},
because it also appears in the quantity $F$, in such a way that the $H$ dependence cancels.
It is, however, essential in the three-particle quantization condition
(in which $\cK_2$ enters through $F_3$),
as it cuts off the sum over the spectator momenta.

As in Ref.~\cite{\threepithreeK}, we explore two choices for the parametrization of the phase shift:
an ``Adler zero'' form, and the effective-range expansion (ERE).
The former is motivated by chiral perturbation theory, which predicts that the scattering amplitude vanishes
below threshold at the position denoted the Adler zero.
For $\ell=0$, the Adler zero form is
\begin{equation}
\frac{q}{M_{1}} \cot \delta_{0}(q) = \frac{M_{1} E^{*}_{2}}{E^{*2}_{2} - z^2 (M_{1}^{2} + M_{2}^{2})} 
\sum_{n=0}^\infty B_n \left( \frac{q^2}{M_{1}^{2}}\right)^n\,,
\label{eq:Adler}
\end{equation}
where $z^2$ and the $B_{n}$ are dimensionless parameters.
Two masses appear in this parameterization: 
$M_{1}$, the mass which we use to set the units, e.g. of $q$, 
and $M_{2}$, the mass of the other particle. 
$M_{1}$ and $M_{2}$  are chosen from the set $\{M_{\pi}, M_{K}\}$ based on the particular
scattering process being considered, and may or may not be distinct.
At leading order in ChPT $z^2=1$, while $B_0$ and $B_1$ take nonzero values to be discussed in the next section,
with all other parameters vanishing.
In practice, we use two choices of parametrization:
\begin{enumerate}
\item
ADLER2, in which $z^2=1$, $B_0$ and $B_1$ are free parameters, and $B_n=0$ for $n\ge 2$; 
\item
ADLER3, in which $z^2$, $B_0$, and $B_1$ are free parameters, and $B_n=0$ for $n\ge 2$.
\end{enumerate}
We use superscripts and subscripts to denote the corresponding channel, e.g. $z_{\pi K}^2$ and
$B_0^{KK}$.
The relation of these parameters to the scattering length $a_0$ and effective range $r_0$ is given by
\begin{align}
a_0 M_1 &=  - \frac{(M_1+M_2)^2 - z^2 (M_1^2+M_2^2)}{M_1(M_1+M_2)B_0}\,,
\\
a_0 r_0 M_1^2 & = \frac{M_1}{M_2} \frac{ (M_1+M_2)^2+ z^2 (M_1^2+M_2^2)}{(M_1+M_2)^2 - z^2 (M_1^2+M_2^2)}
- 2 \frac{B_1}{B_0} \,.
\end{align}
Here we are using the convention in which $a_0$ is positive for repulsive interactions.

The ERE form is simply an expansion in powers of $q^2$,
\begin{equation}
\frac{q}{M} \cot \delta_{0}(q) = \sum_{n=0}^\infty B_n \left( \frac{q^2}{M^{2}}\right)^n\,,
\label{eq:ERE}
\end{equation}
where the $B_n$ are dimensionless. 
We use the same names for the coefficients as in the ADLER forms, 
but it will always be clear from the context which fits are being used.
The mass $M$ sets the scale of $q$ and may be chosen to be either of the masses of the particles
in the scattering pair.
In Ref.~\cite{\threepithreeK}, we found that the Adler zero form was preferred for the $\pi\pi$ and $\pi\pi\pi$
channels, while for kaons the ERE form was slightly preferred. This was not unexpected, as ChPT should
work better for pions than kaons. 
Here we only use the ERE form for $KK$ and $KKK$ channels, where we compare it to the ADLER forms.
Specifically, we use
\begin{enumerate}
\item[3.] ERE3, in which $B_0$, $B_1$, and $B_2$ are free parameters, and $B_n=0$ for $n\ge 3$.
\end{enumerate}
The relation of the ERE3 parameters to the scattering length and effective range is
\begin{equation}
M a_0 = -\frac{1}{B_0} \ \ {\rm and}\ \  a_0 r_0 M^2 = - 2 \frac{B_1}{B_0}\,.
\end{equation}

For $\ell=1$, which is present only for $\pi K$ scattering,
we use a one-parameter form,
\begin{equation}
\frac{q^3}{M_1^3} \cot \delta_{1}(q) = \frac{E^{*}_{2}}{M_1+M_2} \frac{1}{P_0^{\pi K}}\,.
\label{eq:EREp}
\end{equation}
Here $P_0^{\pi K}$ is the $p$-wave scattering length, and the same notation for masses is used as above.
The factor of $E^{*}_{2}$ is adopted from standard continuum analyses~\cite{Yndurain:2007qm,Kaminski:2006qe}.
We find that the signal for nonzero $p$-wave scattering is sufficiently weak that we cannot include
higher-order terms and obtain stable fits.

We now turn to the three-particle K matrix. 
As noted above, $\cK_{\text{df},3}$ is a real, analytic function of Lorentz invariants. 
We use the analog of the ERE, which is an expansion about threshold,
and has been worked out for $2+1$ systems in Ref.~\cite{\BStwoplusone}.
We denote $M_1$ as the mass of the particle that appears twice, while $M_2$ is the mass of the singleton.
Implementing the relevant particle interchange symmetries, as well as parity and time reversal, the resulting form is
\begin{equation}
M_1^2 \cK_{\text{df},3} = \cK_{0} + \cK_{1} \Delta + \cK_{B} \Delta^S_2 + \cK_{E} \tilde{t}_{22} + \cO(\Delta^2).
\label{eq:KdfExpansion}
\end{equation}
Here, $\cK_{0}$, $\cK_{1}$, $\cK_{B}$, and $\cK_{E}$ are real, dimensionless constants,\footnote{%
Previously we have used a more elaborate notation for $\cK_0$ and $\cK_1$,
namely $\Kiso$ and $\Kisoone$, respectively~\cite{\dwave,\threepithreeK,\BRSimplement}.
Here we stick with the simpler notation. We also stress that the quantity $\cK_B$ used here differs from that in Ref.~\cite{\threepithreeK}; the latter involves $d$-wave interactions for identical particles.}
to be determined from fits to the three-particle spectrum,
while $\Delta$, $\Delta_2^S$ and $\tilde{t}_{22}$ are functions of the Mandelstam variables.
The initial (incoming) momenta are $\{k_1, k_{1'}, k_2\}$,
while the (outgoing) final momenta are $\{p_1,p_{1'}, p_2\}$,
where $k_2$ and $p_2$ are the momenta of the singleton.
The kinematic quantities appearing in \Cref{eq:KdfExpansion} are
\begin{multline}
\Delta \equiv \frac{s-M^2_\Sigma}{M^2_\Sigma}, \quad
\Delta_2^S \equiv \frac{(p_1+p_{1'})^2- 4 M_1^2}{M^2_\Sigma} +
                             \frac{(k_1+k_{1'})^2 - 4 M_1^2}{M^2_\Sigma}, \\
\tilde{t}_{22} \equiv \frac{(p_2-k_2)^2-(M_1-M_2)^2}{M^2_\Sigma} \,, 
\label{eq:Mandelstam3}
\end{multline}
where
\begin{equation}
s \equiv (p_1+p_{1'}+p_2)^2 = {E^*}^{2} \quad
\text{and} \quad
M_\Sigma \equiv 2M_1+M_2\,.
\label{eq:Mandelstam2}
\end{equation}
$\Delta$, $\Delta_2^S$ and $\tilde t_{22}$ are all of the same order in the threshold expansion, but
 only $\Delta_2^S$ and $\tilde t_{22}$ have to a nontrivial angular dependence and lead to contributions with both $\ell=0$ and $1$.
 
 The expansion in \Cref{eq:KdfExpansion} can be continued to higher order, with terms of
 $\cO(\Delta^2)$ including  $d$-wave ($\ell=2$) contributions.
However, this leads to a proliferation of parameters, and so we restrict ourselves here to $\ell_{\rm max}=1$.
This is in contrast to Ref.~\cite{\threepithreeK}, where the analysis used $\ell_{\rm max}=2$ 
for the $3\pi^+$ and $3K^+$ spectra, which was possible because there are many fewer allowed forms
in $\kdf$ for identical particles.
Here, where needed, we have redone the identical-particle fits with $\ell_{\rm max}=1$, 
implying that $\kdf$ contains only the $\cK_0$ and $\cK_1$ contributions, 
since the $\cK_B$ and $\cK_E$ terms in \Cref{eq:KdfExpansion} vanish for a fully symmetric system.

The explicit forms for the $\cK_B$ and $\cK_E$ terms in the $\{k\ell m\}$ basis have been worked
out in Ref.~\cite{\BRSimplement}, and we simply use the results.
 
\subsection{Results from chiral perturbation theory}
\label{sec:ChPT}

Analyzing the two- and three-particle spectra for two different pion masses 
allows us to investigate the mass dependence of the scattering parameters we derive from fits. 
In this subsection we collect results from ChPT that will be useful when we fit the dependence of these scattering
 parameters versus $M_{\pi}^2/F_{\pi}^2$.
 Since our focus is on systems including both pions and kaons, we consider results from SU(3) ChPT.
 We note that a criterion for the utility of these results is that $M_{K}^2 / (4 \pi F_{K})^2 \ll 1$,
 where $F_{K}$ is the kaon decay constant in the convention that $F_{\pi} \simeq 92 \text{MeV}$. 
 For ensembles D200 and N203 the ratio $M_{K}^2 / (4 \pi F_{K})^2$ is $0.11$ and $0.13$ respectively.

The next-to-leading order (NLO) ChPT expressions for the $\pi \pi$ and $K K$ scattering lengths 
can be found in Refs.~\cite{Chen:2005ab, Chen:2006wf}.
Both depend on the mass of the $\eta$ meson as well as the pion and kaon, 
but  it is consistent in the NLO contribution
to use the LO result $3 M_{\eta}^2 = 4 M_{K}^2 - M_{\pi}^2$,
as well as to treat $F_{\pi}^2$ and $F_{K}^2$ as interchangeable.
In this way we can express the pion and kaon scattering lengths as functions of 
just $M_{\pi}$, $M_{K}$, $F_{\pi}$, and $F_{K}$:
\begin{align}
M_{\pi} a_{0}^{\pi \pi} &= \frac{M_{\pi}^2}{16 \pi F_{\pi}^2}
\bigg[1 + \frac{M_{\pi}^2}{16 \pi^2 F_{\pi}^2}
\bigg( \frac{3}{2} \log{\frac{M_{\pi}^2}{16 \pi^2 F_{\pi}^2}}
+ \frac{1}{18} \log{\frac{4 M_{K}^2 - M_{\pi}^2}{48 \pi^2 F_{\pi}^2}}
- \frac{4}{9} - 256 \pi^2 L_{\pi \pi} \bigg) \bigg]
\label{eq:a0pp}
\\
\begin{split}
M_{\pi} a_{0}^{KK} &= \frac{M_{K}^2}{16 \pi F_{K}^2}
\bigg[1 + \frac{M_{K}^2}{16 \pi^2 F_{K}^2}
\bigg( \log{\frac{M_{K}^2}{16 \pi^2 F_{K}^2}}
- \frac{M_{\pi}^2}{4(M_{K}^2 - M_{\pi}^2)} \log{\frac{M_{\pi}^2}{16 \pi^2 F_{\pi}^2}}
\\
& \qquad + \frac{20 M_{K}^2 - 11 M_{\pi}^2}{36(M_{K}^2 - M_{\pi}^2)}
\log{\frac{4 M_{K}^2 - M_{\pi}^2}{48 \pi^2 F_{\pi}^2}}
- \frac{7}{9} - 256 \pi^2 L_{\pi \pi} \bigg) \bigg]\,. 
\end{split}
\label{eq:a0KK}
\end{align}
Here the chiral logarithms are given by
\begin{equation}
L_\pi = 2 \log\frac{M_\pi}{4\pi F_\pi}\,, \quad
L_K = 2 \log\frac{M_K}{4 \pi F_\pi}\,, \quad
L_\eta = \log\frac{4 M_K^2 - M_\pi^2}{ 48 \pi^2 F_\pi^2}\,,
\label{eq:chirallogs}
\end{equation}
while $L_{\pi\pi}$ is an LEC, which
is evaluated at the scale $4 \pi F_{\pi}$.

We also need the expression for the $s$-wave $\pi K$ scattering length, which may be found in Ref.~\cite{Chen:2006wf}:
\begin{multline}
\mu_{\pi K} a_0^{\pi K} =  \frac{\mu_{\pi K}^2}{8\pi F_\pi F_K} \bigg[1 
-\frac{16 M_{\pi} M_{K}}{F_{\pi} F_{K}} L_{\pi\pi}
+\frac{4(M_K-M_\pi)^2}{F_{\pi} F_{K}} L_5
\\
 \quad + \frac1{32 \pi^2 F_{\pi} F_{K}} \left(
\kappa_\pi L_\pi + \kappa_K L_K + \kappa_\eta L_\eta
-\frac{86}9 M_\pi M_K + X_{K\pi}\right)\bigg]\,,
\label{eq:a0pK}
\end{multline}
where $\mu_{\pi K}= M_\pi M_K/(M_\pi+M_K)$ is the reduced mass, $L_5$ is another LEC,
and
\begin{align}
\kappa_\pi &= - \frac{M_\pi^2} 4 \frac{(11 M_K^2 + 22 M_K M_\pi - 5 M_\pi^2)}{M_K^2-M_\pi^2}\,,
\\
\kappa_K &= \frac{M_K}{18} \frac{(-9 M_K^3 + 134 M_K^2 M_\pi +55 M_K M_\pi^2 - 16 M_\pi^3)}{M_K^2-M_\pi^2}\,,
\\
\kappa_\eta &= \frac{(-36 M_K^3 -12M_K^2 M_\pi + M_K M_\pi^2+9M_\pi^3)}{36(M_K-M_\pi)}\,,
\\
X_{K\pi} &= \frac{16 M_\pi M_K}9
\frac{\sqrt{2 M_K^2 \!+\! M_K M_\pi \!-\! M_\pi^2}}{M_K\!-\!M_\pi} \;
\arctan\frac{2(M_K\!-\!M_\pi)\sqrt{2 M_K^2\!+\! M_K M_\pi\! -\! M_\pi^2}}
{(2 M_K\!-\!M_\pi)(M_K\!+\!2M_\pi)}\,.
\end{align}

For the effective ranges, we compare only to the LO prediction, since NLO results from SU(3) ChPT are not given
explicitly in the literature (see Refs.~\cite{Bernard:1990kw,Bijnens:2004bu} for the full $\pi K$ scattering amplitude at NLO).
For identical particles, the LO predictions are $M^2_\pi r^{\pi\pi} a_0^{\pi\pi} = M^2_K r^{KK} a_0^{KK} = 3$,
while for non-identical particles, one has
\begin{equation}
M_\pi^2 a_0^{\pi K} r_0^{\pi K} = 1 + \frac{M_\pi}{M_K} + \frac{M_\pi^2}{M_K^2}.
\label{eq:a0r0pK}
\end{equation}

We turn now to the $p$-wave scattering length, which is nonzero only for $\pi K$ scattering.
The NLO ChPT prediction for the $\pi K$ scattering amplitude has been worked out in Refs.~\cite{Bernard:1990kw,Bijnens:2004bu}. However, no closed form for the corresponding scattering length has been provided in the literature, and so we have worked it out and present the results in \Cref{app:pK}---see \Cref{eq:pKNLOpwave}. The expression, which is rather lengthy, depends on two different combinations of LECs.

Since we have only two data points, we opt to fit to the expected leading order chiral behavior. From \Cref{eq:pKNLOpwave}, one can determine that the $p$-wave scattering length is finite in the chiral limit 
(proportional to $M_K/F_\pi^4$), so that 
\begin{equation}
P^{\pi K}_0 = - M_\pi^3 a_1^{\pi K} \propto (M_\pi/F_\pi)^3  \,,
\label{eq:P0ChPT}
\end{equation}
with higher-order terms suppressed by powers of $M_\pi/F_\pi$. Neglecting chiral logs, this is equivalent to considering only the effect of the LECs proportional to $M_K^2$ in \Cref{eq:pKNLOpwave}.

We conclude this section by presenting the LO chiral predictions for $\cK_{\text{df},3}$ 
for $2+1$ systems, which were derived in Ref.~\cite{\BRSimplement}.
At LO, only $\cK_{0}$ and $\cK_{1}$ in the expansion shown in 
\Cref{eq:KdfExpansion} are nonzero:
\begin{align}
M_\pi^2 \cK_{0}^{\pi \pi K} &= 2 x_\pi^4 + 4 x_\pi^3 x_K\,,
\qquad
M_\pi^2 \cK_{1}^{\pi \pi K} = x_\pi^2( 2 x_\pi + x_K)^2\,,
\label{eq:KisoChPTppK}
\\
M_K^2 \cK_{0}^{K K \pi} &= 2 x_K^4 + 4 x_K^3 x_\pi\,,
\qquad
M_K^2 \cK_{1}^{K K \pi} = x_K^2( 2 x_K + x_\pi)^2\,,
\label{eq:KisoChPTKKp}
\end{align}
where
\begin{equation}
x_\pi = \frac{M_\pi}{F_\pi} \ \ \text{and} \ \
x_K = \frac{M_K}{F_K}\,.
\end{equation}
To obtain these forms we have used the interchangeability of $F_\pi$ and $F_K$ in LO terms.
We note that $\Kdf$ is independent of the cutoff function at LO in ChPT, and is in
this sense a physical quantity at this order. Cutoff dependence only enters at NLO.

Predictions are not yet available for $\cK_B$ and $\cK_E$, since these quantities are expected to
appear first at NLO, and no calculation at this order has been done.
Following Ref.~\cite{\BRSimplement}, we choose generic forms for these quantities
\begin{align}
M_\pi^2 \cK_{\text{B}}^{\pi \pi K} &= c^{\pi \pi K}_{\text{B}} x_\pi^4 x_K^2\,,\qquad
M_\pi^2 \cK_{\text{E}}^{\pi \pi K} = c^{\pi \pi K}_{\text{E}} x_\pi^4 x_K^2\,,
\label{eq:KBEChPTppK}
\\
M_K^2 \cK_{\text{B}}^{KK\pi} &= c^{KK \pi}_{\text{B}} x_K^4 x_\pi^2\,,\qquad
M_K^2 \cK_{\text{E}}^{KK\pi} = c^{KK\pi}_{\text{E}} x_K^4 x_\pi^2\,.
\label{eq:KBEChPTKKp}
\end{align}
We stress that the full dependence on $x_\pi$ and $x_K$ will likely be much more complicated, but
this form satisfies the correct chiral power counting and is sufficient given that we have only two 
values of pion masses.

A potentially important, and so far unquantified, source of systematic errors in our results comes from
working at a single lattice spacing. In this regard, it is useful to know the form of the prediction for
discretization effects that is given by Wilson ChPT (WChPT), 
i.e. ChPT including contributions proportional to powers of the lattice spacing $a$~\cite{SS,BRS03}.
In \Cref{app:B} we have worked out the leading, $\cO(a^2)$, terms for the two-particle scattering amplitudes
and $\Kdf$, both for the degenerate and nondegenerate cases. 
The results for scattering lengths are
\begin{align}
M_{\pi} a_{0}^{\pi \pi} &= M_{\pi} a_{0}^{\pi \pi}\bigg|_{a=0} - \frac{(2 w_6'+w_8')}{16 \pi} \,,
\label{eq:a0ppCorrection}
\\
M_{K} a_{0}^{K K} &= M_{K} a_{0}^{K K}\bigg|_{a=0} - \frac{(2 w_6'+w_8')}{16 \pi} \,,
\label{eq:a0KKCorrection}
\\
M_{\pi K} a_{0}^{\pi K} &= 
M_{\pi K} a_{0}^{\pi K} \bigg|_{a=0} - \frac{(2 w_6'+w_8')}{16 \pi} \,,
\label{eq:a0pKCorrection}
\end{align}
where $w_6'$ and $w_8'$ are dimensionless LECs proportional to $a^2$.
Note that we make $a_0^{\pi K}$ dimensionless by multiplying by the average mass
$M_{\pi K}= (M_\pi+M_K)/2$, rather than the reduced mass $\mu_{\pi K}$ used in \Cref{eq:a0pK}.
With this choice we see that all three quantities have the same offset.

The corresponding results for $\kdf$ are
\begin{align}
M_\pi^2 \cK_{\text{df},3}^{\pi \pi K} &= M_\pi^2 \cK_{\text{df},3}^{\pi \pi K}\bigg|_{a=0}  - 6 x_\pi^2 (2 w'_{6} + w'_{8}) \,,
\label{eq:KdfppKCorrection}
\\
M_K^2 \cK_{\text{df},3}^{K K \pi} &= M_K^2 \cK_{\text{df},3}^{K K \pi}\bigg|_{a=0}  - 6 x_K^2 (2 w'_{6} + w'_{8}) \,.
\label{eq:KdfKKpCorrection}
\end{align}
Thus the predicted shifts are proportional to the same combination of LECs as for the scattering lengths.
In \Cref{sec:discretization}, we will use these results to estimate
the magnitude of discretization effects contributing to $\Kdf$.

We close with a comment on the appropriate power counting in WChPT.
In the standard power counting one takes $a^2 \Lambda_{QCD}^2 \sim M_\pi^2/(4\pi F)^2$,
and, using this, it would be inconsistent to include NLO terms proportional $M_\pi^4/F^4$, while not
including discretization contributions proportional to $a^3$, $a^2 M_\pi^2$, etc.
Since such terms have not been calculated, however, we simply attempt fits using the available information,
namely using 
\Cref{eq:a0ppCorrection,eq:a0KKCorrection,eq:a0pKCorrection,eq:KdfppKCorrection,eq:KdfKKpCorrection},
in which we take the continuum NLO
expressions given above for the $a=0$ part.
In effect, we are assuming that $a^2 \Lambda_{QCD}^2 \sim M_\pi^4/(4\pi F)^4$.
We find, in \Cref{sec:discretization}, that the $a^2$ contributions are in fact considerably smaller than the
estimate from standard power counting, providing {\em a posteriori} justification for this approach.

%% file: fits.tex


\section{Extraction of scattering parameters}
\label{sec:fits}

In this section we discuss the extraction of scattering quantities from the energy levels obtained in lattice QCD. 
We start by discussing the different strategies that one can use to fit the spectrum using the quantization conditions. 
We then turn to the results of the fits using different methods. 
Finally, we compare the different approaches and discuss what seems to be the optimal one for this dataset.

\subsection{Fitting strategies}
\label{sec:fitstrategies}

In this work, we will use variations of the so-called spectrum method~\cite{Guo:2012hv}. The main idea is to obtain the best fit parameters by minimizing a $\chi^2$ function that depends only on some spectral quantity $X$:
\begin{equation}
\chi^2\left(\vec p\right)=\sum_{i j}\Delta X_i (C^{-1})_{ij} \Delta X_j , \quad \Delta X_i = X_i-X_i^{\mathrm{QC}}\left(\vec p\right), \label{eq:chi2}
\end{equation}
where $X_i$ is the lattice QCD result for that spectral quantity in the $i$-th energy level, $X_i^{\mathrm{QC}}\left(\vec p\right)$ represents the prediction from the finite-volume formalism assuming a certain parametrization of the K matrices with parameters $\vec p$, and $C$ is the covariance matrix of all the $X_i$ quantities, such that $C_{ij} = \text{cov}(X_i, X_j)$.

Several choices for the spectral quantity $X$ are possible. 
While, in the limit of infinite statistics, all should lead to the same answer, in practice some may be more advantageous.
Specifically, some choices can lead to covariance matrices with larger condition numbers, such that the calculation
of the inverse matrix appearing in \Cref{eq:chi2} may be more unstable. 
Some examples for $X$ are listed below.
\begin{enumerate}
\item 
The original energy levels obtained from lattice QCD,
which are, in general, in a moving (or ``laboratory'') frame. These are denoted $E_\text{lab}$.
\item 
The energy levels boosted to the center-of-mass frame (CMF), assuming the continuum
dispersion relation, i.e. ignoring possible lattice artifacts. The resulting energies are denoted $E_\text{cm}$.\footnote{
Above we have used the quantity $E^*$ to refer to the CMF energy. Here we prefer the more explicit notation $E_\text{cm}$.
} 
This is the choice of Ref.~\cite{Blanton:2021llb}.
\item 
The shift with respect to the non-interacting finite-volume energy, with the latter calculated assuming a continuum
dispersion relation. This can be done either in the lab or cm frame, 
yielding $\Delta E_\text{lab}$ or $\Delta E_\text{cm}$, respectively.
\item 
In the two-particle sector, it is also possible to use the CM momentum, $q^2$. This is the choice used,
for example, in Ref.~\cite{Bulava:2022vpq}.
\end{enumerate}
As we discuss in detail below, we use $\Delta E_{\rm lab}$ for our preferred fits.

Once the best fit parameters have been obtained by minimizing \Cref{eq:chi2}, 
the errors of the best fit parameters (and their covariance) need to be estimated. 
One possibility to do so is to perform a separate fit on each jackknife sample, 
and use the results to estimate the covariance of the parameters. 
While an option in the two-particle sector, present evaluations of the predictions of the three-particle
quantization condition are too slow for this approach to be practical in general.
Instead, we perform a fit only on the mean, while using the jackknife samples to estimate the covariance matrix $C$ of the data,
and then apply the derivative method.
This method, discussed in Ref.~\cite{Blanton:2021llb},
estimates the covariance between the fit parameters $p_n$ and $p_m$ as
\begin{equation}
V_{n m}=\left(\frac{\partial X_i^{\mathrm{QC}}}{\partial p_n} 
(C^{-1})_{ij} \frac{\partial X_j^{\mathrm{QC}}}{\partial p_m}\right)^{-1},
\end{equation}
where the derivatives are evaluated numerically at the minimum of the $\chi^2$ function, $\chi^2_\text{min}$. 
A very similar approach is to find the $1\sigma$ interval by finding the contour such that $\chi^2 = \chi^2_\text{min} +1$,
and assuming that the dependence of $\chi^2$ on $\vec p$ is quadratic.
In practice, we find that these two methods give essentially identical results, and we use them interchangeably
to quote errors in the following.

Another issue to address is how to combine information from the different channels,
e.g. those with different numbers of particles.
In particular, when analyzing three-particle energies, the two-particle interaction parameters are also needed.
In previous work, e.g. in Refs.~\cite{Blanton:2019vdk,Blanton:2021llb}, 
the approach has been to perform a combined fit to both two- and three-particle energy levels.
For example, in the pion sector,
one defines a $\chi^2$ function that combines $\pi\pi$ and $\pi \pi \pi$ levels,
 and uses the covariance matrix of all levels, including cross-correlations between two- and three-particle energies. 
 Fits using this approach will be referred as ``fully correlated fits''. 
 In this work, we take this approach one step further by considering nondegenerate systems, 
 where two different two-particle spectra are needed.
 For example, for the $\pi\pi K$ sector, we must consider also the $\pi\pi$ and $\pi K$ levels.
 This leads to a larger number of total energy levels, and one may be concerned about the reliability of the
 calculation of the covariance matrix between levels.
 We find, in practice, that this is not an issue in the fits done here (either for the $\pi\pi K$ or $KK\pi$ cases), but
 it certainly will become a problem eventually, as the number of channels and levels increases further.
For example, we have not attempted fully correlated simultaneous fits to the 
$\pi\pi$, $\pi K$, $KK$, $\pi\pi K$ and $KK \pi$ levels.

Because the issue of fitting to multiple channels is a generic one in multiparticle physics, it is worthwhile 
investigating alternative approaches. 
We consider two approaches in which varying amounts of information about the correlations between two- and
three-particle levels is dropped.
We can imagine these being relevant in situations where one has limited information on correlations,
because, for example, different ensembles have been used to calculate two- and three-particle quantities,
or the determination of the full covariance matrix is unstable.
The underlying idea here is that two-particle spectra might be able to pin down two-particle scattering quantities
sufficiently well that the three-particle spectra can be used primarily to determine three-particle scattering quantities.

Our first alternative is to use what we refer to as a ``chained fit''. 
Here, we first perform a fit to the two-particle sector. 
Minimizing the two-body $\chi^2$ function will lead to the two-particle best fit parameters, $\vec p^\text{\,fit}_\text{\,2P}$ and their covariance. We will use these values to construct the chained $\chi^2$ function as:
\begin{equation}
\chi^2_\text{chain}\left( \vec p \right)= 
\begin{pmatrix}
\Delta \vec p_\text{\,2P} \\
\Delta \vec X_\text{3P} 
\end{pmatrix}^T
\begin{pmatrix}
\text{cov}(\vec p^\text{\ fit}_\text{\,2P}, \vec p^\text{\ fit}_\text{\,2P} ) & \text{cov}(\vec p^\text{\ fit}_\text{\,2P}, \vec X_\text{3P})  \\ 
\text{cov}(\vec X_\text{3P}, \vec p^\text{\ fit}_\text{\,2P} )  & \text{cov}(\vec X_\text{3P}, \vec X_\text{3P} ) 
\end{pmatrix}^{-1}
\begin{pmatrix}
\Delta \vec p_\text{\,2P} \\
\Delta \vec X_\text{3P}
\end{pmatrix},
 \label{eq:chi2chained}
\end{equation}
where $\vec p = (\vec p_\text{\,2P}, \vec p_\text{\,3P}) $ is a vector that contains the two and three-particle parameters,
 ${\Delta \vec p^\text{ 2P}  = \vec p_\text{\,2P}^\text{ fit} - \vec p_\text{\,2P}}$, 
 and $\vec X_\text{3P}$ represents a specific spectral quantity for all three-particle energy levels. 
 Note that the method requires also covariance between the three-particle energy levels
 and the two-particle best-fit parameters, $\text{cov}(\vec X_\text{3P}, \vec p^\text{\ fit}_\text{\,2P} ) $.
 These can be estimated using a resampling technique, e.g., jackknife.

This approach can be further simplified by neglecting completely the off-diagonal terms in the covariance matrix of \Cref{eq:chi2chained}. In this case, the $\chi^2$ functions becomes:
\begin{align}
\begin{split}
\chi^2_\text{aug}\left(\vec p\right) = \Delta \vec X_\text{3P}^T \left(\text{cov}(\vec X_\text{3P}, \vec  X_\text{3P} ) \right)^{-1} \Delta \vec X_\text{3P} + \, \Delta \vec p_\text{\,2P}^{\ T} \, \left( \text{cov}(\vec p^\text{ fit}_\text{\,2P}, \vec p^\text{ fit}_\text{\,2P} )  \right)^{-1} \, \Delta \vec p_\text{\,2P}. \label{eq:chi2Bayes}
\end{split}
\end{align}
This can be seen as the augmented $\chi^2$ of a ``Bayesian fit'', where $\vec p^\text{ fit}_\text{2P}$ is the prior of those parameters. In the context of nondegenerate spectra, a further simplification of such Bayesian fits is possible.
For example, for the $\pi\pi K$ case, the Bayesian augmentation can include, or not, the correlations between
the fit parameters in the $\pi \pi$ and $\pi K$ channels.

\subsection{Fit results}
\label{sec:fitresults}

In this section we present results from fitting the spectra using the two- and three-particle
quantization conditions. We begin with examples showing the impact of using the different fitting
methods described above, and then present our core new results for the parameters describing
the $\pi\pi K$ and $K K \pi$ systems. Finally, we compare the results for the two-particle scattering
parameters (scattering length and effective range) obtained using different fits.

For every fit we need to choose a maximum value of $E_{\rm cm}$ for the levels to be included.
The quantization conditions that we use formally break down above the lowest inelastic threshold, which
for systems of pions occurs when two additional pions can be created 
(single-pion production being forbidden by G-parity), 
while for systems involving kaons the breakdown occurs when only a single additional pion can be produced.
The issue is discussed in detail in Ref.~\cite{\threepithreeK}, where it is noted that, in practice,
the quantization conditions are likely to remain applicable some distance above the nominal
maximal $E_{\rm cm}$, a conclusion supported by the numerical results of that work.
Thus, here we also work with values of $E_{\rm cm}$ that lie above the nominal maximal,
making the same choices for the $2\pi$, $3\pi$, $2K$, and $3K$ channels as in Ref.~\cite{\threepithreeK}, 
and similar choices for the $\pi K$, $\pi\pi K$ and $KK\pi$ channels.

As discussed above, we include only $s$- and $p$-wave terms in the quantization conditions, but not $d$-wave terms.
For two identical particles, i.e. for $\pi\pi$ and $KK$,
this implies that we must exclude from the fits levels that lie in nontrivial irreps,
since such levels are only shifted by $d$-wave terms ($p$ waves being absent).
For three identical particles, levels in nontrivial irreps are shifted by $s$-wave terms 
because there can be relative angular momentum between the dimer pair and the spectator, 
but these shifts are incomplete due to the absence of $d$-wave terms in the dimer.
Thus we also keep only $\pi\pi\pi$ and $KKK$ levels in trivial irreps in the fits.
This is different from the fits in Ref.~\cite{\threepithreeK}, where were able to include $d$-wave terms, and
thus also levels in nontrivial irreps.

By contrast, for the nondegenerate channels that are of central interest here, the inclusion of $p$ waves implies
that fits to levels in all available irreps are possible, and we include such levels in the fits.

\subsubsection{Comparison of fitting strategies}

We begin by showing examples of the differences between results obtained by fitting
to $E_{\rm cm}$ and $\Delta E_{\rm lab}$. 
As discussed above, the latter fits have the advantage of being to quantities that are closer to those
that are actually obtained from the lattice simulations, and are in this sense preferable.
We used $E_{\rm cm}$ fits in Ref.~\cite{\threepithreeK}, where we studied the $3\pi$ and $3K$ systems,
and our aim here is to study the impact of changing to $\Delta E_{\rm lab}$ fits.
We expect that the latter fits will lead, in general, to larger values of $\chi^2$,
but that this provides a more accurate reflection of the goodness of fit.
We recall that our fits to correlator ratios yield $\Delta E_{\rm lab} a$. This ``primary'' quantity is
then converted to $\Delta E_{\rm lab}/M$, where $M$ is either $M_\pi$ or $M_K$ depending on the
quantity being studied, and we use the value of $M$ from the rest frame fit in the corresponding jackknife
sample. The ``$\Delta E_{\rm lab}$ fits'' are to $\Delta E_{\rm lab}/M$.
When fitting to $E_{\rm cm}/M$, a further conversion is needed, first from $\Delta E_{\rm lab}$ to $E_{\rm lab}$,
and then by a boost to the rest frame. The key point is that this conversion depends on $M L$,
and that fluctuations in this quantity between jackknife samples leads to an increase in the errors 
in $E_{\rm cm}/M$. Thus we expect, in general, that fitting to $E_{\rm cm}/M$ will lead to
a smaller $\chi^2$, but that this reduction is not due to having a better fit, 
but rather due to the errors being overestimated.
What we do not know {\em a priori} is how the results for the fit parameters will change, and
that is the focus of our investigation here.

In a few cases, the difference between $E_{\rm cm}$ and $\Delta E_{\rm lab}$ fits is minimal.
An example is provided by fits to the $\pi K$ spectrum on the D200 ensemble, 
 which are shown in \Cref{tab:piKD200}.
We find good fits in both cases, leading to consistent fit parameters,
although there is a small decrease in the errors in the fit parameters when fitting to $\Delta E_{\rm lab}$.
Most striking is the change in the condition number of the correlation matrix for the values of $E_{\rm cm}$
or $\Delta E_{\rm lab}$.
The correlation matrix is closely related to the covariance matrix, differing by a normalization procedure
that guarantees each element of the principal diagonal is 1,
and each off-diagonal element lies in the range $[-1,1]$: 
\begin{align}
\begin{split}
\text{corr}\left(\vec X \right) = \left(\text{diag}\left[\text{cov}(\vec X, \vec  X ) \right]  \right)^{-\frac{1}{2}}
\text{cov}(\vec X, \vec  X )
\left(\text{diag}\left[\text{cov}(\vec X, \vec  X ) \right]  \right)^{-\frac{1}{2}} .
\end{split}
\end{align}
By providing a more natural normalization, the correlation matrix is better suited to estimate the condition number,
as it avoids issues of significantly different errors among the extracted energies which can lead to artificially large condition numbers.
The reduction in the condition number of the correlation matrix by an order of magnitude
for the $\Delta E_{\rm lab}$ fits implies that the fits will be more stable.
We observe a substantial reduction in the condition number in fits to all quantities.

\begin{table}[h!]
\centering
\begin{tabular}{|c|c|c|}
\hline
Method & $E_{\rm cm}$ & $\Delta E_{\rm lab}$   \\ \hline \hline
Cond. \# & 28630 & 68\\ \hline
$\chi^2$ & 28.3 & 31.0  \\ \hline
DOF     & 26-3=23  & 26-3=23  \\ \hline
 \hline
$B_0^{\pi\pi}$ & -13.24(64) &  -13.13(45) \\ \hline
$B_1^{\pi\pi}$ & -2.51(28)  &  -2.41(23) \\ \hline
$P_0^{\pi\pi}$ & 0.0029(13) & 0.0013(7)  \\ \hline
\end{tabular}
\caption{ Comparison of fitting approaches for the $\pi K$ spectrum on the D200 ensemble. 
All quantities are in units in which $M_\pi=1$.
The fits are to the 26 levels lying below the cutoff $E_{\rm cm}=5 M_\pi = (M_\pi+M_K) + 1.62 M_\pi$,
and use the fit form \Cref{eq:Adler}.
The position of the Adler zero is fixed to its leading order value (i.e. $z_{\pi K}=1)$.
}
\label{tab:piKD200}
\end{table}

In most cases, however, the value of $\chi^2$ increases substantially when fitting to $\Delta E_{\rm lab}$.
As a first illustration of this behavior,
we show, in \Cref{tab:KKD200Ecm},  fits to the $KK$ spectrum on D200.
First, we note that values of $\chi^2_{\rm ref}=\chi^2/{\rm DOF}$ are large in all cases, where DOF stands for degrees of freedom.
One reason for this is that we are fitting without including the $d$-wave interaction,
a choice we make, as explained above, in order to have practical fits when we consider nondegenerate
three-particle systems.
As an example of the impact of this omission, we note that,
were we to include a $d$-wave scattering length in the $E_{\rm cm}$ fit,
$\chi^2$ would be reduced by about 20~\cite{\threepithreeK}.
Our focus here, however, is on the differences between $E_{\rm cm}$ and $\Delta E_{\rm lab}$
fits, and we see that $\chi^2$ increases significantly, consistent with our general expectation discussed above.
However, we note that the fit parameters themselves change little ($B_1^{KK}$ is reduced by $\sim2\sigma$),
and the errors are essentially unchanged.

\begin{table}[h!]
\centering
\begin{tabular}{|c|c|c|}
\hline
Method & $E_{\rm cm}$ & $\Delta E_{\rm lab}$   \\ \hline \hline
Cond. \# & 163 & 28 \\ \hline
$\chi^2$ & 61 & 84  \\ \hline
DOF     &  28-2=26&  28-2=26 \\ \hline 
\hline
$B_0^{KK}$ & -2.865(49) & -2.865(46)  \\ \hline
$B_1^{KK}$ & -2.67(12) & -2.37(11) \\ \hline
\end{tabular}
\caption{Comparison of fitting approaches for the $KK$ spectrum on the D200 ensemble.
All quantities are in units in which $M_K=1$.
The fits are to the 28 levels in trivial irreps lying below the cutoff $E_{\rm cm} = 2.53 M_K = 2 M_K + 1.26 M_\pi$,
using the two-parameter Adler form of \Cref{eq:Adler},  with
the Adler zero fixed to its leading order position ($z_{KK}=1)$.
}
\label{tab:KKD200Ecm}
\end{table}

Now we turn to examples involving three-particle spectra.
The results from a joint fit to $2\pi$ and $3\pi$ levels are shown in \Cref{tab:pppN203},
while those for $2K$ and $3K$ levels are presented in \Cref{tab:KKKD200}.
In the former table, we also include rebinned results to be discussed below.

\begin{table}[h!]
\centering
\begin{tabular}{|c|c|c|c|}
\hline
Method & $E_{\rm cm}$ & $\Delta E_{\rm lab}$ &
$\Delta E_{\rm lab}$, rebin2
\\ \hline \hline
Cond. \# & 1681 & 744 & 717 \\ \hline
$\chi^2$ & 94 & 130 & 143  \\ \hline
DOF   & 27+27-4=50 & 27+27-4=50 & 27+27-4=50 \\ \hline 
\hline
$B_0^{\pi\pi}$ & -4.88(9)   & -4.87(9) & -4.86(8)  \\ \hline
$B_1^{\pi\pi}$ &  -2.27(11) & -1.90(9) & -1.91(9) \\ \hline
$\cK_{0}$ &  240(220)  & 500(210) &  310(180)\\ \hline
$\cK_{1}$ & -1700(330)  & -1400(340) & -1100(320) \\ \hline
\end{tabular}
\caption{Fits to the $\pi\pi+\pi\pi\pi$ spectrum on ensemble N203.
All quantities are in units in which $M_\pi=1$.
We use cutoffs of $E_{\rm cm}=3.46 M_\pi$ and $4.46 M_\pi$ respectively from the $\pi\pi$ and
$\pi\pi\pi$ spectra, and fit only to levels in trivial irreps, leading to 27 levels for each channel.
In the two-particle channel, the fit model used is the Adler zero form given in \Cref{eq:Adler}, 
with the Adler zero fixed to its leading order position ($z_{\pi\pi}=1$).
The fit model used in the three-particle channel includes only the $\cK_{0}$ and $\cK_{1}$ terms of \Cref{eq:KdfExpansion}.
}
\label{tab:pppN203}
\end{table}

\begin{table}[h!]
\centering
\begin{tabular}{|c|c|c|}
\hline
Method & $E_{\rm cm}$ & $\Delta E_{\rm lab}$\\  \hline \hline
Cond. \# & 7075 &  4044 \\ \hline
$\chi^2$ & 92  &   118  \\ \hline
DOF    & 28+26-4=50  & 28+26-4=50 \\ \hline 
\hline
$B_0^{KK}$     & -2.89(4)  &-2.91(4) 
\\ \hline
$B_1^{KK}$    & -2.58(13) & -2.29(11) 
\\ \hline
$\cK_{0}$    & -880(900)  &  -240(920) 
\\  \hline
$\cK_{1}$    & -10000(3500) & -8300(3600)
\\  \hline
\end{tabular}
\caption{Fits to the $KK+KKK$ spectrum on ensemble D200.
All quantities are in units in which $M_K=1$.
We use cutoffs of $E_{\rm cm}=2.53 M_K = 2 M_K + 1.26 M_\pi$ and $3.53 M_K=3M_K+1.26 M_\pi$
for the $KK$ and $KKK$ spectra, respectively, leading to 28 and 26 levels in the trivial irreps.
In the two-particle channel, the fit model used is the Adler zero form given in \Cref{eq:Adler}, 
with the Adler zero fixed to its leading order position ($z_{KK}=1$).
The fit model used in the three-particle channel includes only the $\cK_{0}$ and $\cK_{1}$ terms of \Cref{eq:KdfExpansion}.
}
\label{tab:KKKD200}
\end{table}

Both tables show the same pattern as for the $KK$ results above: there is an increase in $\chi^2$ when
using $\Delta E_{\rm lab}$ fits,  while fit parameters and errors are largely consistent.
The largest change is for $B_1$, which decreases by more than $2 \sigma$.
The conclusions drawn in Ref.~\cite{\threepithreeK}, namely that $\kdf$ is significantly different from zero,
and that $\Kisoone$ is negative, remain valid for the $\Delta E_{\rm lab}$ fits.
Since the fit parameters are highly correlated, one cannot judge the significance of a nonzero $\kdf$ from
the tables alone; using the full covariance matrices we find this to be 
7.0$\sigma$ and 4.4$\sigma$ for the $E_{\rm cm}$ and $\Delta E_{\rm lab}$ fits to $\pi\pi+\pi\pi\pi$,
while the corresponding results for the $KK+KKK$ fits are
$6.4\sigma$ and $4.5\sigma$, respectively.
Thus the significance of the nonzero $\kdf$ is somewhat reduced, but remains high.

\bigskip
In the remainder of this section we only consider fits to $\Delta E_{\rm lab}$.
Our first task is to compare the results of fits using
the different choices for $\chi^2$ described in \Cref{sec:fitstrategies} above. 
We do so only for the $\pi\pi+\pi K + \pi\pi K$ fits, on ensemble D200, as the pattern we find is
the same in all other fits.

In \Cref{tab:ppKD200} we compare the results when using
(a) the standard choice of $\chi^2$, given in \Cref{eq:chi2},
(b) chained fits, using the $\chi^2$ from \Cref{eq:chi2chained}, 
and (c) Bayesian fits using the $\chi^2$ given in \Cref{eq:chi2Bayes}. 
(The final column will be discussed in \Cref{sec:corefits} below.)
We recall that there are two types of Bayesian fits, depending on whether one
keeps the correlations between the results of the parameters for the $\pi\pi$ and $\pi K$ fits, or not.
We find little difference in the results of these two approaches, and present results only for the latter.
 
The standard fit is to a total of 69 levels,\footnote{We note here that one level (the second level in the $A_2(8)$ irrep) shows $\sim 3 \sigma$ fluctuations due to the choice of GEVP parameters. We opt to keep it in the fit, but have checked that removing this level from the fit barely impacts the best fit parameters in \Cref{tab:ppKD200}.} using 9 parameters. 
This is a challenging fit, but we see no signs of numerical instability in the calculation of $\chi^2$.
Indeed, the main challenge is finding the minimal $\chi^2$ with a large number of parameters,
and our minimizer takes $\sim 10^3$ iterations to converge. In all cases we have checked the fits
by repeating them with different initial conditions, and by using three independent codes.
The final $\chi^2$ is high, but a large part of this arises from the fit to the $\pi\pi$ sector, where
a fit to the $22$ levels alone leads to $\chi^2\approx 52$ due to the absence of $d$-wave terms
(as discussed above in the context of $KK$ fits).
In addition, we have seen above that moving from $E_{\rm cm}$ fits to $\Delta E_{\rm lab}$ fits leads
to increased $\chi^2$.

\begin{table}[h!]
\centering
\begin{tabular}{|c|c|c|c|c|}
\hline
Fit &  (a) Standard & (b) Chained & (c) Bayesian & (d) Standard${}^*$ \\ \hline \hline
Cond. \# & 2027 & 792 & 670 & 1881 \\ \hline
$\chi^2$ & 129  & 32 & 31 & 112
\\ \hline
DOF & 22+26+21-9=60 & 21+5-9=17 & 21+5-9=17 & 22+16+21-9=50
\\ \hline
\hline
$B_0^{\pi\pi}$     & -11.7(6) & -11.5(6) & -11.3(6) & -11.5(6)
\\ \hline
$B_1^{\pi\pi}$    & -2.4(4) & -2.5(4) & -2.4(4)   & -2.5(4)
\\ \hline
$B_0^{\pi K}$     & -13.0(4) & -13.2(4) & -13.1(4) & -12.9(4)
\\ \hline
$B_1^{\pi K}$    & -2.58(20) & -2.43(21) & -2.45(23) & -2.8(3)
\\ \hline
$P_0^{\pi K}$     & 0.0010(6)  & 0.0017(7) & 0.0014(7)  & 0.0007(6)
\\ \hline
$\cK_{0}$    & 220(70) & 500(200) & 650(310) & 190(80)
\\ \hline
$\cK_{1}$    & -620(340) & -300(500) & -70(620) & -690(340)
\\ \hline
$\mathcal K_B$    & 140(640) & -500(1100) & -500(1200) & 160(650)
\\ \hline
$\mathcal K_E$    & {290(410)} & 1200(800) & 2200(1400) & 170(420)
\\ \hline
\end{tabular}
\caption{Fits to the  $\pi\pi+\pi K+\pi\pi K$ spectrum on ensemble D200, using $\Delta E_{\rm lab}$.
All quantities are in units in which $M_\pi=1$.
For the first three columns,
we use cutoffs $E_{\rm cm}=3.74 M_\pi$, $5 M_\pi = (M_\pi+M_K)+1.62 M_\pi$,
and $5.4 M_\pi = (2 M_\pi+ M_K) + 1.02 M_\pi$ for the $\pi\pi$, $\pi K$ and $\pi\pi K$ channels,
respectively, leading to 22, 26 and 21 levels.
For fit (d), the cutoff for the $\pi K$ spectra is reduced to $4.64 M_\pi=(M_\pi+M_K)+ 1.26 M_\pi$,
so that there are 16 $\pi K$ levels.
In the two-particle channel, the fits use the functions of \Cref{eq:Adler} and \Cref{eq:EREp}, 
with the $\pi\pi$ and $\pi K$ Adler zeros fixed to their lowest-order values.
The fit model used in the three-particle channel is given by \Cref{eq:KdfExpansion}.
In fits (b) and (c), the number of DOF is given by the number of levels (21) plus the number of fit parameters
in the $\pi\pi+\pi K$ fits (2+3), minus the number of parameters fit (9).
}
\label{tab:ppKD200}
\end{table} 

The large number of levels that must be fit motivates investigating the alternatives provided by
using chained and Bayesian fits. We stress that the values of $\chi^2$ in the three fits cannot be compared.
A rough comparison can be obtained by adding to the $\chi^2$ for fits (b) and (c) the values of
the $\chi^2$ from the individual $\pi\pi$ and $\pi K$ fits, which are $52$ and $31$, respectively,
leading to total values of $\sim 115$  for fits (b) and (c). However, this ignores the impact on the $\chi^2$
in fit (a) of including the full correlations between the levels.

Comparing the results, we see that the central values for all two-particle scattering parameters are
very similar, and have essentially the same errors, in all three fits.
By contrast, while the values of the three-particle parameters are consistent within errors,
those errors are significantly larger for the chained fit than the standard fit, and larger still for
the Bayesian fit. In other words, the information on correlations between levels that is lost
when using ``sequential'' fits makes it harder to pin down the (already challenging) three-particle parameters.
It is thus no surprise that the significance of the nonzero $\kdf$ is reduced when moving from the standard fit
to the chained and Bayesian fits:
it is $3.4\sigma$, $2.4\sigma$ and $2.8\sigma$, respectively for the three fits.

A similar pattern is observed for all other channels, and thus we conclude that standard fits are clearly
preferable if they are possible, as is the case here. We use only standard fits for our central results to be
presented shortly. 

The final issue that we address in this subsection is whether to rebin the results on the D200 ensemble.
As discussed above, the results on the D200 ensemble are rebinned by $N_{\rm rebin}= 3$,
 leading to 771 jackknife
samples. As discussed in Ref.~\cite{\threepithreeK}, this rebinning is useful to account for
autocorrelations, and indeed the errors in the energy levels increase as one increases the rebinning factor.
However, for N203, where we have only 666 configurations, rebinning can lead to unstable fits
when the number of samples becomes too close to the number of levels being considered.
We have tested this in several cases, two examples being given in \Cref{tab:pppN203} above and
\Cref{tab:ppKN203} below. What we find is that the fits are stable for both $N_{\rm rebin} = 2$ or 3,
that they lead to essentially the same results for all fit parameters, with no change in errors, but that
$\chi^2$ increases significantly (as do the condition numbers).
We interpret these results as indicating that any autocorrelations have minimal effect on the scattering
parameters, while the reduction in the number of samples leads to less well conditioned fits.
Thus we choose to use no rebinning for our central fits N203.
The choice of using $N_{\rm rebin} = 3$ for D200 is, therefore, conservative.

\subsubsection{Fits to nondegenerate channels}
\label{sec:corefits}

As noted above, our central values are obtained from fits to $\Delta E_{\rm lab}$,
using the standard, fully-correlated $\chi^2$,
and without rebinning on ensemble N203. Results for $\pi\pi+\pi K + \pi\pi K$ are shown in
\Cref{tab:ppKD200} (above) and \Cref{tab:ppKN203}, while those for $KK+\pi K+KK\pi$ are shown in
\Cref{tab:KKpD200} and \Cref{tab:KKpN203}.

\begin{table}[h!]
\centering
\begin{tabular}{|c|c|c|c|}
\hline
Fit & 72 level & 82 level & 82 level, rebin 2
\\ \hline \hline
Cond. \# & 612 & 834 & 985\\ \hline
$\chi^2$ &  117 & 119 & 155 \\ \hline
DOF &$27\!+\!19\!+\!26\!-\!9=63$ & $27\!+\!19\!+\!36\!-\!9=73$ & $27\!+\!19\!+\!36\!-\!9=73$\\ \hline
\hline
$B_0^{\pi\pi}$     & -5.05(10) & -5.05(10) & -5.01(10) 
\\ \hline
$B_1^{\pi\pi}$  & -1.77(9) & -1.78(9) & -1.82(9) 
\\ \hline
$B_0^{\pi K}$   &-5.40(11) & -5.39(11) & -5.39(11)
\\ \hline
$B_1^{\pi K}$    & -1.88(17) & -1.89(17) & -1.80(16)
\\ \hline
$P_0^{\pi K}$     & 0.005(4) & 0.006(4) & 0.007(4)
\\ \hline
$\cK_{0}$    & -250(160) & -240(150) & -380(160)
\\ \hline
$\cK_{1}$    & -1400(600) & -1300(600) & -1080(580)
\\ \hline
$\mathcal K_B$    & 1100(800) & 990(740) & 1020(680)
\\ \hline
$\mathcal K_E$    & -3300(1000) & -3200(1000) &-3200(1000)
\\ \hline
\end{tabular}
\caption{$\Delta E_{\rm lab}$ fits to the $\pi\pi+\pi K+\pi\pi K$ spectrum on ensemble N203.
All quantities are in units in which $M_\pi=1$.
In all fits, the cutoffs for $\pi \pi$ and $\pi K$ are 
$E_{\rm cm}=3.464 M_\pi=2 M_\pi + 1.464 M_\pi$ and $3.4 M_\pi=(M_\pi+M_K) + 1.12 M_\pi$,
corresponding to 27 and 19 levels, respectively.
For the first column, the cutoff for $\pi \pi K$ is $4.1 M_\pi = (2M_\pi+M_K)+ 0.82 M_\pi$,
while for the remaining two columns it is $4.3 M_\pi = (2 M_\pi+M_K) + 1.02 M_\pi$,
corresponding to 26 and 36 levels, respectively.
In the two-particle channel, the
fits are to \Cref{eq:Adler} and \Cref{eq:EREp} with $\pi\pi$ and $\pi K$ Adler zeros fixed to their lowest-order values.
The fit model used in the three-particle channel is given by \Cref{eq:KdfExpansion}.
}
\label{tab:ppKN203}
\end{table}

\begin{table}[h!]
\centering
\begin{tabular}{|c|c|c||c|}
\hline
Fit & ADLER2 & ADLER3 & ERE3  \\ \hline \hline
Cond. \# & 10200 & 10200 & 10200 \\ \hline
$\chi^2$ & 163  &   162 & 161 \\ \hline
DOF     & $28\!+16\!+\!29\!-\!9 = 64$  & $28\!+16\!+\!29\!-\!10 = 63$ & $28\!+16\!+\!29\!-\!10 = 63$ 
\\ \hline 
\hline
$B_0^{KK}$     & -2.90(4)  &-3.55(78) & -2.87(4) \\ \hline
$B_1^{KK}$    & -2.32(11) & -1.95(45) & 1.40(25) \\ \hline
$z^2_{KK}/B_2^{KK}$     & $1$ (fixed) & 0.76(28)  & -1.25(38) \\ \hline
$B_0^{K\pi}$     & -2.44(7)  &  -2.42(7) & -2.41(7) \\ \hline
$B_1^{K\pi}$    & -2.18(31) & -2.28(32) & -2.28(32)\\ \hline
$P_0^{K\pi}$     & 0.027(8)  &  0.027(8) & 0.027(8) \\ \hline
$\cK_{0}$    & 180(270)  &  170(270) & 150(270)\\  \hline
$\cK_{1}$    & -6600(1700) & -6800(1700) & -6800(1700)
\\  \hline
$\mathcal K_B$    & 2800(1300) & 2800(1300) & 2900(1200)
\\  \hline
$\mathcal K_E$    & -5700(3800) & -5900(3700) & -6000(3700)
\\  \hline
\end{tabular}
\caption{$\Delta E_{\rm lab}$ fits to the $KK+\pi K +KK\pi$ spectrum on ensemble D200.
All quantities are in units in which $M_K=1$. Note that this means that $B_0^{K\pi}$ differs from $B_0^{\pi K}$ in \Cref{tab:ppKD200} by powers of $(M_\pi/M_K)$, and similarly for other parameters. 
Cutoffs are given by $E_{\rm cm}=2.53 M_K$, $1.95 M_K = (M_K+M_\pi) + 1.26 M_\pi$,
and $2.832 M_K = (2 M_K+M_\pi)+ 0.98 M_\pi$ for the $KK$, $\pi K$ and $KK\pi$ channels, respectively,
corresponding to 28, 16, and 29 levels, respectively.
In the $KK$ channel, the ADLER2 and ADLER3 fits use \Cref{eq:Adler},
with the Adler zero fixed to its leading-order position for the former fit and allowed to vary for the latter,
while the ERE3 fit uses \Cref{eq:ERE}.
In all three fits the $\pi K$ channel is fit using the ADLER2 form for the s-wave K matrix and \Cref{eq:EREp}
for the p wave, while
the fit model used in the three-particle channel is given by \Cref{eq:KdfExpansion}.
}
\label{tab:KKpD200}
\end{table}

\begin{table}[h!]
\centering
\begin{tabular}{|c|c|c||c|}
\hline
Fit & ADLER2 & ADLER3 & ERE3
\\ \hline \hline
Cond. \# & 770 & 770 & 770 \\ \hline
$\chi^2$ & 181  &  173 &  187\\ \hline
DOF     & $23\!+\!19\!+\!32\!-\!9 = 65$  & $23\!+\!19\!+\!32\!-\!10 = 64$ & $23\!+\!19\!+\!32\!-\!10 = 64$
\\ \hline 
\hline
$B_0^{KK}$     & -3.41(5)  & -2.8(2) & -3.39(5)
\\ \hline
$B_1^{KK}$    & -2.08(8) &   -2.39(13) & 2.20(15)
\\ \hline
$z^2_{KK}/B_2^{KK}$     & $1$ (fixed)  &  1.19(6) & -1.53(13)
\\ \hline
$B_0^{K\pi}$     & -3.25(8)  & -3.31(8) & -3.24(8)
\\ \hline
$B_1^{K\pi}$    & -2.17(19) & -2.07(19) & -2.18(19)
\\ \hline
$P_0^{K\pi}$     & 0.018(8)  &  0.020(8) & 0.021(8)
\\ \hline
$\cK_{0}$    & 170(300)  &  260(310) & 90(310)
\\  \hline
$\cK_{1}$    & -3900(1700) & -3900(1700) & -3400(1700)
\\  \hline
$\mathcal K_B$    & 3500(1600) &  3500(1600) & 3500(1700)
\\  \hline
$\mathcal K_E$    & -1100(2000) & -400(2000) & -100(2100)
\\  \hline
\end{tabular}
\caption{$\Delta E_{\rm lab}$ fits to the $KK+\pi K+KK\pi$ spectrum on ensemble N203.
All quantities are in units in which $M_K=1$. Note that this means that $B_0^{K\pi}$ differs from $B_0^{\pi K}$ in \Cref{tab:ppKN203} by powers of $(M_\pi/M_K)$, and similarly for other parameters. 
Cutoffs are given by $E_{\rm cm}=2.9 M_K$, $2.66 M_K = (M_K+M_\pi) + 1.12 M_\pi$,
and $3.521 M_K = (2 M_K+M_\pi)+ 0.95 M_\pi$ for the $KK$, $\pi K$ and $KK\pi$ channels, respectively,
corresponding to 23, 19, and 32 levels, respectively. For the $KK$ channel, only the ground state
in the frame with $\bm d^2=9$ is kept.
In the $KK$ channel, the ADLER2 and ADLER3 fits use \Cref{eq:Adler},
with the Adler zero fixed to its leading-order position for the former fit and allowed to vary for the latter,
while the ERE3 fit uses \Cref{eq:ERE}.
In all three fits the $\pi K$ channel is fit using the ADLER2 form for the s-wave K matrix and \Cref{eq:EREp}
for the p wave, while
the fit model used in the three-particle channel is given by \Cref{eq:KdfExpansion}.
}
\label{tab:KKpN203}
\end{table}

For the $\pi\pi+\pi K + \pi\pi K$ fits we show examples of the results of fitting to different levels.
Table~\ref{tab:ppKD200} compares our ``standard" fit, in the first column, with an alternative, in the final column,
in which the cutoff on the $\pi K$ levels is reduced from $0.62 M_\pi$ above the inelastic threshold
($2M_\pi+M_K$) to $0.26 M_\pi$ above this threshold. The fits are of similar quality and give consistent
results, and errors, for all parameters, indicating that our more aggressive cutoff is acceptable.
Table~\ref{tab:ppKN203} compares using a cutoff below the inelastic threshold for the $\pi\pi K$ channel (72 level fit)
to one slightly above this threshold (82 level fit). Here we are investigating whether fits to 82 levels are stable,
and we find that they are. The results from the two fits are consistent within errors, and we use the results
from the 82-level fit henceforth.

For the $KK+\pi K+KK\pi$ fits we show comparisons between different fit choices.\footnote{%
Note that in the $KK+\pi K+KK\pi$ case we fit to $\Delta E_{\rm lab}/M_K$ rather than to $\Delta E_{\rm lab}/M_\pi$, as in the ${\pi\pi+\pi K + \pi\pi K}$ fits.
This makes a small difference as there are fluctuations of the single-particle masses between jackknife samples.}
In Ref.~\cite{\threepithreeK} it was observed that the $KK$ interactions were slightly better described on the
D200 ensemble (for which the kaon is heavier) if,
compared to our standard two-parameter Adler-zero fit,
one either allowed the position of the Adler zero to float, or used a three-parameter ERE fit.
Thus we have done all three fits for both ensembles and compare the results in the tables.
Freeing the position of the Adler zero leads to
no significant improvement in $\chi^2_{\rm ref}$ on D200, but a mild improvement on N203.
Switching to an ERE fit, leads to a slight improvement on D200, but a worse fit on N203.
Results for the $\pi K$ and $KK\pi$ fit parameters are essentially unchanged.

One of our major aims is to study how well three-particle interactions can be determined by using a
large collection of energy levels. As can be seen from the tables, the significance of the nonzero values
for individual terms in $\kdf$ varies, with the most significant parameter being $\cK_{1}$ on the D200
ensemble. The significance of the entire $\kdf$ being nonzero, including the correlations between the
parameters, is $3.4\sigma$ and $3.6\sigma$ for the $\pi\pi+\pi K+\pi\pi K$ standard 
fits on the D200 and N203 ensembles,  respectively,
and $5.0\sigma$ and $2.4\sigma$ for these ensembles in the $KK+\pi K+KK\pi$ fits.

\subsubsection{Visualization of fits}
\label{sec:plots}

We have shown several global views of the fits in \Cref{fig:n203_kpp_spectrum,fig:n203_kkp_spectrum,fig:d200_kpp_spectrum,fig:d200_kkp_spectrum},
which illustrate that the fits match the
spectrum well both in the fit range and also above the nominal inelastic threshold.
To investigate this more carefully, we need to zoom in and show results for the quantities $\Delta E_{\rm lab}$
to which we actually fit. This is the purpose of this section.

\begin{figure}[h!]
\centering
\includegraphics[width=0.9\linewidth]{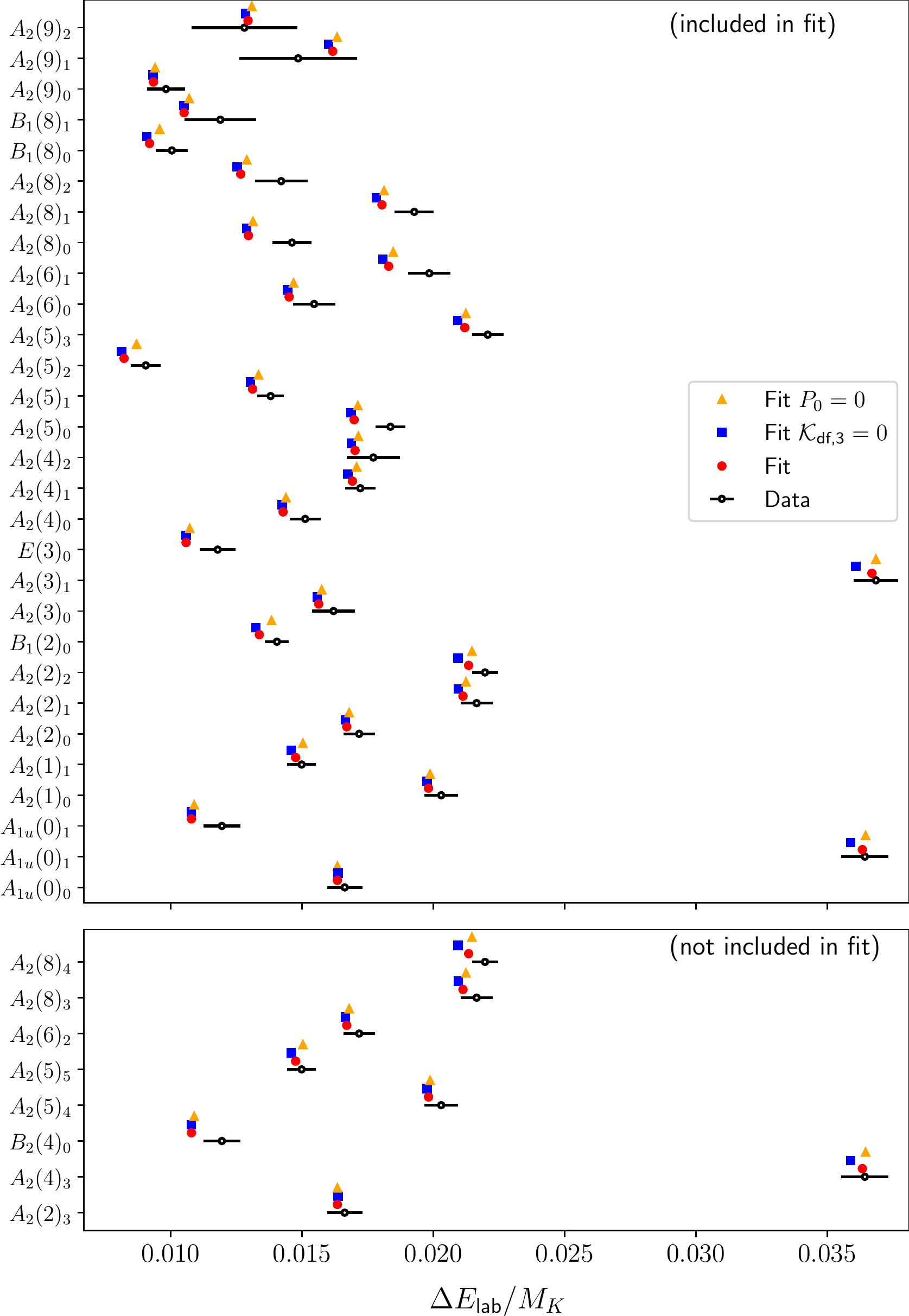}  
\caption{Comparison of values for $\Delta E_{\rm lab}/M_K$ to the predictions of various fits for $K^+K^+\pi^+$
levels on D200. The upper panel shows 29 levels included in the fits, 
while the lower panel shows eight that lie above our maximal $E_{\rm cm}$
and are thus not included in the fits. 
Level are denoted by their irrep, followed in parenthesis by the value of total momentum-squared parametrized
by $\bm d_{\rm ref}^2$, with the subscript indicating the level number for the given irrep and total momentum,
starting at $0$. Above each data point we show, using red dots, blue squares, and orange triangles, respectively,
the fit values from the ADLER3 fit of \Cref{tab:KKpD200}, the values predicted by the quantization condition if
$\kdf=0$ but all other parameters are unchanged, and the values predicted if $P_0=0$ with all other parameters unchanged.
}
\label{fig:KKpD200}
\end{figure}

We focus first on the $KK+\pi K +KK\pi$ fit on the D200 ensemble, as this is the fit for which both $\kdf$
and $P_0$ are determined to be nonzero with the greatest significance. Specifically, we consider the
ADLER3 fit to 73 levels in \Cref{tab:KKpD200}. 
In the upper panel of \Cref{fig:KKpD200} we compare the results for $\Delta E_{\rm lab}$ from our simulations
to those obtained using the quantization condition with the best fit parameters.
The latter is shown by the red dots just above each of the data points.
We see that the shifts are small (of order 1\% of $M_K$), but are determined with errors ranging from
a few percent to $10-15\%$. We note that all levels, whether in trivial or nontrivial irreps, are shifted by
the same order of magnitude. This differs from the situation with the two particle spectrum, as will be seen below,
and is due to the fact that the two-particle K matrix, $\cK_2$, contributes to all irreps. Indeed, as is well known,
and was clearly seen in Ref.~\cite{\threepithreeK},
the dominant physical effect leading to these energy shifts is the two-particle interaction,
and it is nontrivial to determine the subdominant contribution of $\kdf$.
To illustrate the impact of $\kdf$, 
we also show, as blue squares, the results predicted by the quantization condition if we set $\kdf=0$ while keeping
all other parameters unchanged. The shifts in the levels are small, reaching the size of the error bars in the data
only for the higher energy levels. From this figure alone, it would appear that there is little chance of determining
$\kdf$, but this is misleading because all levels are correlated, and the fit includes not only these 29 levels, but also
the 28 $K^+K^+$ and 16 $K^+\pi^+$ levels. One illustration of the claim
that the nonzero value of $\Kdf$ leads to a significant  improvement in the fit is 
that $\chi^2$ increases from $162$ to $188$ when $\Kdf$ is turned off.

We also include, using orange triangles,
 the result of turning off the $p$-wave $\pi K$ scattering amplitude by setting $P_0=0$, with all other
parameters unchanged from the ADLER3 fit. This changes the levels by an amount that is typically smaller
than that caused by setting $\kdf$ to zero. 

We next investigate how the fit works for the levels that are not included in the fit, because they lie above the
maximal CMF energy, $E_{\rm cm}= (2 M_K+M_\pi) + 0.98 M_\pi$. These levels thus lie at or above the
inelastic threshold. We have determined eight of them, and the comparison of their lab shifts to the ADLER3 fit
is shown in the lower panel of \Cref{fig:KKpD200}, along with the predictions if $\Kdf$ or $P_0$ are set to zero.
The values of the shifts are comparable to those for the fitted levels, and we see that the fit continues
to work at a similar level of accuracy even in the inelastic regime.

\begin{figure}[h!]
\centering
\includegraphics[width=0.9\linewidth]{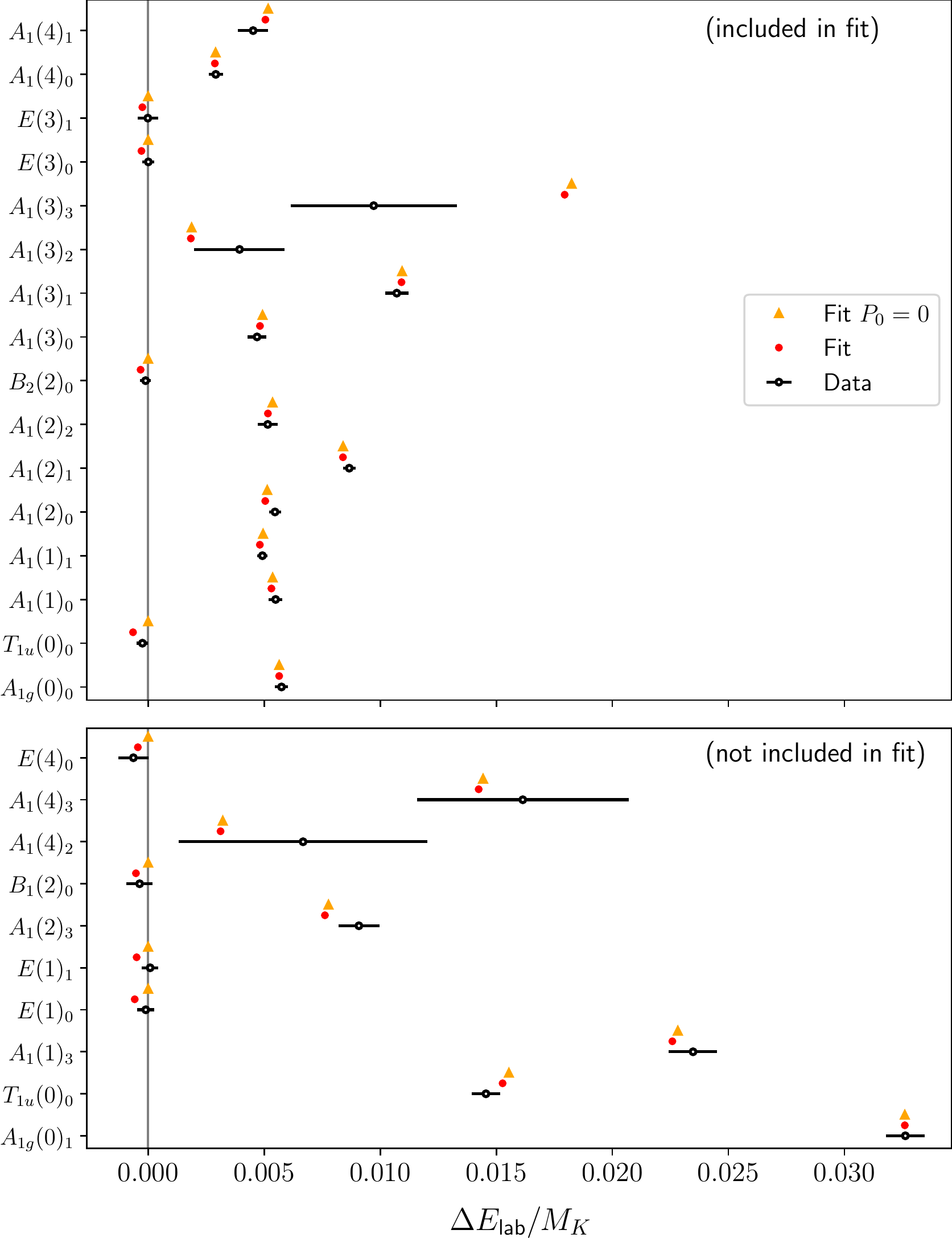}  
\caption{Same as for \Cref{fig:KKpD200} except for the $\pi^+K^+$ levels in the ADLER3 $KK+\pi K+KK\pi$ fits. The upper panel shows the 16 levels included in the fit, 
while the lower panel shows 10 levels lying above our maximal $E_{\rm cm}$ and which thus are not included
in the fit.
}
\label{fig:KpD200}
\end{figure}

We next display the corresponding results for the $K^+\pi^+$ channel in the same fit. 
These are shown in \Cref{fig:KpD200}: the upper panel shows 16 levels that are
included the fit, and the lower panel shows 10 that are not. 
We recall from \Cref{tab:KKpD200} that the cutoff for the fit lies at
$E_{\rm cm} = (M_K+M_\pi) + 1.26 M_\pi$ in this channel, 
and thus lies slightly above the inelastic threshold.
Since two-particle levels do not depend on $\Kdf$, we show only the impact of setting $P_0$ to zero.

For the trivial irreps, the situation is similar to that for the $KK\pi$ levels: the fit works well, and continues to
do so in the inelastic regime. The shifts due to the nonzero $P_0$ are small, although they increase for the
higher-lying levels. The levels in nontrivial irreps, however, are shifted only by the $p$-wave interactions
(as can be seen by the fact that the yellow triangles lie at $\Delta E_{\rm lab}=0$.
This is as expected based on group-theoretical considerations. 
For some of these levels, there is evidence that the energy shift is negative, and it is this that
leads to the result that the scattering length is slightly attractive.

The situation for the other fits and ensembles is qualitatively similar, and we do not show the corresponding
plots for these other cases.

\clearpage

\subsubsection{Derived results for two-particle scattering quantities}
\label{sec:twoparticle}
 
In this section we collect the results for the scattering lengths and effective ranges for
$\pi\pi$, $\pi K$ and $KK$ scattering. These quantities are obtained from the fits presented above,
and from additional fits to the following systems: $\pi\pi$ alone, $\pi\pi+\pi\pi\pi$, $KK$ alone, and
$KK+KKK$. We stress that these additional fits are different from those presented in
Ref.~\cite{\threepithreeK}, as here we fit to $\Delta E_{\rm lab}$, rather than $E_{\rm cm}$.
Our overall aim in this section is to compare the results, and in particular the errors, obtained by using
fits to different sets of levels. We discuss the chiral behavior of these results in the
following section.

One of the methods used here is to determine the scattering length from the $1/L$ expansion of
the energy shift of the ground state (i.e. for which the noninteracting state has all particles at rest).
For two nondegenerate scalars the result is~\cite{Beane:2003yx}
\begin{equation}
\Delta E_{\rm g.s.}^{(2)} = \frac{2 \pi a_0}{\mu_{12} L^3} \left[
1 - \cI \frac{a_0}{\pi L} + (\cI^2 - \cJ) \left(\frac{a_0}{\pi L}\right)^2 + \cO(1/L^3) \right]\,,
\end{equation}
where $\mu_{12}$ is the reduced mass of the pair, $a_0$ the corresponding scattering length, and the constants 
are~\cite{Beane:2007qr}
\begin{equation}
\cI = -8.91363291759\,,\qquad
\cJ = 16.532315960\,.
\end{equation}
For three degenerate particles of mass $M$ the result is~\cite{Beane:2007qr}
\begin{equation}
\Delta E_{\rm g.s.}^{(3)} = \frac{12 \pi a_0}{M L^3} \left[
1 - \cI \frac{a_0}{\pi L} + (\cI^2 +\cJ) \left(\frac{a_0}{\pi L}\right)^2 + \cO(1/L^3) \right]\,.
\end{equation}
We observe that the first two orders are a factor of $3$ larger for three particles than for two particles, reflecting
the number of two-particle pairs. This implies an increased sensitivity to $a_0$ in the three-particle channel.

The above-described truncated $1/L$ expansions of the ground state energy shifts are  
sometimes used to determine scattering lengths from two-particle energy shifts 
(see, e.g., Ref.~\cite{PhysRevD.96.054516}).
Truncation at the $1/L^5$ term is needed to have a  one-to-one relation between the energy shift and $a_0$,
since the $1/L^6$ terms include the effective range and, for three particles, also
a subtracted version of the three-particle amplitude at threshold~\cite{\HSTH}.
Thus, in this approach, one must proceed by assuming the $1/L^6$ term is numerically small,
and then estimate the resulting systematic error due to truncation~\cite{PhysRevD.96.054516}.
Here our interest is less in the central value obtained in this fashion, but rather in the size of the error
obtained in this method compared to those from global fits.

\begin{table}[h!]
\centering
\begin{tabular}{|c|c|c|c|c|c|}
\hline
Fit & $B_0^{\pi\pi}$ & $B_1^{\pi\pi}$ & $\chi^2/\text{DOF}$ & $M_\pi a_0^{\pi\pi}$ & $M_\pi^2 a_0^{\pi\pi} r_0^{\pi\pi}$
\\ \hline
\multicolumn{6}{|c|}{ADLER2 fits}
\\ \hline
$\pi\pi+\pi\pi\pi$ 
& -11.5(6)  &  -2.4(4) & 74/37 & 0.0869(47) & 2.587(86)
\\ \hline
$\pi\pi+\pi K+\pi\pi K$ 
& -11.5(6)  & -2.5(4) & 112/50 & 0.0869(45) & 2.561(84)
\\ \hline
$\pi\pi$ (22 levels) 
& -11.6(7) & -2.3(4)  & 52/20 & 0.0862(50) & 2.604(90)
\\ \hline
$\pi\pi$ (12 levels) 
& -11.5(8) & -3.0(7) & 25/10 & 0.0873(60) & 2.48(15)
\\ \hline
\multicolumn{6}{|c|}{Fitting ground-state energy shift to $1/L$ expansion}
\\ \hline
$\pi\pi\pi$ 
& N/A & N/A & N/A & 0.0885(76) & N/A
\\ \hline
$\pi\pi$ 
& N/A & N/A & N/A & 0.0816(76) & N/A
\\ \hline 
\multicolumn{6}{|c|}{$E_{\rm cm}$ ADLER2 fit, including $d$ waves~\cite{\threepithreeK}}
\\ \hline
$\pi\pi+\pi\pi\pi$ ($E_{\rm cm}$)
& N/A &  N/A & N/A & 0.0859(41)(28) & 2.62(8)(26)
\\ \hline 
\end{tabular}
\caption{
Comparison of $\pi\pi$ scattering parameters from different fits to $\Delta E_{\rm lab}$ on the D200 ensemble,
using units in which $M_\pi=1$. 
The $\pi\pi+\pi\pi\pi$ fits are to 22 $\pi\pi$ and 19 $\pi\pi\pi$ levels, all in trivial irreps,
with cutoffs at $E_{\rm cm}=3.74$ and $4.74$,
respectively (the same values as used in Ref.~\cite{\threepithreeK}).
The $\pi\pi+\pi K+\pi\pi K$ fits are from \Cref{tab:ppKD200} (59 level fit).
The next two rows give the results of fits to the $\pi\pi$ data alone,
with cutoff for the two fits being $E_{\rm cm}=3.74 M_\pi$ and $3.0 M_\pi$, respectively.
The next block of two rows give the results obtained using the ground-state energy shifts alone,
as described in the text.
The final row gives the results from Ref.~\cite{\threepithreeK} from $E_{\rm cm}$ fits, with the first error
being statistical and the second a systematic error due to the variation between fits.
}
\label{tab:ppD200}
\end{table}

\begin{table}[h!]
\centering
\begin{tabular}{|c|c|c|c|c|c|}
\hline
Fit & $B_0^{\pi\pi}$ & $B_1^{\pi\pi}$ & $\chi^2/\text{DOF}$ & $M_\pi a_0^{\pi\pi}$ & $M_\pi^2 a_0^{\pi\pi} r_0^{\pi\pi}$
\\ \hline
\multicolumn{6}{|c|}{ADLER2 fits}
\\ \hline
$\pi\pi+\pi\pi\pi$ & -4.87(9)  &  -1.9(9) & 130/50 & 0.2052(38) 
& 2.222(46)
\\ \hline
$\pi\pi+\pi K+\pi\pi K$ & -5.05(10)  & -1.78(9) & 119/73 & 0.1981(39) 
& 2.296(45)
\\ \hline 
$\pi\pi$ (27 level) & -4.94(11) & -1.83(9)  & 66/25 & 0.2024(44)
& 2.258(51)
\\ \hline
$\pi\pi$ (16 level) & -4.87(11) & -2.00(15) & 19/14 & 0.2055(48)
& 2.178(74)
\\ \hline 
\multicolumn{6}{|c|}{Fitting ground-state energy shift to $1/L$ expansion}
\\ \hline
$\pi\pi\pi$ & N/A & N/A & N/A & 0.2125(56) & N/A
\\ \hline
$\pi\pi$ & N/A & N/A & N/A & 0.2095(54) & N/A
\\ \hline 
\multicolumn{6}{|c|}{$E_{\rm cm}$ ADLER2 fit, including $d$ waves~\cite{\threepithreeK}}
\\ \hline
$\pi\pi+\pi\pi\pi$ ($E_{\rm cm}$)
& N/A &  N/A & N/A & 0.2059(34)(21) & 2.15(5)(12)
\\ \hline 
\end{tabular}
\caption{
Comparison of $\pi\pi$ scattering parameters from different fits to $\Delta E_{\rm lab}$ on the N203 ensemble, 
using units in which $M_\pi=1$.
The $\pi\pi\pi$
fits are from Table~\ref{tab:pppN203}, while those for
$\pi\pi K$ fits are from Table~\ref{tab:ppKN203}.
The next two rows give results from fits to the $\pi\pi$ data alone, 
using cutoffs $E_{\rm cm}=3.436 M_\pi$ (which is the same as that used in the $\pi\pi+\pi\pi\pi$
and $\pi\pi+\pi K+\pi\pi K$ fits) and $3.0 M_\pi$.
The next block of two rows give the results obtained using the ground-state energy shifts alone,
as described in the text.
The final row gives the results from Ref.~\cite{\threepithreeK} from $E_{\rm cm}$ fits, with the first error
being statistical and the second a systematic error due to the variation between fits.
}
\label{tab:ppN203}
\end{table}

We begin with results for $\pi\pi$ scattering, which are shown in \Cref{tab:ppD200} and \Cref{tab:ppN203},
respectively, for the D200 and N203 ensembles.
These include the results of the core fits presented above, as well as those to different numbers of $\pi\pi$ levels,
and the results of fitting the truncated $1/L$ expansions to the ground-state energy shifts.
We see that all central values are consistent within $1-2\sigma$, including results obtained from in
Ref.~\cite{\threepithreeK} using $E_{\rm cm}$ fits.
Our focus here is on a comparison of the errors. In particular,
we want to know if anything is gained by increasing the number of levels in the fits, i.e.
 moving from a single level (in the ground-state-only fits), 
 to the $\pi\pi$ levels alone, and finally to the fits involving two- and three-particle levels. 
 Our results indicate that, for the scattering lengths, the errors decrease slightly as the number of
levels in the fit increases. This trend is what we would have naively expected, but the size of the change
is relatively small. We attribute this smallness to the fact that the higher levels constrain the phase shift
at values away from threshold, and because of the strong correlations between two- and three-particle levels.
A similar pattern is observed for the errors in $a_0 r_0$.

In light of this discussion,
in the chiral fits below we will take the values from the fits involving two- and three-particle levels.
Specifically, we use the $\pi\pi+\pi K +\pi\pi K$ fits, since the
$\chi^2_{\rm ref}$ of this fit is significantly smaller on the N203 ensemble, while the values and errors
are essentially the same as those from the $\pi\pi+\pi\pi\pi$ fits on D200.

We next turn to the results for $KK$ scattering parameters, which are collected in \Cref{tab:KKD200}
and \Cref{tab:KKN203}. Here we also show results with the ADLER3 and ERE3 choices for the $KK$ phase
shift, as these can lead to improved fits, as noted above.
For a given choice of fit type,
the sizes of the errors follow a similar pattern to those for $\pi\pi$ scattering,
 with those from fits involving two and three particles being smallest.
As expected, 
errors increase for the fits in which the $KK$ phase shift is described with three parameters.

All fits using $\Delta E_{\rm lab}$ lead to consistent results for $a_0^{KK}$.
Additionally, these fits are consistent with those using $E_{\rm cm}$ at the $\lesssim 2\sigma$ level.
There is, however, much more variation in the results for $a_0^{KK} r_0^{KK}$.

\begin{table}[h!]
\centering
\resizebox{\columnwidth}{!}{
\begin{tabular}{|c|c|c|c|c|c|c|}
\hline
Fit & $B_0^{KK}$ & $B_1^{KK}$ & $z^2_{KK}/B_2^{KK}$
&$\chi^2/\text{DOF}$ & $M_K a_0^{KK}$ & $M_K^2 a_0^{KK} r_0^{KK}$
\\ \hline
\multicolumn{7}{|c|}{ADLER2}
\\ \hline
$KK\!+\!KKK$ & -2.91(4)  &  -2.29(11) & N/A & 118/50 & 0.3438(50)
& 1.43(9)
\\ \hline
{$KK\!+\!\pi K\!+\!KK\pi$} & -2.90(4)  & -2.32(11) & N/A & 163/64 & 0.3444(48)
& 1.40(9)
\\ \hline
$KK$ (28) & -2.87(5) & -2.37(11)  & N/A  & 84/26 & 0.3489(56)
& 1.34(10)
\\ \hline
$KK$ (15) &  -2.84(5) & -2.50(18) & N/A &25/13 & 0.3521(62)
& 1.24(15)
\\ \hline
\multicolumn{7}{|c|}{Fitting ground-state energy shift to $1/L$ expansion}
\\\hline
$K KK$ & N/A & N/A & N/A  & N/A & 0.3541(83) & N/A
\\ \hline
$K K$  & N/A & N/A & N/A  & N/A & 0.3531(83) & N/A
\\ \hline
\multicolumn{7}{|c|}{ADLER3}
\\\hline
$KK\!+\!KKK$  & -3.4(8)  & -2.0(5)  & 0.8(3)  & 117/49 & 0.3467(62)
& 1.20(29)
\\ \hline
{$KK\!+\!\pi K\!+\!KK\pi$}& -3.6(8)& -2.0(5) & 0.8(3) & 162/63 & 0.3478(58)
& 1.14(27)
\\ \hline
$K K$ (28 levels) & -3.3(8) & -2.2(5) & 0.9(3)  & 84/25 & 0.3512(69)
& 1.17(32)
\\ \hline
$K K$ (15 levels) & -2.6(1.2) & -2.7(8) & 1.1(4)  & 25/12 & 0.3515(71)
& 1.31(48)
\\ \hline
\multicolumn{7}{|c|}{ERE3}
\\\hline
$KK\!+\!KKK$  & -2.88(5)  & 1.47(25) & -1.32(38)  & 116/49 & 0.3472(57) & 1.02(17)
\\ \hline
{ $KK\!+\!\pi K\!+\!KK\pi$} & -2.87(4) & 1.40(25)  & -1.25(38) & 161/63 & 0.3486(54) & 0.98(16)
\\ \hline
$K K$ (28 levels) & -2.84(5) & 1.35(29) & -1.25(44)  
& 84/25 & 0.3524(65) & 0.95(19)
\\ \hline
$K K$ (15 levels) & -2.84(5) & 1.65(41) & -2.4(1.0)
& 24/12 & 0.3518(67) & 1.16(27)
\\ \hline 
\multicolumn{7}{|c|}{$E_{\rm cm}$ fit, ADLER3, including $d$ waves~\cite{\threepithreeK}}
\\ \hline
$KK\!+\!KKK$ & N/A  & N/A & N/A  & N/A
& 0.3648(59)(29) & 0.77(24)(23)
\\ \hline
\end{tabular}
}
\caption{
Comparison of $K K$ scattering parameters from different fits to $\Delta E_{\rm lab}$ 
on the D200 ensemble, using units in which $M_K=1$. The exception is the final line,
which is a fit to $E_{\rm cm}$ from Ref.~\cite{\threepithreeK}.
Blocks are divided according to the form of the fit function used for the $KK$ phase shift, as discussed in
the text. Note that the meaning of the entry in the $z_{KK}^2/B_2^{KK}$ column depends on
which fit function is used: it is $z_{KK}^2$ in the ADLER3 block and $B_2^{KK}$ in the ERE3 block.
$KK+\pi K+KK\pi$ fits are the same as those in \Cref{tab:KKpD200}.
$KK+\pi K+KK\pi$ fits are the same as those in \Cref{tab:KKpN203}.
The ADLER2 $KK+KKK$ fit is the same as that in \Cref{tab:KKKD200};
the ADLER3 and ERE3 $KK+KKK$ fits use the same set of energy levels.
}
\label{tab:KKD200}
\end{table}

\begin{table}[h!]
\centering
\resizebox{\columnwidth}{!}{
\begin{tabular}{|c|c|c|c|c|c|c|}
\hline
Fit & $B_0^{KK}$ & $B_1^{KK}$ & $z^2_{KK}/B_2^{KK}$ 
&$\chi^2/\text{DOF}$ & $M_K a_0^{KK}$ & $M_K^2 a_0^{KK}r_0^{KK}$
\\ \hline
\multicolumn{7}{|c|}{ADLER2}
\\ \hline
$KK\!+\!KKK$ 
& -3.38(4)  &  -2.12(7) & N/A & 153/42& 0.2959(37) & 1.75(5)
\\ \hline
$KK\!+\!\pi K\!+\!KK\pi$
& -3.41(5)  & -2.08(8) & N/A & 181/65 & 0.2931(39) & 1.78(6)
\\ \hline
$K K$ (23 levels) 
& -3.36(5)& -2.14(8) & N/A & 90/21 & 0.2973(42) & 1.73(6)
\\ \hline
\multicolumn{7}{|c|}{Fitting ground-state energy shift to $1/L$ expansion}
\\\hline
$K K K$ 
& N/A & N/A & N/A & N/A & 0.3060(51) & N/A
\\ \hline
$K K$ 
& N/A & N/A & N/A & N/A & 0.3022(54) & N/A
\\ \hline
\multicolumn{7}{|c|}{ADLER3}
\\ \hline
$KK\!+\!KKK$ 
& -2.72(19)  &  -2.45(12) & 1.21(6) & 144/41 &  0.2896(43)& 2.28(20)
\\ \hline
$KK\!+\!\pi K\!+\!KK\pi$
&-2.81(20) & -2.39(13) & 1.19(6) & 173/64 &  0.2872(45)& 2.26(20)
\\ \hline
$K K$ (23 levels) & -2.75(21) & -2.47(13)  & 1.20(7) & 88/20 & 0.2926(49) & 2.18(21)
\\ \hline 
\multicolumn{7}{|c|}{ERE3}
\\ \hline
$KK\!+\!KKK$ 
&-3.32(4) & 1.94(13)  & -1.33(12) & 175/41 & 0.3008(39) & 1.17(7)
\\ \hline
$KK\!+\!\pi K\!+\!KK\pi$
&-3.39(5) & 2.20(15) & -1.53(13) & 187/64  & 0.2953(41) & 1.30(7)
\\ \hline
$K K$ (23 levels) 
&-3.33(5) &2.04(17) &-1.40(14)  &96/20  & 0.2999(46) & 1.22(8)
\\ \hline
\multicolumn{7}{|c|}{$E_{\rm cm}$ fit, ADLER3, including $d$ waves~\cite{\threepithreeK}}
\\ \hline
$KK\!+\!KKK$ & N/A  & N/A & N/A  & N/A
& 0.3012(44)(18) & 1.92(19)(41)
\\ \hline
\end{tabular}
}
\caption{
Comparison of $K K$ scattering parameters from different fits to $\Delta E_{\rm. lab}$ 
on the N203 ensemble, using units in which $M_K=1$. The exception is the final line which
is a fit to $E_{\rm cm}$ from Ref.~\cite{\threepithreeK}.
Blocks are divided according to the form of the fit function used for the $KK$ phase shift, as discussed in
the text. Note that the meaning of the entry in the $z_{KK}^2/B_2^{KK}$ column depends on
which fit function is used: it is $z_{KK}^2$ in the ADLER3 block and $B_2^{KK}$ in the ERE3 block.
The $KK$ and $KKK$ fits use cutoffs of $2.9 E_K$ and $3.9 M_K$, respectively, keeping only levels in trivial
irreps, and discarding all but the lowest $\bm d^2=9$ level in the $KK$ channel.
This leads to 23 levels in both channels.
}
\label{tab:KKN203}
\end{table}

Finally, in \Cref{tab:pKD200} and \Cref{tab:pKN203}, we present results for $\pi K$ scattering parameters.
In both cases we show results with two choices of $E_{\rm cm}$ cutoff for the $\pi K$ fits: the 16-level
fit with excellent $\chi^2_{\rm ref}$ and the more aggressive 32-level fit with slightly poorer fit quality.
The results for $P_0^{\pi K}$, $a_0^{\pi K}$ and $r_0^{\pi K}$ are consistent within $1-2\sigma$.
The pattern of the sizes of errors is consistent with that described above.

\begin{table}[h!]
\centering
\begin{tabular}{|c|c|c|c|c|c|c|}
\hline
Fit & $B_0^{\pi K}$ & $B_1^{\pi K}$ & $P_0^{\pi K}$ &$\chi^2/\text{DOF}$
& $ M_\pi a_0^{\pi K}$ & $M_\pi a_0^{\pi K} r_0^{\pi K}$
\\ \hline \hline
$\pi\pi\!+\!\pi K\!+\!\pi\pi K$ & -12.9(4)  &  -2.8(3) & 0.0007(6) & 112/50 & 0.110(3)
& 1.154 (53)
\\ \hline
$KK\!+\!\pi K\!+\!KK\pi$ 
& -13.7(4)  & -2.3(3) & 0.0020(6)& 162/63 & 0.103(3) & 1.263(54)
\\ \hline
$\pi K$ (16 level) 
& -13.1(5)& -2.5(3) & 0.0012(7) & 15/13 & 0.107(4) & 1.217(62)
\\ \hline
$\pi K$ (26 level) 
& -13.1(5)& -2.4(2) & 0.0013(6) & 31/23 & 0.107(4) & 1.229(45)
\\ \hline
\multicolumn{7}{|c|}{Fitting ground-state energy shift to $1/L$ expansion}
\\\hline
$\pi K$ & N/A & N/A & N/A & N/A & 0.106(5) & N/A
\\ \hline
\end{tabular}
\caption{
Comparison of $\pi K$ scattering parameters from different fits on the D200 ensemble, 
using units in which $M_\pi=1$.
All fits use the ADLER2 $+$ ERE1 form for the $\pi K$ phase shift.
 The first row gives the results from the 59-level fit in \Cref{tab:ppKD200},
 while the second gives the result from the ADLER3-$KK$ fit in \Cref{tab:KKpD200}, 
 with the latter converted  to units in which $M_\pi=1$.
The next two give the results of fits to the $\pi K$ data alone, 
The cutoff energies for the these fits are, respectively, $E_{\rm cm}=3.64 M_\pi$ 
and  $5.0 M_\pi$, with the former also used for the $\pi K$ channel in the $\pi\pi+\pi K+\pi\pi K$ 
and $KK+\pi K+KK\pi$ fits. 
The final row shows the result of fitting the ground-state energy shift to the $1/L$ expansion.
}
\label{tab:pKD200}
\end{table}

\begin{table}[h!]
\centering
\begin{tabular}{|c|c|c|c|c|c|c|}
\hline
Fit & $B_0^{\pi K}$ & $B_1^{\pi K}$ & $P_0^{\pi K}$ &$\chi^2/\text{DOF}$ &
$M_\pi a_0^{\pi K}$ & $M_\pi^2 a_0^{\pi K} r_0^{\pi K}$
\\ \hline \hline
$\pi\pi\!+\!\pi K\!+\!\pi\pi K$ & -5.39(11)  &  -1.89(17) & 0.006(4) & 119/73 & 0.208(4)
& 1.693(74)
\\ \hline
$KK\!+\!\pi K\!+\!KK\pi$ & -5.40(13)  & -2.07(19) & 0.010(4)& 173/64 & 0.208(5)
& 1.630(83)
\\ \hline
$\pi K$ (19 level) & -5.42(15)& -2.07(21) & 0.004(4) & 21.1/16 
& 0.207(6) & 1.630(94)
\\ \hline
$\pi K$ (36 level) & -5.45(15)& -2.02(21) & 0.002(3) & 36.1/33 
& 0.206(6) & 1.653(80)
\\ \hline
\multicolumn{7}{|c|}{Fitting ground-state energy shift to $1/L$ expansion}
\\\hline
$\pi K$ & N/A & N/A & N/A & N/A & 0.213(7) & N/A
\\ \hline
\end{tabular}
\caption{
Comparison of $\pi K$ scattering parameters from different fits on the N203 ensemble, 
using units in which $M_\pi=1$.
All fits use the ADLER2 $+$ ERE1 form for the $\pi K$ phase shift.
 The first row gives the results from the 82-level fit in \Cref{tab:ppKN203},
 while the second gives the result from the ADLER3-$KK$ fit in \Cref{tab:KKpN203}, 
 with the latter converted  to units in which $M_\pi=1$.
The next two rows give the results of fits to the $\pi K$ data alone,  using cutoff energies 
$3.4 M_\pi$ and $4.3 M_\pi$, respectively,
with the former also used for the $\pi K$ channel in the $\pi\pi+\pi K+\pi\pi K$ 
and $KK+\pi K+KK\pi$ fits. 
The final row shows the result of fitting the ground-state energy shift to the $1/L$ expansion.
}
\label{tab:pKN203}
\end{table}

%% file: results.tex

\clearpage
\section{Discussion of results}
\label{sec:discussion}

In this section, we present and discuss our final results for different two- and three-meson scattering quantities. First, in \Cref{sec:scatteringparams}, we provide our final numbers for these quantities by combining results from different fits. Then, in \Cref{sec:ChPTRs}, we compare these numbers to expectations and predictions from ChPT. 
Finally, in \Cref{sec:discretization}, we discuss the size of discretization errors, based on
the LO Wilson-ChPT results presented above.

\subsection{Final results for scattering parameters}
\label{sec:scatteringparams}

In \Cref{sec:fitresults}, we have presented results for the two- and three-meson scattering parameters 
using different fit forms and strategies. While we find overall consistency, 
it is useful to have a set of final results that contain both statistical uncertainties as well an estimate of the systematic spread due to different fit forms. Note that we always use lab-frame shift fits for this set of final results.

We first consider two-particle parameters. 
Results for scattering lengths and for the combination $M_X a^{XY}_0 r^{XY}_0$
are given in \Cref{tab:a0summary} and \Cref{tab:r0summary}, respectively.
In each case, the central values are obtained by averaging the results from a pair of three-particle fits
(which pair will be explained shortly), 
and include a systematic error that is obtained from the spread of the fit results
(and which we call the ``fit systematic'').
For the $\pi\pi$ case, the central values are the average of those from fits to 
$\pi\pi+\pi\pi\pi$ and $\pi\pi+\pi K + \pi\pi K$—see  \Cref{tab:ppD200,tab:ppN203}---while
the fit systematic is the standard deviation obtained from the two results. 
For the $\pi K$ case, the same procedure is used but now taking the $\pi\pi+\pi K + \pi \pi K$
and $KK+ \pi K +KK\pi$ fit results from \Cref{tab:pKD200,tab:pKN203}.
For the $KK$ case, the central value and statistical error are obtained using the same procedure
but taking the $KK+KKK$ and $KK+\pi K+KK\pi$ fits obtained using the ADLER3 parametrization for the
$KK$ phase shift from \Cref{tab:KKD200,tab:KKN203}.
The fit systematic  is given by the standard deviation of the results from these two fits 
together with those obtained with the ADLER2 and ERE3 parametrizations (six fits in total). 
In all cases, we take the largest of the statistical errors when combining results.

\begin{table}[h!]
\centering
\begin{tabular}{|c|c|c|c|c|}
\hline
Ensemble & $M_\pi a_0^{\pi\pi}$ & $M_\pi a_0^{\pi K}$ & $M_K a_0^{KK}$ & $P_0^{\pi K}=-M_\pi^3 a_1^{\pi K}$
\\ \hline
D200 & 0.0869(47)(0) & 0.107(3)(5) & 0.3473(62)(19) & 0.0014(6)(9)
\\ \hline
N203 & 0.2017(39)(50) & 0.208(5)(0) & 0.2884(45)(48) & 0.008(4)(3)
\\ \hline
\end{tabular}
\caption{
Final results for $s$- and $p$-wave scattering lengths obtained by combining fits in \Cref{sec:twoparticle}, 
as explained in the text.
The errors are respectively statistical and systematic, with the latter obtained from variations between fits.
}
\label{tab:a0summary}
\end{table}

\begin{table}[h!]
\centering
\begin{tabular}{|c|c|c|c|}
\hline
Ensemble & $M_\pi^2 a_0^{\pi\pi} r_0^{\pi\pi}$& $M_\pi^2 a_0^{\pi K} r_0^{\pi K}$ & $M_K^2 a_0^{KK} r_0^{KK}$ 
\\ \hline
D200 & 2.574(86)(18) & 1.209(54)(77) & 1.17(29)(19)
\\ \hline
N203 & 2.259(46)(52) & 1.662(83)(45) & 2.27(20)(46)
\\ \hline
\end{tabular}
\caption{
Final results for products $a_0 r_0$ obtained by combining fits in \Cref{sec:twoparticle}, as explained in the text.
The errors are respectively statistical and systematic, with the latter obtained from variations between fits.
}
\label{tab:r0summary}
\end{table}

For the three-particle parameters, we quote the final values in \Cref{tab:Kdfsummary}. In this case we take the results from a single fit—the one we view as most reliable, based on the discussion in the previous section—and do not quote a fit systematic as any such error would be dwarfed by the statistical errors.

\begin{table}[h!]
\centering
\begin{tabular}{|c|c|c|c|c|}
\hline
Ensemble & $\cK_{0}$ & $\cK_{1}$ & $\cK_B$ & $\cK_E$
\\ \hline
\multicolumn{5}{|c|}{$\pi\pi + \pi K   + K K \pi$ fits}
\\ \hline
D200 & 190(80) & -690(340) & 160(650) & 170(420)
\\ \hline
N203 & -240(150) & -1300(600) & 990(740) & -3200(1000)
\\ \hline
\multicolumn{5}{|c|}{$KK + \pi K + KK\pi$ fits}
\\ \hline
D200 & 170(270) & -6800(1700) & 2800(1300) & -5900(3700)
\\ \hline
N203 &260(310) & -3900(700) & 3500(1600) & -400(2000)
\\ \hline
\end{tabular}
\caption{
Final results for parameters in $\kdf$ for $2+1$ systems. Results are from the Standard${}^*$ fit in
\Cref{tab:ppKD200}, the 82-level fit in \Cref{tab:ppKN203}, and the ADLER3 fits in
\Cref{tab:KKpD200} and \Cref{tab:KKpN203}.
}
\label{tab:Kdfsummary}
\end{table}

\subsection{Comparison with chiral perturbation theory}
\label{sec:ChPTRs}

We now turn to the comparison of the final results of \Cref{tab:a0summary,tab:r0summary,tab:Kdfsummary} to ChPT. 
We perform fits to ChPT expressions wherever they are available,
and compare with the form of the expected dependence on $M_\pi^2$ in other cases.
We also compare to previous results in the literature.

\begin{figure}[h!]
\centering
\includegraphics[width=0.8\linewidth]{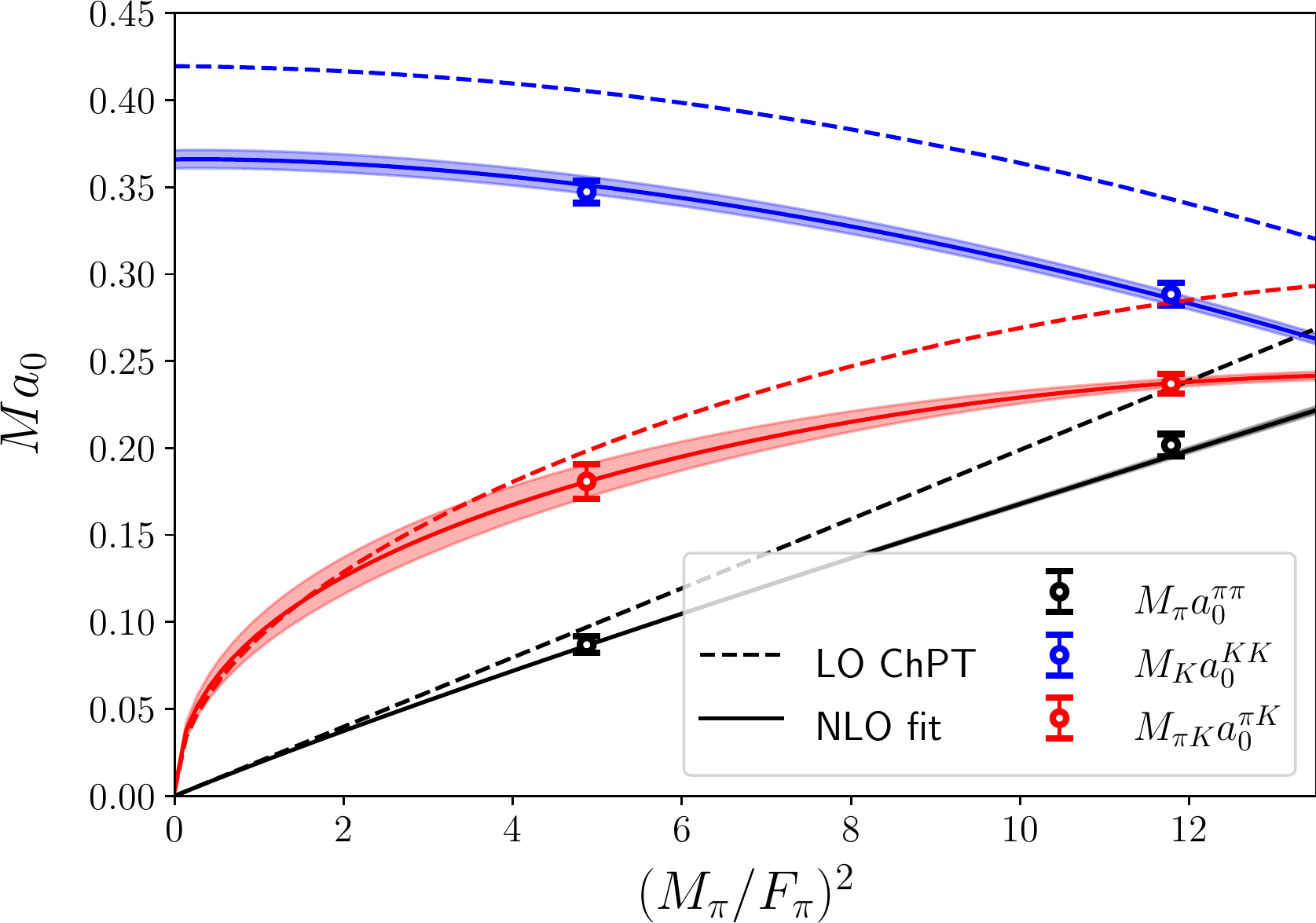}  
\caption{Results for $M_\pi a^{\pi\pi}$, $M_{\pi K} a^{\pi K}$ and $M_K a^{KK}$
 as a function of $M_\pi^2/F_\pi^2$, where ${M_{\pi K} = (M_{\pi} + M_{K})/2}$. 
 The LO ChPT result is shown, along with a fit to NLO SU(3) ChPT. 
 The shaded bands show the $1\sigma$ uncertainties in the fit.
}
\label{fig:a0_all}
\end{figure}

We start with the scattering lengths. 
Figure \ref{fig:a0_all} shows the results for each of the dimensionless $s$-wave scattering 
lengths,\footnote{%
We use $M_{\pi K} a^{\pi K} = [(M_\pi+M_K)/2] a^{\pi K}$ rather than the quantity $M_\pi a^{\pi K}$ quoted earlier so as to
separate the curves, and because the $a^2$ dependence predicted by WChPT is simpler, as seen
in \Cref{eq:a0pKCorrection}.}
along with the LO chiral prediction, and
a fit to the NLO expressions in \Cref{eq:a0pp,eq:a0pK,eq:a0KK}.
We have not determined the correlations between the three quantities on a given ensemble,
and thus use an uncorrelated fit.
In order to plot the result as a function of $(M_\pi/F_\pi)^2$, we use an interpolating function for 
$M_K/F_K$ as a function of $M_\pi/F_\pi$ extracted from the results of Ref.~\cite{\threepithreeK}.
We find a good fit with $\chi^2/\text{DOF} = 1.5/4$ (6 data points and 2 parameters). 
We determine the two LECs
(evaluated at a renormalization scale $\mu = 4 \pi F_\pi$) to be
\begin{equation}
L_{\pi \pi} = -8.77(36)\times 10^{-4}, \quad L_5 = 0.0(1.5)\times 10^{-3}\,.
\label{eq:LpipiL5}
\end{equation}
We can compare these values to previous determinations. 
In Ref.~\cite{\threepithreeK}, a larger value of $L_{\pi \pi}=-1.13(3) \cdot 10^{-3}$ was found, which is many
standard deviations away from our new result. 
However,  an important difference that may explain the discrepancy
is that Ref.~\cite{\threepithreeK} included $d$-wave  interactions, which is not possible here. 
For $L_5$, we can compare to results from lattice QCD, which are summarized in the FLAG report~\cite{FlavourLatticeAveragingGroupFLAG:2021npn}, 
and based on Refs.~\cite{MILC:2010hzw,Dowdall:2013rya}), 
or from phenomenological determinations~\cite{Bijnens:2011tb}. 
These are, however, quoted at a different renormalization scale, $\mu =770$ MeV. 
Changing the renormalization scale using
\begin{equation}
L_i^r\left(\mu_2\right)=L_i^r\left(\mu_1\right)+\frac{\Gamma_i}{16 \pi^2} \ln \left(\frac{\mu_1}{\mu_2}\right)\,, \label{eq:LECrunning}
\end{equation}
with $ \Gamma_5=3/8$, we find that the result from our fit yields $L_5(770 \text{ MeV}) = 1.0(1.5) \cdot 10^{-3} $. 
Varying the choice of $4\pi F_\pi$ to take for the initial scale 
(using the physical value of $F_\pi$, or the value on either of the ensembles)
leads to changes in $L_5$ that are significantly smaller than the error.
Our result for $L_5$ is in agreement with all values in the literature, 
although we note that our error is much larger than that in the other values.

Next, we discuss our results for the effective range parameters, which are presented in
 \Cref{tab:r0summary} in the combination $M^2_X r^{XY} a_0^{XY}$. 
For the case of identical particles ($X=Y=\pi$ or $K$), 
the LO ChPT prediction from \Cref{sec:ChPT} is that this quantity equals $3$.
For two pions, the results lie 15\% and 25\% below this prediction on the D200 and N203 ensembles,
respectively, which is consistent with being due to an NLO correction.
For two kaons, the results lie very far away from the LO prediction.
Both findings are qualitatively similar to those obtained in Ref.~\cite{\threepithreeK}.

For the $\pi K$ channel, which is a novel result of this work,
the LO ChPT prediction—given in \Cref{eq:a0r0pK}—depends on the ensemble:
\begin{equation}
M_\pi^2 a_0^{\pi K} r_0^{\pi K} \bigg \rvert^\text{LO ChPT}_\text{D200} =  1.597, \quad M_\pi^2 a_0^{\pi K} r_0^{\pi K}\bigg \rvert^\text{LO ChPT}_\text{N203} =   2.395. 
\end{equation}
Our results in \Cref{tab:r0summary} lie $\sim 25\%$ and $\sim 30\%$, respectively,
below the LO ChPT prediction. 
Again we view this as reasonable consistency, given the absence of NLO corrections.

\begin{figure}[tbh!]
\centering
\includegraphics[width=0.8\linewidth]{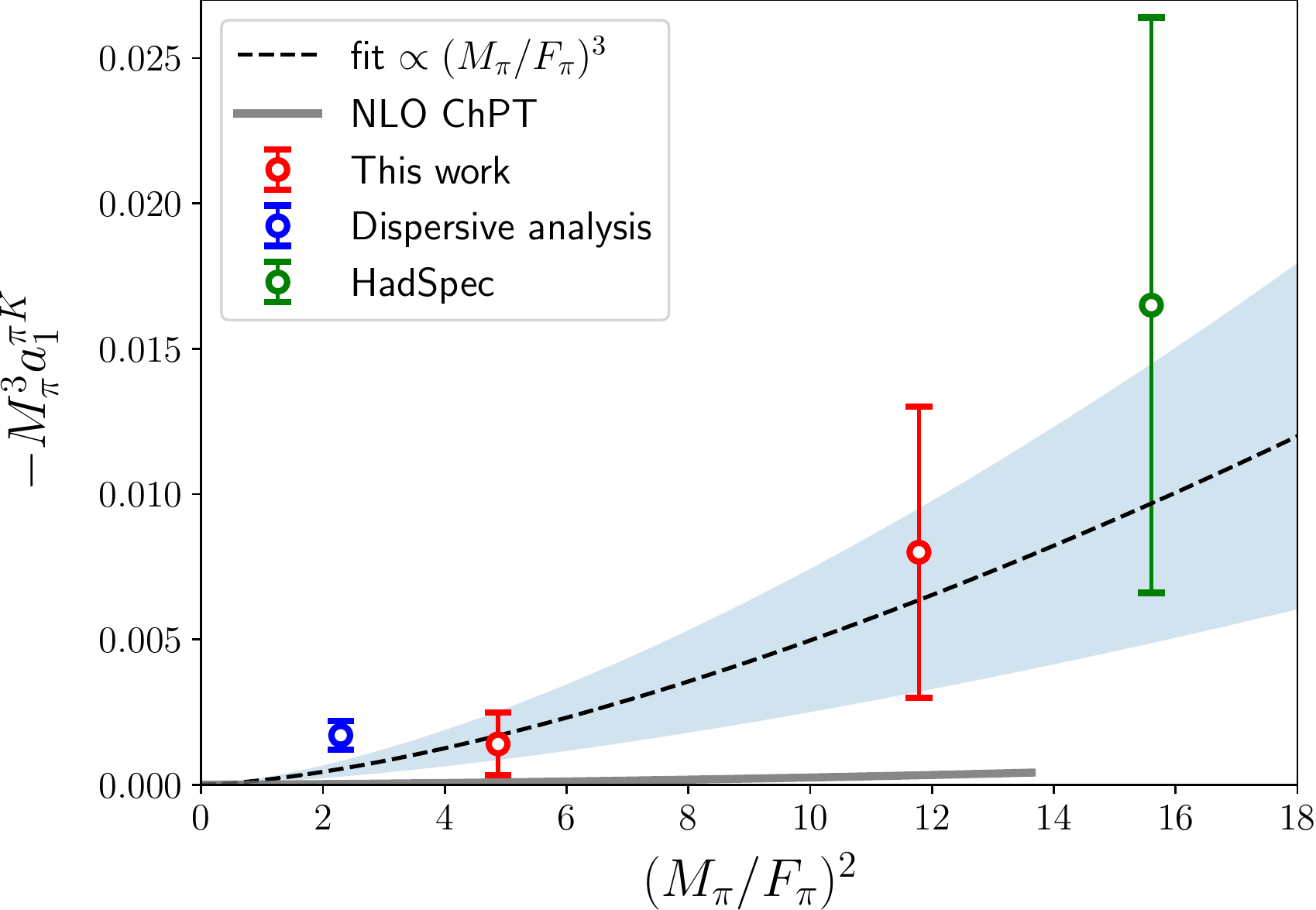}  
\caption{Results for the $p$-wave scattering parameter, $P_{0}^{\pi K} = - M_\pi^3 a_1^{\pi K}$, 
plotted as a function of $M_\pi^2/F_\pi^2$. A fit to the leading chiral scaling of $P^{\pi K}_0 \propto (M_\pi/F_\pi)^3$
is shown with the corresponding error band, as well as the NLO ChPT prediction as described in \Cref{app:pK}. Also included are the result at the physical point from the dispersive analysis of Ref.~\cite{Pelaez:2020gnd}
(see Table 29 of that work), and the lattice QCD determination of the HadSpec collaboration at 
a heavier pion mass~\cite{Wilson:2014cna}. 
}
\label{fig:P0piK}
\end{figure}

We now turn to the $p$-wave $\pi^+ K^+$ scattering length, reported in the rightmost column of \Cref{tab:a0summary}
through the dimensionless combination $P^{\pi K}_0 = - M_\pi^3 a_1^{\pi K}$. 
Note that, in contrast to all the $s$ wave results, the value of $P^{\pi K}_0$ corresponds to slightly attractive interactions.
We plot the results for the two ensembles in \Cref{fig:P0piK}, including a fit to the leading chiral behavior
given by \Cref{eq:P0ChPT}, which shows reasonable consistency.
 
We also plot the NLO ChPT prediction given in \Cref{app:pK}. 
To do so we use values for the requisite LECs determined in Ref.~\cite{Amoros:2001cp} from experimental
data (specifically, fit 10 to $\mathcal O(p^4)$ from that work).
As can be seen, the NLO ChPT result has the same sign as our results, 
but its magnitude is significantly smaller. 
The failure of NLO ChPT for this quantity was, in fact, expected,
based on the observation of Ref.~\cite{Bijnens:2004bu} that the
NNLO contribution is two orders of magnitude larger than the NLO one at the physical point 
(see table 2 of that work).

We can also compare to the expectations and results in the literature from experiment and dispersive analyses.
The current understanding is summarized in figure~10 of Ref.~\cite{Pelaez:2020gnd}. 
Experimental results~\cite{Estabrooks:1977xe} for the $p$-wave phase shift point 
to a negative (repulsive) value at high energies. 
By contrast, the dispersive analysis indicates a change of sign for the phase at around 
$\sqrt{s}\simeq M_K+3M_\pi$ (physical values of the masses), resulting in an attractive scattering length. 
The value from analysis of Ref.~\cite{Pelaez:2020gnd} is also shown in \Cref{fig:P0piK}. 
As can be seen, our results for the two ensembles of this work are in qualitative agreement 
with the low-momentum behavior found by the dispersive analysis. 
We note that our fits only include levels in the region where the phase shift is expected to stay positive. 

We are aware of two other LQCD results concerning $p$-wave $\pi^+ K^+$ scattering. 
First, Ref.~\cite{Lang:2012sv}, 
reports a single energy level far from threshold (at much higher energy than our levels,
and in the inelastic regime) that is dominated by $p$-wave interactions. 
There, the $p$-wave $\pi K$ interactions seems repulsive, 
which is consistent with what experiments find at those high energies.
This result therefore gives no information concerning the scattering length.

Second, Ref.~\cite{Wilson:2014cna} computed the $p$-wave scattering length at heavy meson masses,
$M_\pi\simeq 391$ MeV and $M_K\simeq 549$ MeV, and its sign and magnitude are consistent with our results at lighter pion masses. We include this result with the label ``HadSpec" in the plot, although it is not strictly speaking
comparable as Ref.~\cite{Wilson:2014cna} does not follow the same chiral trajectory.
We conclude that, overall, the results from LQCD are in qualitative agreement with 
dispersive and experimental results.

\begin{figure}[h!]
\centering
\includegraphics[width=0.8\linewidth]{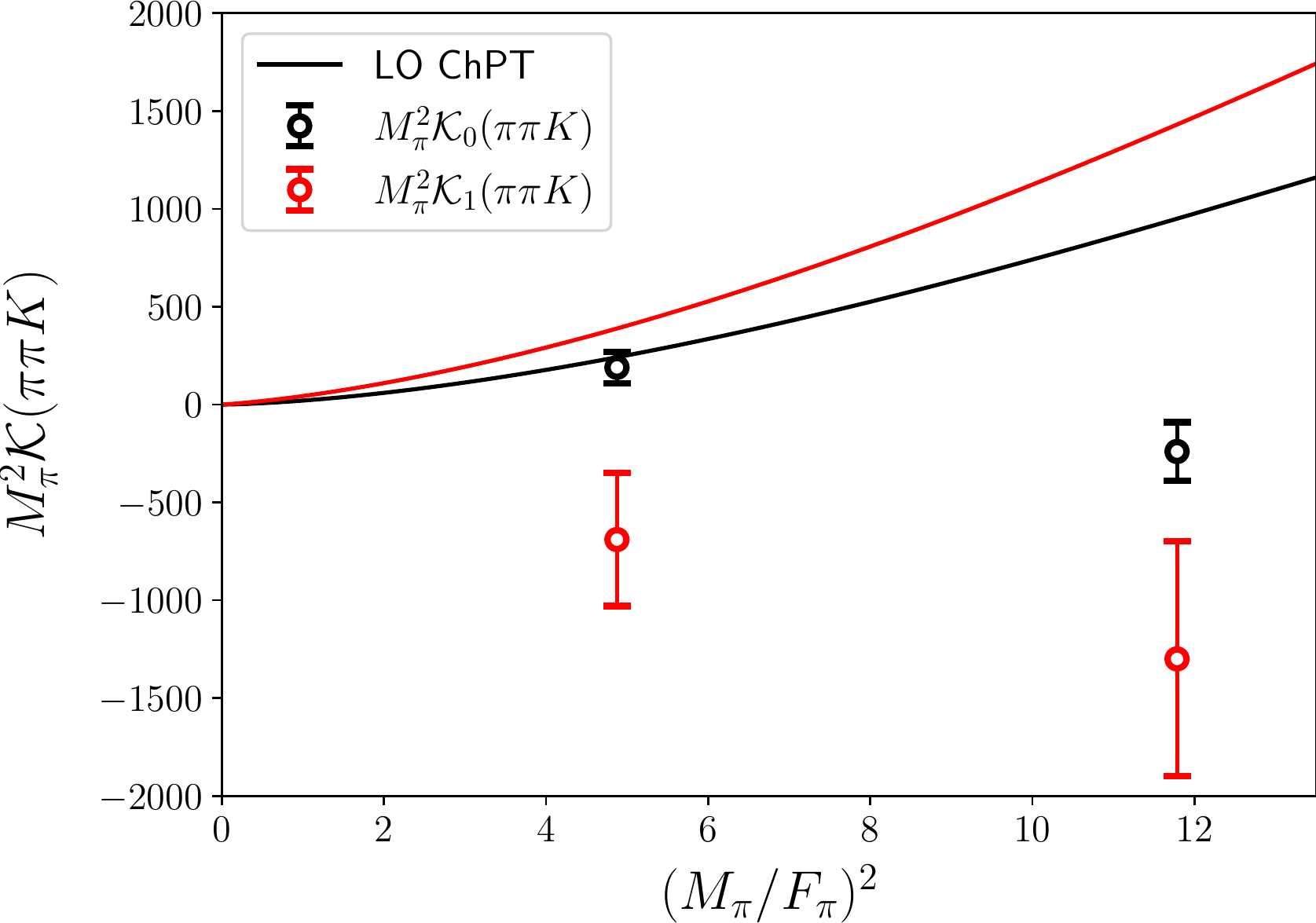}  
\caption{Results for $\cK_{0}$ and $\cK_{1}$ for $\pi \pi K$ scattering as a function of $M_\pi^2/F_\pi^2$. The LO SU(3) ChPT predictions given in \Cref{eq:KisoChPTKKp} are also shown.}
\label{fig:KisoppK}
\end{figure}

\begin{figure}[h!]
\centering
\includegraphics[width=0.8\linewidth]{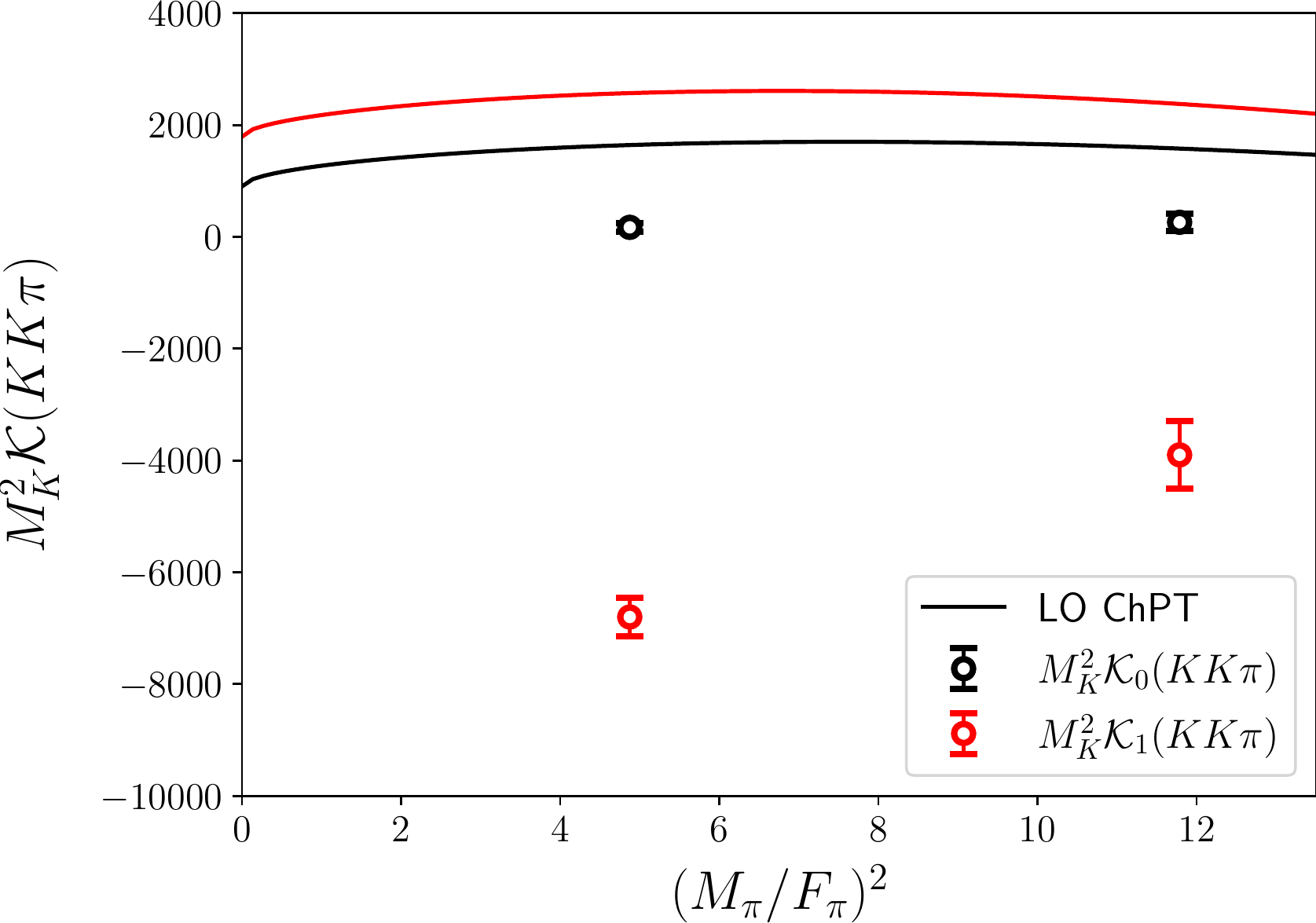}  
\caption{Results for $\cK_{0}$ and $\cK_{1}$ for $K K \pi$ scattering as a function of $M_\pi^2/F_\pi^2$.  The LO SU(3) ChPT predictions given in \Cref{eq:KisoChPTppK} are also shown.}
\label{fig:KisoKKp}
\end{figure}

Finally, we compare our results for $\kdf$ for $2+1$ systems to ChPT.
In \Cref{fig:KisoppK,fig:KisoKKp} we plot the results for $\cK_{0}$ and $\cK_{1}$ 
for $\pi \pi K$ and $K K \pi$ scattering, respectively.
We compare to the LO ChPT predictions of \Cref{eq:KisoChPTppK,eq:KisoChPTKKp},
and find substantial disagreement, most notably in the sign of $\cK_1$, while the magnitudes
are better matched.
Similar disagreement has been observed for $3\pi$ and $3K$ systems~\cite{\threepithreeK}. 
There are two possible interpretations for this disagreement.
First, it may be that we have underestimated the errors in the determinations of $\cK_0$ and $\cK_1$.
One possibility is that discretization errors might be large, although we present evidence against this
option in \Cref{sec:discretization}.
Second, NLO terms in ChPT may be substantial, and invalidate the LO result,
such as in the case of $M_K^2 a_0^{KK} r_0^{KK}$ discussed above.
To address the latter possibility, a NLO ChPT calculation would be needed, but, while
NLO results are available for the three-particle scattering amplitude~\cite{Bijnens:2021hpq,Bijnens:2022zsq},
the relation to $\kdf$ has yet to be worked out.

\begin{figure}[h!]
\centering
\includegraphics[width=0.8\linewidth]{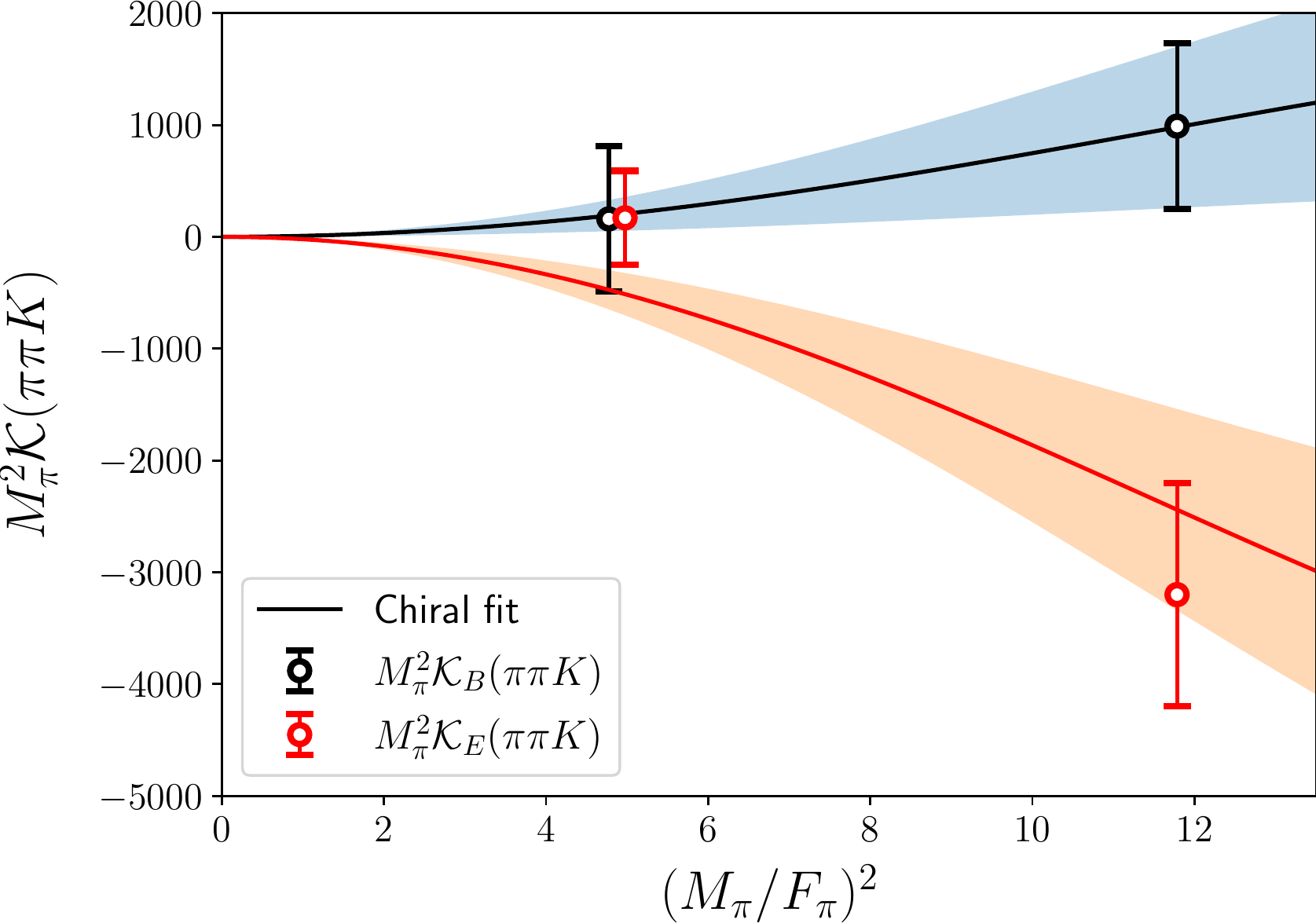}  
\caption{Results for $\cK_{B}$ and $\cK_{E}$ for $\pi \pi K$ scattering as a function of $M_\pi^2/F_\pi^2$. 
Fits to the expected leading chiral behavior given in \Cref{eq:KBEChPTppK} are plotted alongside the data.
For better visibility, the x-coordinates of the left-most datapoints have been shifted.}
\label{fig:KBEppK}
\end{figure}

\begin{figure}[h!]
\centering
\includegraphics[width=0.8\linewidth]{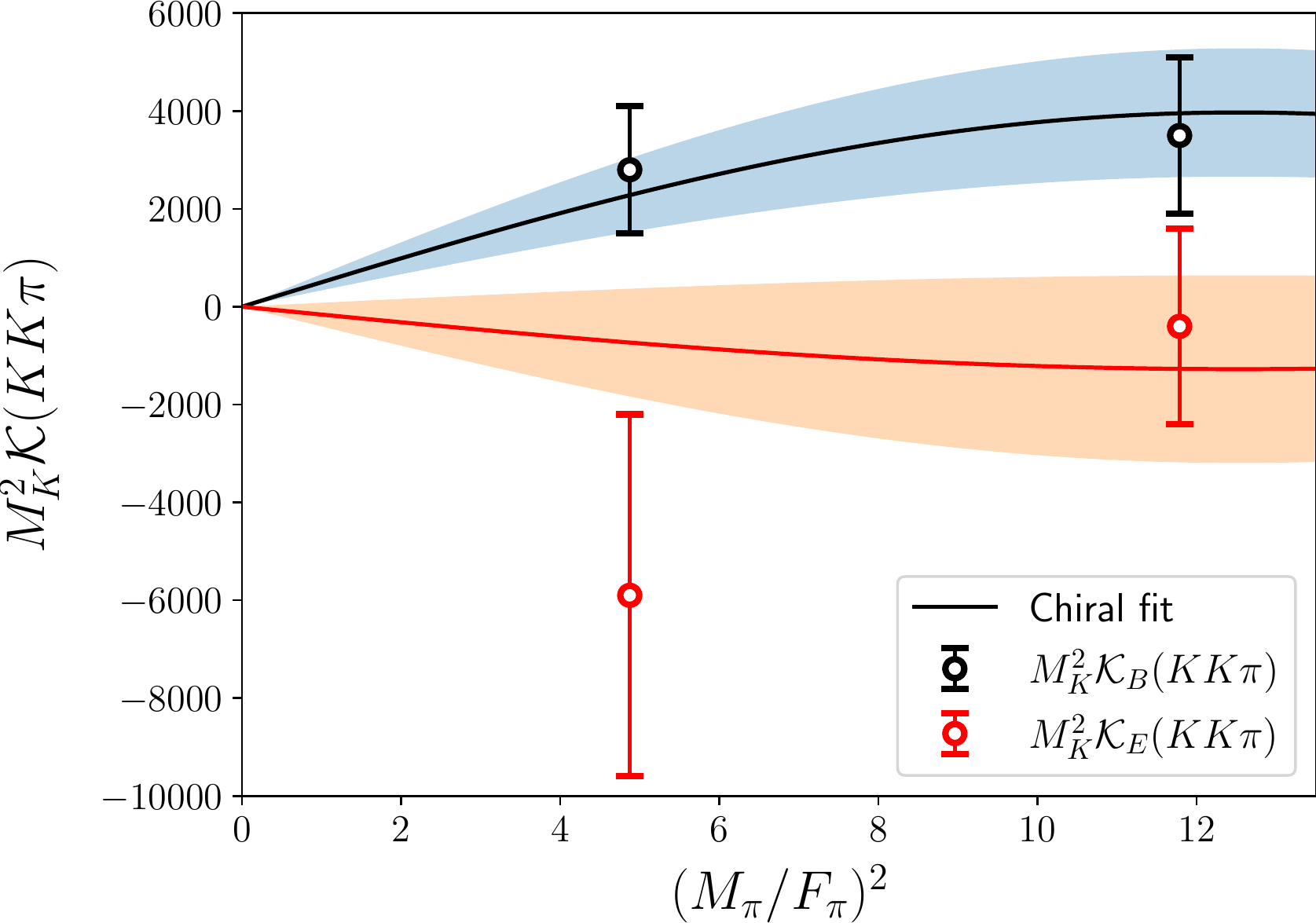}  
\caption{Results for $\cK_{B}$ and $\cK_{E}$ for $K K \pi$ scattering as a function of $M_\pi^2/F_\pi^2$. Fits to the expected leading chiral behavior given in \Cref{eq:KBEChPTKKp} are plotted alongside the data.}
\label{fig:KBEKKp}
\end{figure}

In \Cref{fig:KBEppK,fig:KBEKKp} we plot the results for $\cK_{\text{B}}$ and $\cK_{\text{E}}$ for $\pi \pi K$ and
$K K \pi$ scattering, respectively. 
These quantities vanish at LO in ChPT; their first nontrivial contribution is expected to appear at NLO in ChPT.
Since a NLO calculation has yet to be done, we have fit to the
expected chiral scaling given in \Cref{eq:KBEChPTppK,eq:KBEChPTKKp}, finding parameters
\begin{equation}
\begin{split}
c_B^{\pi \pi K} &= 0.41(30), \quad \chi^2/\text{DOF}=0.0036, \qquad
c_E^{\pi \pi K} = -1.02(38), \quad \chi^2/\text{DOF}=3.1, 
\\
c_B^{K K \pi} &= 1.13(37), \quad \chi^2/\text{DOF}=0.24, \ \ \qquad
c_E^{K K \pi} = -0.36(54), \quad \chi^2/\text{DOF}=2.1.
\end{split}
\end{equation}
We find a reasonable description of the data based on these fit forms.

\subsection{Discretization errors}
\label{sec:discretization}

Up to this point we have neglected the effects of discretization errors in our two- and three-particle fits. Since the ensembles used in this work are $\cO(a)$ improved, these errors are of $\cO(a^2)$.
Here we extend the fits by including the leading $a^2$ terms predicted by WChPT.
As explained in \Cref{sec:ChPT}, this is only consistent with chiral power counting if we
assume $a^2\Lambda_{QCD}^2 \sim M_\pi^4/(4\pi F_\pi)^4$.

We begin with the two-particle scattering lengths.
The WChPT results of \Cref{eq:a0ppCorrection,eq:a0KKCorrection,eq:a0pKCorrection}
predict that each of these quantities receive a common offset proportional to $a^2$.
Repeating the global fit to the six $s$-wave scattering lengths shown in \Cref{tab:a0summary},
allows us to find the value of this offset, which we denote as follows,
\begin{equation}
\delta_a(M a_0) = \lim_{M_\pi \to 0} M_\pi a_0^{\pi \pi} = - \frac{(2 w_6'+w_8')}{16\pi}\,.
\end{equation}
The new fit has $\chi^2/\text{DOF} = 1.1/3$ (6 data points and 3 parameters), and yields
\begin{equation}
L_{\pi \pi} = -8.47(59)\times 10^{-4}, \quad L_5 = -0.3(1.6)\times 10^{-3}, \quad \delta_a(M a_0) = 2.7(4.5)\times 10^{-3}\,.
\label{eq:LpipiL5disc}
\end{equation}
The results for $L_{\pi\pi}$ and $L_5$ are consistent with those found in the fit without discretization errors,
given in \Cref{eq:LpipiL5},
and the offset is found to be consistent with zero.
The result for $\delta_a(M a_0)$ can be converted into a determination of the $\cO(a^2)$ LECs,
\begin{equation}
2 w’_6 + w’_8 = -0.14(23)\,.
\label{eq:wLECs}
\end{equation}
As noted above, the magnitude of this effect is very small. For example, at the physical pion mass
$M_\pi a_0^{\pi\pi} \sim 0.04$ in our fits, which is about an order of magnitude larger than the central
value of, or error in, $\delta_a(M a_0)$.

Using the WChPT results given in \Cref{eq:KdfppKCorrection,eq:KdfKKpCorrection},
we can use \Cref{eq:wLECs} to predict the size of the $\cO(a^2)$ terms in the leading isotropic
contribution to $\kdf$.
This is because they are proportional to the same combination of LECs.
The resulting predictions are presented in \Cref{tab:Kdfdiscretization}.
We observe that the size of the discretization errors is an order of magnitude smaller than the central
values and errors we obtain from fits to the spectrum
This indicates that, given the precision with which we are able to calculate
the parameters in $\Kdf$, discretization effects can be viewed as negligible.
Similar results hold also for the $3\pi$ and $3K$ channels, for which the WChPT results are
presented in \Cref{app:B}.

\begin{table}[h!]
\centering
\begin{tabular}{|c|c|c|}
\hline
Ensemble & $\cK_{0}$ & $\delta_a(\cK_0) $
\\ \hline
\multicolumn{3}{|c|}{$\pi\pi + \pi K + \pi \pi K$ fits}
\\ \hline
D200 & 190(80) & 4(7)
\\ \hline
N203 & -240(150) & 10(16)
\\ \hline
\multicolumn{3}{|c|}{$KK + \pi K + KK\pi$ fits}
\\ \hline
D200 & 170(270) & 17(27)
\\ \hline
N203 &260(310) & 14(23)
\\ \hline
\end{tabular}
\caption{
Final results for $\cK_{0}$ compared to the estimated discretization effects based on WChPT, 
${\delta_a(\cK_0) = -6 (2 w'_{6} + w'_{8})M_X^2/F_X^2}$, 
where $X=\pi$ for the $\pi\pi + \pi K + \pi \pi K$ fits, 
and $X=K$ for the $KK + \pi K + KK\pi$ fits.
}
\label{tab:Kdfdiscretization}
\end{table}

%% file: conclu.tex

This works represents the first determination, theoretical or experimental, of quantities related to  $\pi\pi K$ and $K K \pi$ scattering at maximal isospin. In particular, we have studied these systems using LQCD for two different ensembles along a trajectory of approximately constant trace of the quark mass matrix, $2 m_{\ell} + m_s \simeq $ const. 

We have extracted more than 200 finite-volume energies across the $\pi \pi K$ and $KK \pi$ systems,
leading to roughly 35 -- 45 and 20 -- 30 levels below the inelastic thresholds on N203 and D200, respectively.
As with our previous work in Ref.~\cite{Blanton:2021llb}, the ability to determine such a large number of energies was enabled by advanced algorithms, like stochastic LapH and common subexpression elimination.
In order to facilitate the use of our extracted spectrum by other collaborations, we have made the jackknife samplings for all energies determined in this work available in HDF5 format as ancillary files with the arXiv submission.

As explained in \Cref{sec:fitstrategies}, we have tested several strategies to extract scattering quantities 
from the finite-volume spectra. Our findings can be summarized as follows. 
First, in the systems of this work, it is better to perform fits to energy shifts with respect 
to the noninteracting energies, rather than to the energy levels themselves. 
This leads to less correlation between energy levels (better-conditioned correlation matrices), 
and to more constrained best-fit parameters. 
Second, it seems that the best approach when dealing with multiple processes 
is to fit all the available information at the same time 
(which here means a simultaneous fit to the two- and three-hadron spectra). 
Other fitting strategies, such as first fitting the two-particle spectra and then using the results as Bayesian priors 
for a fit to the three-particle levels, produce consistent results, but with significantly larger uncertainties. 
We expect that these conclusions will be relevant for other finite-volume systems.

The main results of this work are summarized in \Cref{tab:a0summary,tab:r0summary,tab:Kdfsummary}. 
First, we are able to extract two-particle threshold parameters, including 
 the $p$-wave $\pi^+ K^+$ scattering length. 
Second, for each of the 2+1 systems, we extract the four parameters of $\kdf$ 
corresponding to first and second order in the threshold expansion. 
We observe the expected hierarchy, that is, interactions in the $KK\pi$ channel are stronger than in $\pi\pi K$. 
Moreover, the values for the different terms of the K matrix have the correct order of magnitude, 
at least based on chiral expectations. 
We are also able to determine their values to be nonzero with greater than 
$2\sigma$ significance in some cases, and we find the entire $\cK_{\text{df},3}$ 
to be nonzero with significance between $2.4$ and $5.0 \sigma$ in all channels on each ensemble. 

Another novel feature of this work is the extraction of discretization errors in $\kdf$,
which at leading order in the appropriate power counting enter only in $\cK_0$.
By comparing the two-particle scattering lengths computed on the lattice to those calculated in WChPT,
we are able to determine a numerical value for the magnitude of discretization effects
in a WChPT calculation of $\cK_0$. We find the discretization errors to be small
compared to both the central values and uncertainties of $\cK_0$
for each scattering channel and on each ensemble.

Natural extensions of this work are to three pseudo-Goldstone bosons at nonmaximal isospin, 
for which two- and three-meson resonances are present, or to the doubly charmed tetraquark $T_{cc}$. 
While there is a long way to go to determine properties of particles coupling to both
two- and three-particle channels, and involving particles with spin, such as the Roper resonance,
the strategies and tools of this work bring us one step closer.

%% file: AppA.tex
\begin{figure}
  \centering
  \includegraphics{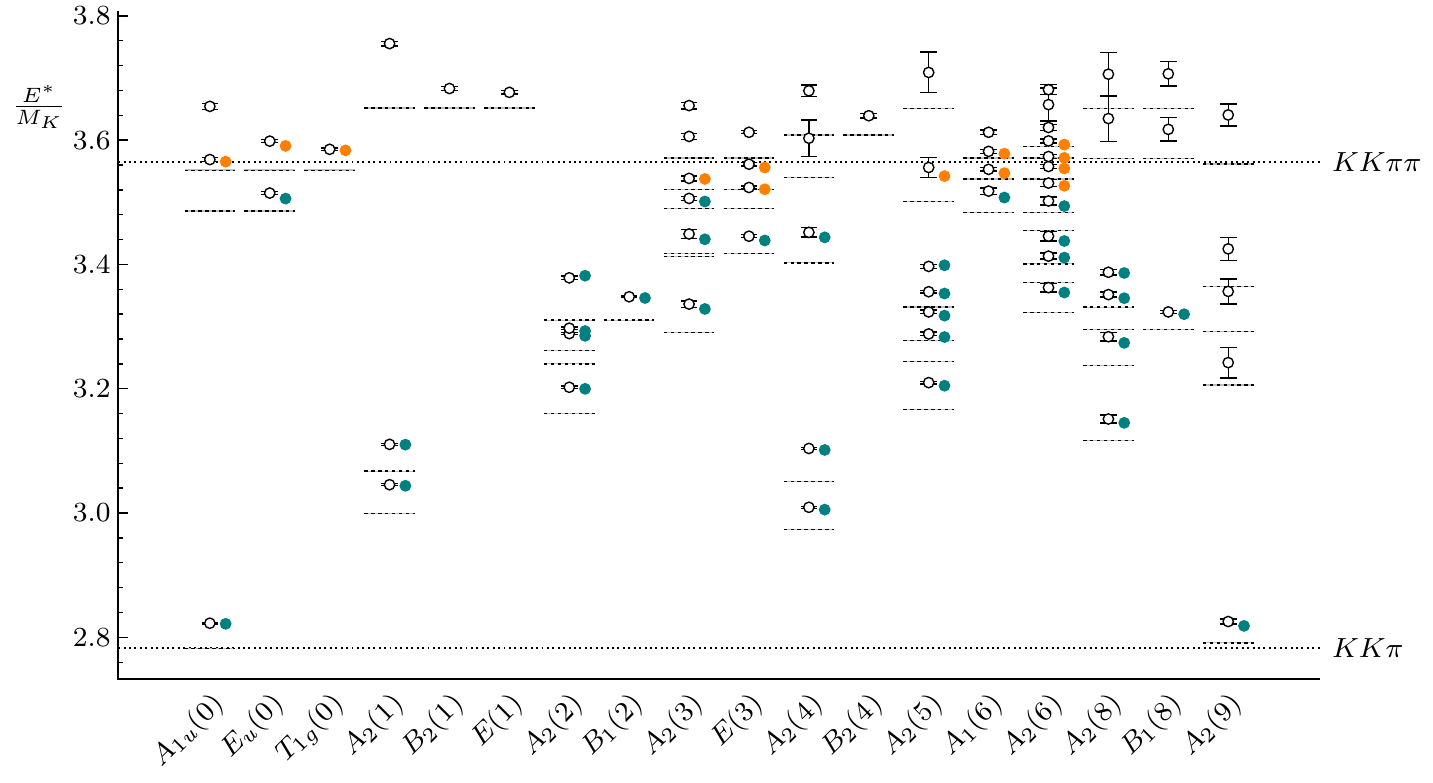}
  \caption{$KK\pi$ CMF spectrum on ensemble N203. Notation as in \Cref{fig:n203_kpp_spectrum}, except
  energies here are in units of $M_K$. Colored symbols are from the ADLER3 fit in \Cref{tab:KKpN203}.
  }
  \label{fig:n203_kkp_spectrum}
\end{figure}
\begin{figure}
  \centering
  \includegraphics{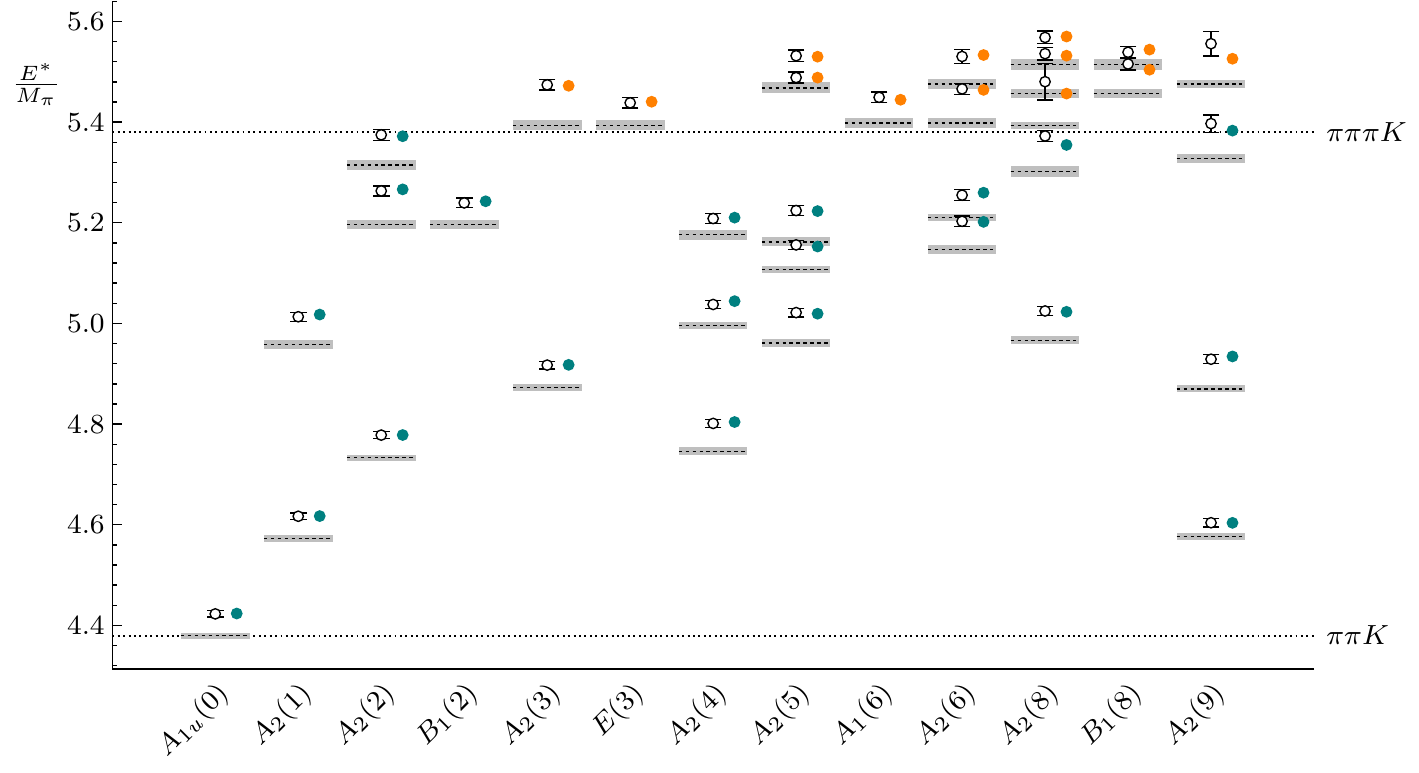}
  \caption{$\pi\pi K$ CMF spectrum on ensemble D200. Notation as in \Cref{fig:n203_kpp_spectrum}.
  Colored symbols are from the Standard fit in \Cref{tab:ppKD200}.
  }
  \label{fig:d200_kpp_spectrum}
\end{figure}
\begin{figure}
  \centering
  \includegraphics{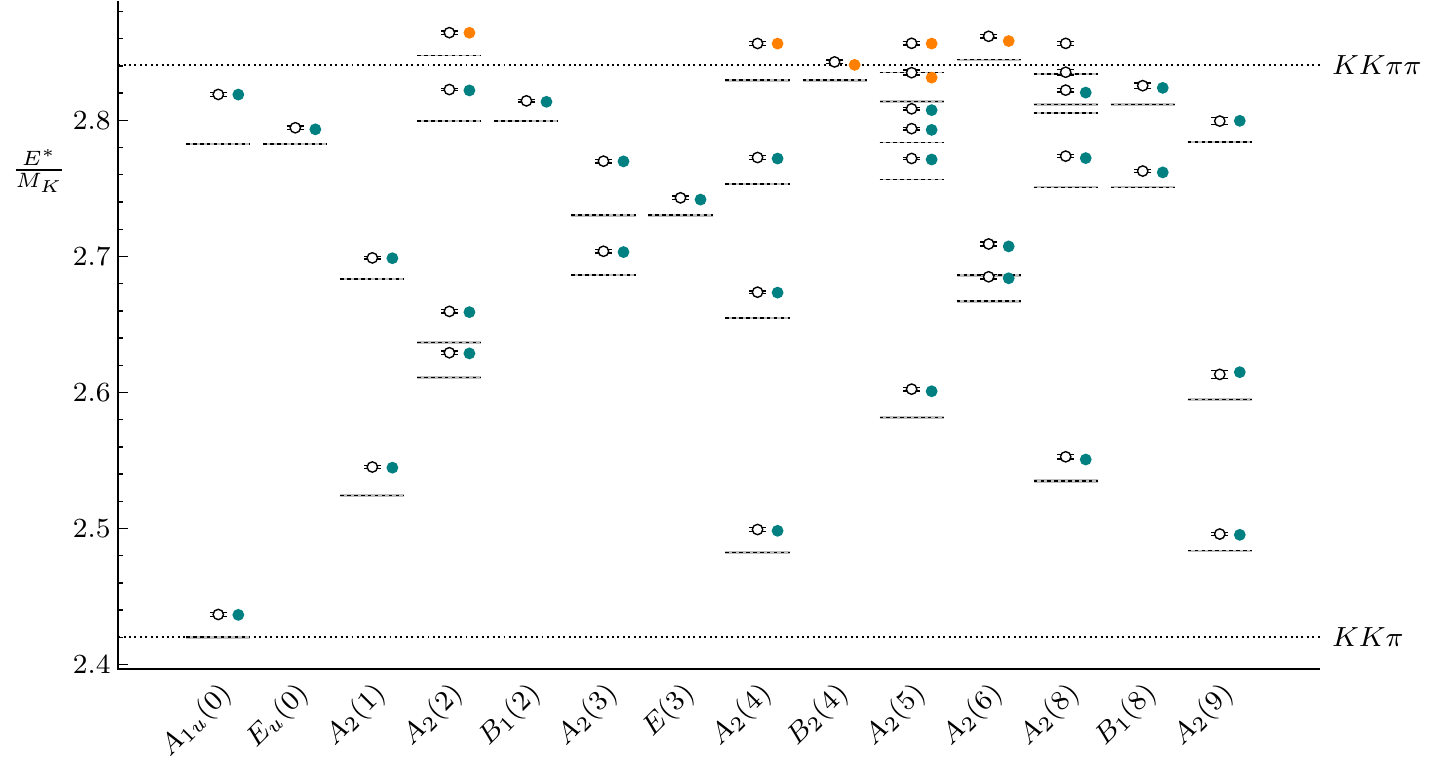}
  \caption{$KK\pi$ CMF spectrum on ensemble D200. Notation as in \Cref{fig:n203_kkp_spectrum},
  except energies here are in units of $M_K$.
  Colored symbols are from the ADLER3 fit in \Cref{tab:KKpD200}.
  }
  \label{fig:d200_kkp_spectrum}
\end{figure}

In this appendix we give further details of the fits described in the main text.
First, in \Cref{tab:energies_used} we list the energy levels used in the fits.
Second, in \Cref{fig:n203_kkp_spectrum,fig:d200_kpp_spectrum,fig:d200_kkp_spectrum}
we show the CMF energy levels for the N203 $K K \pi$, D200 $\pi\pi K$ and D200 $KK\pi$ spectra, respectively,
along with the predictions from the quantization condition using our standard fit parameters.
The corresponding result for the N203 $\pi\pi K$ spectrum is shown in the main text in \Cref{fig:n203_kpp_spectrum}.

\begin{table}[bph!]
\centering
\begin{tabular}{c|c|ccc}
$\vec{d}_{\rm ref}$&{Type}&N203&D200&\\%
\hline%
\multirow{3}{*}{(0,0,0)}& $\pi K$& $2A_{1g} + [A_{1g}]$&$A_{1g} + T_{1u} +[A_{1g}]$ \\%
&$\pi \pi K$& $A_{1u}$  & $A_{1u}$\\%
&$KK\pi$&$A_{1u}+E_u$& $2A_{1u} + E_u$\\ \hline
\multirow{3}{*}{(0,0,1)} &$\pi K$& $2A_1 + [6A_1+2E]$ &$2A_{1} +[2A_{1} +2E]$ \\%
&$\pi \pi K$&$2A_2$& $2 A_{2}$\\%
&$KK\pi$&$2A_{2}$& $2A_2$\\ \hline
\multirow{3}{*}{(0,1,1)} &$\pi K$& $3A_1+B_2 + [3A_1+3B_1+B_2]$&$3A_{1} + B_2 + [A_1 + B_1]$ \\%
&$\pi \pi K$&$4A_2+B_1$& $3 A_{2} + B_1$\\%
&$KK\pi$&$4A_{2}+B_1$& $3A_2 + B_1$\\ \hline
\multirow{3}{*}{(1,1,1)}& $\pi K$&$4A_1+2E$ &$4A_{1} + 2E $ \\%
&$\pi \pi K$&$A_2 + [2A_2+2E]$& $A_2$\\%
&$KK\pi$&$3A_{2}+E$&  $2A_2+E$\\ \hline
\multirow{3}{*}{(0,0,2)}& $\pi K$& $4A_1+E + [A_1]$&$2A_1+[2A_{1} + E] $ \\%
&$\pi \pi K$&$3A_2$& $3 A_{2} $\\%
&$KK\pi$&$3A_{2}$&$3A_2$\\ \hline
\multirow{2}{*}{(0,1,2)}& $\pi \pi K$&$4A_2+[2A_2]$& $ 3A_{2} $\\%
&$KK\pi$&$5A_{2}$&$4A_2$\\ \hline
\multirow{2}{*}{(1,1,2)}& $\pi \pi K$&$A_2+[3A_2+A_1]$& $ 2A_{2} $\\%
&$KK\pi$&$4A_{2}+A_1$&$2A_2$\\ \hline
\multirow{2}{*}{(0,2,2)}& $\pi \pi K$&$4A_2+B_1$& $ 2A_{2}$\\%
&$KK\pi$&$4A_{2}+B_1$&$3A_2 +2B_1$\\ \hline
\multirow{2}{*}{(0,0,3)}& $\pi \pi K$&$4A_2$& $ 3A_{2} $\\%
&$KK\pi$&$A_{2}$&$3A_2$\\ \hline
\end{tabular}
\caption{\label{tab:energies_used} Energy levels used in the fits of this work. Notation is as follows: ``$A_{1u} +E_u$'' means the lowest level in the $A_{1u}$ irrep, and the lowest in the $E_u$ irrep. When two different energy ranges have been chosen, the levels in brackets contribute only to the fit with the higher energy cutoff. For instance, in D200 a fit with 16 $\pi K$ levels is shown in the last column of \Cref{tab:ppKD200}  {(``(d) Standard *'')}, whereas another with 26 levels is shown in the first column of the same table (``(a) Standard''). In this case, ``$A_{1g} + T_{1u} +[A_{1g}]$'' means that the lowest $A_{1g}$ and $T_{1u}$ levels contribute to the fit with 16 levels, and the second $A_{1g}$ level is additionally included in the fit with 26 levels. The energy levels included in $\pi\pi, KK, \pi\pi\pi$ and $KKK$ systems can be read off from Table 32 of Ref.~\cite{Blanton:2021llb} by selecting only those in trivial irreps ($A_1$ or $A_{1g}$ for two particles and $A_2$ or $A_{1u}$ for three).
}
\end{table}

%% file: appB.tex

In this appendix we extend previous SU(3) ChPT results for
two- and three-particle K matrices by including discretization errors. 
We work at LO, and use SU(3), rather than SU(2), ChPT due to the presence of both kaons and pions.
The methodology for including the effects of nonzero $a$ into ChPT
was developed in Refs.~\cite{SS, BRS03}, and is generally referred to as Wilson ChPT, or WChPT for short.
Previous work has considered $\pi\pi$ scattering in SU(2) WChPT~\cite{ABB} (working at NLO),
but, to our knowledge, no previous calculations of two- or three-particle 
scattering using SU(3) WChPT have been performed.

For present-day lattice calculations, with light quark masses close to their physical values, and with
nonperturbative $\cO(a)$ improvement, the appropriate
power counting to use is $m_q  \sim p^2  \sim a^2$ (leaving factors of $\Lambda_{\rm QCD}$ implicit).
In other words, discretization errors, which begin at $\cO(a^2)$, 
are comparable to the squared masses of the psuedo-Goldstone bosons (PGBs),
$M_{\rm PGB}^2 = \cO(m_q)$.
For an example of the appropriateness of this power counting, we point to Ref.~\cite{Draper:2021wga},
where it is noted, in the related context of simulations with twisted-mass fermions,
that an $\cO(a^2)$ contribution to PGB mass-squareds ($w_8' f^2$ in the notation below) is indeed of
the same order as the physical $M_\pi^2$.

The LO chiral Lagrangian (in Euclidean space), including discretization terms, is given by~\cite{BRS03}
\begin{equation}
\begin{split}
\cL_{\rm LO} &= \frac{F^2}4 {\rm Tr}\!\left(\partial_\mu \Sigma^\dagger \partial_\mu \Sigma \right)
-\frac{F^2}{4} {\rm Tr}\!\left(\chi \Sigma^\dagger+\Sigma \chi^\dagger\right) 
+ \cV_{a^2} \,,
\\
\cV_{a^2} &=  - \hat a^2 W_6' \,{\rm Tr}\!\left(\Sigma+\Sigma^\dagger\right)^2
- \hat a^2 W_8'\, {\rm Tr}\!\left(\Sigma^2+\Sigma^{\dagger2}\right)\,,
\end{split}
\label{eq:LchptLO}
\end{equation}
where $\Sigma \in \SU(3)$ is the field that contains the PGBs,
$\chi = 2 B_0 M$ is proportional to $M$, the renormalized mass matrix,
and $F\approx F_{\pi} \simeq 92\;$MeV and $B_0$ are the usual continuum LO LECs. 
Additionally, we have $\hat a= 2 W_0 a$, where $W_0$, $W_6'$, and $W_8'$
are LECs associated with discretization errors.
We assume exact isospin symmetry, so that $M={\rm diag}(m_\ell, m_\ell, m_s)$.

With the Lagrangian in hand, we can now compute the LO pion and kaon masses,
\begin{align}
M_{\pi}^2 &= 2 B_0 m_{\ell} + F^2 (3 w'_{6} + w'_{8})\,, 
\label{eq:Mp}
\\
M_{K}^2 &= B_0 (m_{\ell} + m_{s}) + F^2 (3 w'_{6} + w'_{8})\,,
\label{eq:MK}
\end{align}
where we have converted to dimensionless quantities by defining 
\begin{equation}
w'_6 = \frac{16 \hat a^2 W'_6}{F^4} \quad \text{and} \quad w'_8 = \frac{16 \hat a^2 W'_8}{F^4}\,.
\end{equation}
We observe in \Cref{eq:Mp,eq:MK} the violation of chiral symmetry with Wilson fermions, which here
leads to an offset in the PGB masses in the (naive) chiral limit.
These offsets can be effectively absorbed into shifts in the quark masses by writing subsequent
expressions in terms of $M_\pi$ and $M_K$, rather than $m_\ell$ and $m_s$.

The LO $\pi \pi$, $\pi K$, and $K K$ scattering amplitudes are given by
\begin{align}
\cM_{\pi \pi} &= \frac{1}{F^2} \left[-s + 2 M_{\pi}^2 + 2 F^2 (2 w'_{6} + w'_{8}) \right]\,, 
\label{eq:Mpp}
\\
\cM_{\pi K} &= \frac{1}{2 F^2} \left[-s + M_{\pi}^2 + M_{K}^2 + 2 F^2 (2 w'_{6} + w'_{8}) \right]\,, 
\label{eq:MpK}
\\
\cM_{K K} &= \frac{1}{F^2} \left[-s + 2 M_{K}^2 + 2 F^2 (2 w'_{6} + w'_{8}) \right]\,,
\label{eq:MKK}
\end{align}
where $s$ is the usual two-particle Mandelstam variable.
We thus find that, when we express the results in terms of the PGB masses, only a single
linear combination of $w'_{6}$ and $w'_{8}$ appears in all three expressions.
Indeed, if one carries out the calculation in ${\rm SU}(N)$ WChPT, the same result is obtained, with all the
$N$ dependence absorbed into the PGB masses [$3$ becoming $N$ in \Cref{eq:Mp} and \Cref{eq:MK}].
Thus we can compare the result for $\cM_{\pi\pi}$ with that obtained in SU(2) WChPT in Ref.~\cite{BRS03},
and find complete agreement.

The $\cO(a^2)$ terms in \Cref{eq:Mpp}, \Cref{eq:MpK}, and \Cref{eq:MKK} 
again display a violation of chiral symmetry, for the scattering amplitudes
do not vanish at threshold in the chiral limit. To see this explicitly, we give the result for the scattering lengths
\begin{align}
M_\pi a_0^{\pi\pi} &= -\frac1{32\pi} \cM_{\pi\pi}\bigg|_{s=4M_\pi^2} 
=\frac{M_\pi^2}{16\pi F^2} -  \frac{(2 w_6'+w_8')}{16\pi} + \cO(M^4/F^4)\,,
\label{eq:a0ppWChPT}
\\
\begin{split}
M_{\pi K} a_0^{\pi K} &= - \frac{1}{16\pi}  \cM_{\pi K}\bigg|_{s=(M_\pi+M_K)^2} 
= \frac{M_\pi M_K}{16\pi F^2} - \frac{(2 w_6'+w_8')}{16 \pi} + \cO(M^4/F^4)\ \,,
\end{split}
\label{eq:a0pKWChPT}
\\
M_K a_0^{KK} &= -\frac1{32\pi} \cM_{KK}\bigg|_{s=4M_K^2} =
\frac{M_K^2}{16\pi F^2} - \frac{(2 w_6'+w_8')}{16\pi} + \cO(M^4/F^4)\,.
\label{eq:a0KKWChPT}
\end{align}
These results agree with the LO parts of \Cref{eq:a0pp}, \Cref{eq:a0pK}, and \Cref{eq:a0KK}, in the limit that
$a^2\to 0$.
By taking appropriate linear combinations of the scattering lengths, one can cancel the $\cO(a^2)$ term.
This  prediction could be tested in simulations with several lattice spacings.
We note that there are no $\cO(a^2)$ corrections to the effective range at LO.

Finally, we turn to $\kdf$. This has previously been calculated without discretization effects
for three pions in Ref.~\cite{\HHanal}, 
three kaons in Ref.~\cite{\threepithreeK}, and for $\pi\pi K$ and $KK\pi$ scattering in Ref.~\cite{\BRSimplement}.
We refer to those works for the methodology, and simply quote the final results.
We find that [see \Cref{eq:Mandelstam3} for notation]
\begin{equation}
\cK_{\text{df},3}^{\pi \pi \pi} = \frac{18}{F^4}\left[M_{\pi}^{2} - F^2 (2 w'_{6} + w'_{8})\right] 
+ \frac{27 M_{\pi}^{2}}{F^4} \Delta + \cO(M^4/F^6)\,,
\label{eq:KdfpppWChPT}
\end{equation}
and a nearly identical expression for $\cK_{\text{df},3}^{K K K}$ obtained by simply substituting $M_{K}$ for $M_{\pi}$.
For the nondegenerate case we obtain
\begin{equation}
\cK_{\text{df},3}^{\pi \pi K} = \frac{1}{F^4}\left[2 M_{\pi}(M_{\pi} + 2 M_{K}) - 6 F^2 (2 w'_{6} + w'_{8})\right]
+ \frac{1}{F^4}(2 M_{\pi} + M_{K})^2 \Delta + \cO(M^4/F^6)\,,
\label{eq:KdfppKWChPT}
\end{equation}
with the result for $KK\pi$ obtained by switching the roles of the masses $M_{\pi}$ and $M_{K}$.
We observe that both results are isotropic. For three identical particles this is guaranteed by the fact that
angular dependence only enters at relative order $p^4$, i.e. at NNLO in ChPT.
For the nondegenerate cases, this result, first observed in Ref.~\cite{\BRSimplement}, is not required by
symmetries or power counting, as there could be terms proportional to $\Delta_2^S$ and $\tilde t_{22}$
[see \Cref{eq:KdfExpansion}]. Indeed, such terms enter at intermediate stages but cancel in the final result.

At LO in WChPT, the $\cO(a^2)$ terms enter only in $\cK_0$ (given by the above expressions with $\Delta\to 0$), 
since power-counting requires that at $\cO(a^2)$ correction to $\cK_1$ requires a NLO contribution.
As with the case of the two-particle amplitudes, discretization effects lead to $\cK_0$ having nonvanishing values
in the chiral limit.
What is particularly striking, however, is that this offset is given by 
the same combination of LECs that enters in $\kdf$ as in the two-particle amplitudes.
Thus the size of the offset in $\cK_0$ can be determined from a fit to the two-particle amplitude.
We use this result in the main text, as described in \Cref{sec:discretization}.

%% file: app_pK.tex
This Appendix discusses briefly how to obtain an analytic expression for the $I=3/2$ $\pi K$ $p$-wave scattering length, $a_1^{\pi K}$. We use the results from NLO ChPT presented in Ref.~\cite{\Bijnensb}, which refers to functions presented in Ref.~\cite{\Bijnensa}.

The $I=3/2$ $\pi K$ amplitude is related to the $s$- and $p$-wave scattering lengths by
\begin{equation}
T^{3/2}(s,t,u) = -16 \pi \frac{M_\pi+M_K}2 ( a_0^{\pi K} + q^2 \cos\theta  \, a_1^{\pi K} + \dots )\,,
\end{equation}
where we have used our convention for the sign of the scattering lengths,
and have expanded about threshold showing only the terms that do not vanish at threshold for the partial-wave projected amplitudes.
Here $\theta$ is the scattering angle in the CMF and $q$ the spatial momentum of both particles in this frame.
Thus we must pick out the coefficient of $q^2 \cos\theta$. 

The dependence on the angle $\theta$ only enters through the two-particle Mandelstam variables $t$ and $u$,
\begin{align}
\begin{split}
t &= - 2 q^2 (1-\cos\theta) = 2 q^2 \cos\theta + \dots\,, 
\\
u &= 2 M_\pi^2 + 2 M_K^2 -s + 2 q^2 (1-\cos\theta)
= u_0 - 2 q^2 \cos\theta + \dots \,,
\\
u_0 &= (M_\pi-M_K)^2\,.
\end{split}
\end{align}
where here and below the ellipsis consists of $\cO(q^2)$ terms that are independent of $\cos\theta$,
and $\cO(q^4)$ terms that do depend on $\cos\theta$.
We now expand $T^{3/2}$ about threshold in powers of $s-s_0$ [with $s_0=(M_\pi+M_K)^2$], $t$ and $u-u_0$, 
which are all proportional to $q^2$, with only $t$ and $u-u_0$ depending on $\cos\theta$.
Specifically, we write
\begin{equation}
T^{3/2}(s,t,u) = T^{3/2}(s_0,0,u_0) + c_t t + c_u (u-u_0) + \dots\,,
\end{equation}
Then we have that
\begin{equation}
a_1^{3/2} = -\frac{c_t-c_u}{4 \pi (M_\pi+M_K)}\,.
\label{eq:masterscatlen}
\end{equation}

From Ref.~\cite{\Bijnensb}, we take the result
\begin{equation}
T^{3/2}(s,t,u) = G_1(s) + G_2(t) + G_3(u) + (s-u) G_4(t) + (s-t)G_5(u) + G_6(s,t,u)\,,
\end{equation}
where each of these functions can be expressed in terms of three one-loop functions:
$\bar A(m_1^2)$
$\bar B(m_1^2, m_2^2, t)$, $\bar B_1(m_1^2, m_2^2, t)$, and
$\bar B_{21}(m_1^2, m_2^2, t)$.
These in turn are defined as the finite parts of the corresponding functions in $d=4 - 2\epsilon$ dimensions
presented in Ref.~\cite{\Bijnensa}. For $\bar A$, $\bar B$ and $\bar B_1$, these finite parts can be read off
from the results in Ref.~\cite{\Bijnensa}, but for $\bar B_{21}$ one must account for the explicit
$d$ dependence, which leads to finite ``$\epsilon/\epsilon$'' contributions.
All these functions are given in terms of chiral logarithms and an integral $\bar J(m_1^2, m_2^2,t)$, 
which is given explicitly in Ref.~\cite{\Bijnensa}.

Using the leading order SU(3) relation $M_\eta^2=(4 M_K^2-M_\pi^2)/3$, and we obtain,
 after much algebra,
\begin{multline}
c_t - c_u = \frac{\kappa}{F_\pi^4} \frac{1}{288} \bigg\{
M_K M_\pi \left( 16 \bar  L_2 +8 \bar L_3\right) -
\left(M_K^2+M_\pi^2\right)
\left(  16 \bar  L_1  + 4 \bar L_3 - 8 \bar L_4\right)
\\
+L_\pi \frac{ M_\pi \left(4 M_K^4+87
M_K^3 M_\pi-24 M_K^2 M_\pi^2+56 M_K M_\pi^3-9 M_\pi^4\right)}{24 (M_K-M_\pi)^3}
\\
-L_K \frac{ \left(M_K^5+279
M_K^4 M_\pi-125 M_K^3 M_\pi^2+478 M_K^2 M_\pi^3-27 M_K M_\pi^4
-8 M_\pi^5\right)}{108 (M_K-M_\pi)^3}
\\
+L_\eta \frac{
\left(56 M_K^5+36 M_K^4 M_\pi+155 M_K^3 M_\pi^2-16 M_K^2 M_\pi^3-72 M_K M_\pi^4+11 M_\pi^5\right)}{216
(M_K-M_\pi)^3}
\\
+\frac{-80 M_K^6+524 M_K^5 M_\pi+458 M_K^4 M_\pi^2+397 M_K^3 M_\pi^3-96 M_K^2 M_\pi^4-131
M_K M_\pi^5-4 M_\pi^6}{36 (M_K-M_\pi)^2 \left(4 M_K^2-M_\pi^2\right)}
\\\
+\frac{\sqrt{(2 M_K-M_\pi) (M_K+M_\pi)}
\left(8 M_K^5-12 M_K^4 M_\pi-7 M_K^3 M_\pi^2+2 M_K^2 M_\pi^3-24 M_K M_\pi^4-4 M_\pi^5\right) }
{27 (M_K-M_\pi)^3 (M_K+M_\pi)}
\\
\times \tan^{-1}\left[\frac{2
\sqrt{(M_K-M_\pi)^2 (2 M_K-M_\pi) (M_K+M_\pi)}}{2 M_K^2+3 M_K M_\pi-2 M_\pi^2}\right]
\bigg\}\,, \label{eq:pKNLOpwave}
\end{multline}
where $L_\pi = \log(M_\pi^2/\mu^2)$, etc., and $\bar L_i^r = 16\pi^2 L_i^r$. Note that only two linearly independent combinations of LECs are accessible by varying the meson masses in this quantity.

We have performed a number of checks on this expression: (i) the expansion has been done 
independently by two of us, (ii) we have checked that the dependence on the scale $\mu$ cancels in $c_t-c_u$, and (iii) we have taken the chiral limit $M_\pi \to 0$ and seen that the combination  $c_t-c_u$ remains finite and proportional to $M_K^2/F_\pi^4$, as expected.

For the numerical evaluation of this expression, we use values of the LECs from 
Ref.~\cite{Amoros:2001cp}. In particular,  the ones from fit 10 to $\mathcal O(p^4)$ reported in that work.
 The results in Ref.~\cite{Amoros:2001cp} are quoted at a renormalization scale $\mu = 770$ MeV, and we convert to $\mu = 4 \pi F_\pi$ using the running as in \Cref{eq:LECrunning}, with 
\begin{equation}
\Gamma_1 = \frac{3}{32}, \quad \Gamma_2 = \frac{3}{16}, \quad \Gamma_3 = 0, \quad \Gamma_4 = \frac{1}{8}.
\end{equation}

%% file: app_operators.tex
\section{Tables of interpolating operators}
\label{app:operators}

Here we list all two- and three-hadron operators for nondegenerate channels used in this work.
Our operator construction follows the procedure detailed in Ref.~\cite{Morningstar:2013bda},
where the notation for the irreps is given.
Those for degenerate channels ($\pi\pi$, $\pi\pi\pi$, $KK$, $KKK$) are listed in Ref.~\cite{\threepithreeK}.
$\pi^+K^+$ operators are given in \Cref{tab:kp-ops},
$\pi^+\pi^+K^+$ operators in \Cref{tab:kpp-ops1,tab:kpp-ops2},
and $K^+K^+\pi^+$ operators in \Cref{tab:kkp-ops1,tab:kkp-ops2}.
Note that for the $\pi^+ K^+$ channel we only use operators with total momentum-squared up to $4 (2\pi/L)^2$,
while the three-particle operators have momentum-squared values up to $9 (2\pi/L)^2$.

\begin{table}[!bp]
\centering
\begin{tabular}{c|c|cc|c}
\multirow{2}{*}{$\bm {d}_{\rm ref}$}&\multirow{2}{*}{$[d_{K}^2,d_{\pi}^2]$}&\multicolumn{2}{c}{$E^{\rm free}/M_{\pi}$}&\multirow{2}{*}{operators}\\%
&&N203&D200&\\%
\hline%
(0, 0, 0)&{[}0, 0{]}&2.278&3.3798&$A_{1g}$\\%
&{[}1, 1{]}&3.2609&4.6104&$A_{1g} \oplus T_{1u}^\ast$\\%
&{[}2, 2{]}&4.0064&5.5243&$A_{1g}^\dagger \oplus T_{1u}^\ast$\\%
&{[}3, 3{]}&4.6327& &$A_{1g}$\\%
\hline%
(0, 0, 1)&{[}1, 0{]}&2.4675&3.505&$A_1$\\%
&{[}0, 1{]}&2.5598&3.9023&$A_1$\\%
&{[}2, 1{]}&3.4236&4.7539&$A_1 \oplus E$\\%
&{[}1, 2{]}&3.4618&4.929&$A_1 \oplus E$\\%
&{[}4, 1{]}&4.0217& &$A_1$\\%
&{[}1, 4{]}&4.0966& &$A_1$\\%
&{[}3, 2{]}&4.1491& &$A_1$\\%
&{[}2, 3{]}&4.1713& &$A_1$\\%
\hline%
(0, 1, 1)&{[}2, 0{]}&2.6072&3.61&$A_1$\\%
&{[}0, 2{]}&2.748&4.2192&$A_1$\\%
&{[}1, 1{]}&2.8162&4.0963&$A_1 \oplus B_2$\\%
&{[}3, 1{]}&3.5565&4.8786&$A_1 \oplus B_1$\\%
&{[}1, 3{]}&3.6198& &$A_1 \oplus B_1$\\%
&{[}2, 2{]}&3.6536&5.1032&$A_1^\dagger \oplus B_1^\dagger \oplus B_2$\\%
\hline%
(1, 1, 1)&{[}3, 0{]}&2.7208&3.7013&$A_1$\\%
&{[}0, 3{]}&2.8947&4.459&$A_1$\\%
&{[}2, 1{]}&3.0032&4.2571&$A_1 \oplus E$\\%
&{[}1, 2{]}&3.0466&4.4519&$A_1 \oplus E$\\%
\hline%
(0, 0, 2)&{[}1, 1{]}&2.2867&3.5077&$A_1$\\%
&{[}4, 0{]}&2.8177&3.7828&$A_1$\\%
&{[}0, 4{]}&3.0169&4.656&$A_1$\\%
&{[}2, 2{]}&3.2629&4.644&$A_1 \oplus E$\\%
&{[}3, 3{]}&4.0072& &$A_1$\\%
\end{tabular}
\caption{\label{tab:kp-ops} 
$\pi^+ K^+$ operators used in this work.
  For each set of momenta that are equivalent up to allowed rotations, 
  one representative choice of $\bm d_{\rm ref}$ is given,
  where the total momentum is $\bm P=(2\pi/L) \bm d_{\rm ref}$.
  The momentum squared (in units of $(2\pi/L)^2$) of the individual single particles in each operator is 
  listed as $d_{\rm sh}^2$,
  where ``sh" indicates the particular single hadron.
  The lab-frame free energies in units of $M_\pi$ are listed for those operators that are used on each ensemble
  together with irreps that are included.
  Daggered and starred irreps were only used for the N203 and D200 ensembles, respectively.
}
\end{table}

\begin{table}[!bp]
\centering
\begin{tabular}{c|c|cc|c}
\multirow{2}{*}{$\bm{d}_{\rm ref}$}&\multirow{2}{*}{$[d_{K_1}^2,d_{K_2}^2,d_{\pi}^2]$}&\multicolumn{2}{c}{$E^{\rm free}/M_{K}$}&\multirow{2}{*}{operators}\\%
&&N203&D200&\\%
\hline%
(0, 0, 0)&{[}0, 0, 0{]}&2.7824&2.4202&$A_{1u}$\\%
&{[}1, 1, 0{]}&3.4859&2.7826&$A_{1u} \oplus E_u$\\%
&{[}1, 0, 1{]}&3.5515& &$A_{1u} \oplus E_u \oplus T_{1g}$\\%
\hline%
(0, 0, 1)&{[}1, 0, 0{]}&2.9993&2.5243&$A_2$\\%
&{[}0, 0, 1{]}&3.0678&2.6835&$A_2$\\%
&{[}2, 1, 0{]}&3.6519& &$A_2 \oplus B_2 \oplus E$\\%
\hline%
(0, 1, 1)&{[}2, 0, 0{]}&3.1599&2.6111&$A_2$\\%
&{[}1, 1, 0{]}&3.2399&2.6367&$A_2$\\%
&{[}0, 0, 2{]}&3.261&2.8478&$A_2$\\%
&{[}1, 0, 1{]}&3.3104&2.7996&$A_2 \oplus B_1$\\%
\hline%
(1, 1, 1)&{[}3, 0, 0{]}&3.2907&2.6863&$A_2$\\%
&{[}0, 0, 3{]}&3.4129& &$A_2$\\%
&{[}2, 1, 0{]}&3.4178&2.7304&$A_2 \oplus E$\\%
&{[}2, 0, 1{]}&3.4899& &$A_2 \oplus E$\\%
&{[}1, 0, 2{]}&3.5209& &$A_2 \oplus E$\\%
&{[}1, 1, 1{]}&3.5712& &$A_2 \oplus E$\\%
\hline%
(0, 0, 2)&{[}1, 1, 0{]}&2.9737&2.4823&$A_2$\\%
&{[}1, 0, 1{]}&3.0503&2.6547&$A_2$\\%
&{[}4, 0, 0{]}&3.4025&2.7532&$A_2$\\%
&{[}0, 0, 4{]}&3.5402& &$A_2$\\%
&{[}2, 2, 0{]}&3.6083&2.8296&$A_2 \oplus B_2$\\%
\hline%
(0, 1, 2)&{[}2, 1, 0{]}&3.1666&2.5816&$A_2$\\%
&{[}2, 0, 1{]}&3.2442&2.7565&$A_2$\\%
&{[}1, 0, 2{]}&3.2776&2.8352&$A_2$\\%
&{[}1, 1, 1{]}&3.3315&2.7838&$2 A_2$\\%
&{[}5, 0, 0{]}&3.5008&2.8137&$A_2$\\%
&{[}0, 0, 5{]}&3.6508& &$A_2$\\%
\end{tabular}
\caption{\label{tab:kkp-ops1} 
{
$K^+K^+\pi^+$ operators with ${d}_{\rm ref}^{\,2}\leq 5$ used in this work, where $\bm{P} = (2\pi/L)\bm{d}_{\rm ref}$.
Notation as in \Cref{tab:kp-ops}, except that free energies are in units of $M_K$,
and if an operator irrep is multiplied by an integer (as in $2 A_2$), this integer indicates the number
of linearly-independent operators in that irrep that is used.
}
}
\end{table}

\begin{table}[!bp]
\centering
\begin{tabular}{c|c|cc|c}
\multirow{2}{*}{$\bm{d}_{\rm ref}$}&\multirow{2}{*}{$[d_{K_1}^2,d_{K_2}^2,d_{\pi}^2]$}&\multicolumn{2}{c}{$E^{\rm free}/M_{K}$}&\multirow{2}{*}{operators}\\%
&&N203&D200&\\%
\hline%
(1, 1, 2)&{[}3, 1, 0{]}&3.3222&2.6672&$A_2$\\%
&{[}2, 2, 0{]}&3.3713&2.6863&$A_2$\\%
&{[}3, 0, 1{]}&3.4009&2.8445&$A_2$\\%
&{[}1, 0, 3{]}&3.4546& &$A_2$\\%
&{[}2, 0, 2{]}&3.4833& &$A_1 \oplus A_2$\\%
&{[}2, 1, 1{]}&3.5377& &$2 A_1 \oplus 3 A_2$\\%
&{[}1, 1, 2{]}&3.5711& &$A_1 \oplus 2 A_2$\\%
&{[}6, 0, 0{]}&3.5892& &$A_2$\\%
\hline%
(0, 2, 2)&{[}2, 2, 0{]}&3.1163&2.5349&$A_2$\\%
&{[}2, 0, 2{]}&3.2371&2.8054&$A_2$\\%
&{[}2, 1, 1{]}&3.2956&2.7508&$A_2 \oplus B_1$\\%
&{[}1, 1, 2{]}&3.3314&2.8342&$A_2$\\%
&{[}5, 1, 0{]}&3.5702&2.8114&$A_2 \oplus B_1$\\%
&{[}5, 0, 1{]}&3.6507& &$A_2 \oplus B_1$\\%
\hline%
(0, 0, 3)&{[}1, 1, 1{]}&2.7911&2.4836&$A_2$\\%
&{[}4, 1, 0{]}&3.206&2.5949&$A_2$\\%
&{[}4, 0, 1{]}&3.2917&2.784&$A_2$\\%
&{[}1, 0, 4{]}&3.3646& &$A_2$\\%
&{[}2, 1, 2{]}&3.5614& &$A_2 \oplus B_2 \oplus E$\\%
\end{tabular}
\caption{\label{tab:kkp-ops2} 
{
$K^+K^+\pi^+$ operators with $6 \leq {d}_{\rm ref}^{\,2}\leq 9$ used in this work, where $\bm{P} = (2\pi/L)\bm{d}_{\rm ref}$.
Notation as in \Cref{tab:kkp-ops1}.
}
}
\end{table}

\begin{table}[!bp]
\centering
\begin{tabular}{c|c|cc|c}
\multirow{2}{*}{$\bm{d}_{\rm ref}$}&\multirow{2}{*}{$[d_{K}^2,d_{\pi_1}^2,d_{\pi_2}^2]$}&\multicolumn{2}{c}{$E^{\rm free}/M_{\pi}$}&\multirow{2}{*}{operators}\\%
&&N203&D200&\\%
\hline%
(0, 0, 0)&{[}0, 0, 0{]}&3.278&4.3798&$A_{1u}$\\%
&{[}1, 1, 0{]}&4.2609& &$A_{1u} \oplus E_u \oplus T_{1g}$\\%
&{[}0, 1, 1{]}&4.3447& &$A_{1u} \oplus E_u$\\%
\hline%
(0, 0, 1)&{[}1, 0, 0{]}&3.5417&4.5724&$A_2$\\%
&{[}0, 1, 0{]}&3.6298&4.9585&$A_2$\\%
&{[}2, 1, 0{]}&4.4668& &$A_2 \oplus B_2 \oplus E$\\%
\hline%
(0, 1, 1)&{[}2, 0, 0{]}&3.7366&4.7329&$A_2$\\%
&{[}0, 2, 0{]}&3.8673&5.3142&$A_2$\\%
&{[}1, 1, 0{]}&3.931&5.1962&$A_2 \oplus B_1$\\%
&{[}0, 1, 1{]}&4.0217& &$A_2$\\%
\hline%
(1, 1, 1)&{[}3, 0, 0{]}&3.8951&4.872&$A_2$\\%
&{[}0, 3, 0{]}&4.0536& &$A_2$\\%
&{[}2, 1, 0{]}&4.1533&5.3936&$A_2 \oplus E$\\%
&{[}1, 2, 0{]}&4.1934& &$A_2 \oplus E$\\%
&{[}0, 2, 1{]}&4.2862& &$A_2 \oplus E$\\%
&{[}1, 1, 1{]}&4.351& &$A_2 \oplus E$\\%
\hline%
(0, 0, 2)&{[}1, 1, 0{]}&3.5708&4.746&$A_2$\\%
&{[}0, 1, 1{]}&3.6704&5.1762&$A_2$\\%
&{[}4, 0, 0{]}&4.0305&4.9956&$A_2$\\%
&{[}0, 4, 0{]}&4.2095& &$A_2$\\%
&{[}2, 2, 0{]}&4.4339& &$A_2 \oplus B_2 \oplus E$\\%
\hline%
(0, 1, 2)&{[}2, 1, 0{]}&3.8141&4.9613&$A_2$\\%
&{[}1, 2, 0{]}&3.8578&5.1619&$A_2$\\%
&{[}0, 2, 1{]}&3.9584& &$A_2$\\%
&{[}1, 1, 1{]}&4.0286&5.4677&$2 A_2$\\%
&{[}5, 0, 0{]}&4.1496&5.1073&$A_2$\\%
&{[}0, 5, 0{]}&4.3448& &$A_2$\\%
&{[}4, 1, 0{]}&4.488& &$A_2$\\%
\end{tabular}
\caption{\label{tab:kpp-ops1} 
{
$\pi^+\pi^+K^+$ operators with ${d}_{\rm ref}^{\,2}\leq 5$ used in this work, where $\bm{P} = (2\pi/L)\bm{d}_{\rm ref}$.
Notation as in \Cref{tab:kkp-ops1}, except that free energies quoted here are in units of $M_\pi$.
}
}
\end{table}

\begin{table}[!bp]
\centering
\begin{tabular}{c|c|cc|c}
\multirow{2}{*}{$\bm{d}_{\rm ref}$}&\multirow{2}{*}{$[d_{K}^2,d_{\pi_1}^2,d_{\pi_2}^2]$}&\multicolumn{2}{c}{$E^{\rm free}/M_{\pi}$}&\multirow{2}{*}{operators}\\%
&&N203&D200&\\%
\hline%
(1, 1, 2)&{[}3, 1, 0{]}&4.01&5.1463&$A_2$\\%
&{[}1, 3, 0{]}&4.0804&5.4753&$A_2$\\%
&{[}2, 2, 0{]}&4.1179&5.3981&$A_1 \oplus A_2$\\%
&{[}0, 3, 1{]}&4.1821& &$A_2$\\%
&{[}6, 0, 0{]}&4.2564&5.2097&$A_2$\\%
&{[}0, 2, 2{]}&4.2629& &$A_2$\\%
&{[}2, 1, 1{]}&4.2901& &$A_1 \oplus 2 A_2$\\%
&{[}1, 2, 1{]}&4.3334& &$2 A_1 \oplus 3 A_2$\\%
&{[}0, 6, 0{]}&4.4652& &$A_2$\\%
\hline%
(0, 2, 2)&{[}2, 2, 0{]}&3.7755&4.9662&$A_2$\\%
&{[}0, 2, 2{]}&3.9331& &$A_2$\\%
&{[}2, 1, 1{]}&3.9626&5.3016&$A_2$\\%
&{[}1, 2, 1{]}&4.0094&5.5141&$A_2 \oplus B_1$\\%
&{[}5, 1, 0{]}&4.3212&5.4563&$A_2 \oplus B_1$\\%
&{[}1, 5, 0{]}&4.4247& &$A_2 \oplus B_1$\\%
&{[}8, 0, 0{]}&4.4433&5.3928&$A_2$\\%
\hline%
(0, 0, 3)&{[}1, 1, 1{]}&3.29&4.5765&$A_2$\\%
&{[}4, 1, 0{]}&3.8389&4.8698&$A_2$\\%
&{[}1, 4, 0{]}&3.9356&5.3271&$A_2$\\%
&{[}0, 4, 1{]}&4.0469& &$A_2$\\%
&{[}2, 2, 1{]}&4.3044& &$A_2 \oplus B_2 \oplus E$\\%
&{[}1, 2, 2{]}&4.3508& &$A_2 \oplus B_2$\\%
&{[}9, 0, 0{]}& &5.4757&$A_2$\\%
\end{tabular}
\caption{\label{tab:kpp-ops2} 
{
$\pi^+\pi^+K^+$ operators with $6 \leq {d}_{\rm ref}^{\,2}\leq 9$ used in this work, 
where $\bm{P} = (2\pi/L)\bm{d}_{\rm ref}$.
Notation as in \Cref{tab:kpp-ops1}.
}
}
\end{table}